\documentclass[aps,prc,showpacs,amssymb,amsmath,amsfonts,nofootinbib,floatfix]{revtex4-2}
\usepackage{graphicx}

\makeatletter
\def\p@subsection{}
\def\p@subsubsection{}
\makeatother
\usepackage{natbib}
\usepackage[usenames]{color}
\usepackage{epsfig}
\usepackage{epstopdf}
\usepackage{amssymb,amsmath}
\usepackage{amsmath}
\usepackage{subfig}
\usepackage{epstopdf}
\usepackage{enumitem}
\usepackage{verbatim}
\usepackage[bookmarks={false}]{hyperref}
\usepackage{lineno,hyperref}
\usepackage{enumitem}
\usepackage{bbm}
\usepackage{overpic}

\definecolor{grey}{rgb}{0.9,0.9,0.9}
\definecolor{black}{rgb}{0,0,0}

\hypersetup{pdfstartview={XYZ null null 1.2}}

\def\thesection{\Roman{section}}
\def\thesubsection{\Roman{section}. \Alph{subsection}}
\def\thesubsubsection{ \textit{\textbf{\Roman{section}. \Alph{subsection}. \arabic{subsubsection}}}}

\newcommand{\be}{\begin{eqnarray}}
\newcommand{\ee}{\end{eqnarray}}
\newcommand{\bc}{\begin{center}}
\newcommand{\ec}{\end{center}}
\newcommand{\Ocal}{\mathcal{O}}

\begin{document}

\title{{Amplitude- and truncated partial-wave analyses combined: A novel, almost theory-independent single-channel method for extracting photoproduction multipoles directly from measured data}}

\email{Corresponding author: alfred.svarc@irb.hr}
\author{ A. \v{S}varc\,$^{1,2}$, Y. Wunderlich$\,^3$, and   L. Tiator$\,^4$   \vspace*{0.3cm}  }
\affiliation{$\,^1$ Rudjer Bo\v{s}kovi\'{c} Institute, Bijeni\v{c}ka cesta 54, P.O. Box 180, 10002 Zagreb, Croatia}
\affiliation{$\,^2$ Tesla Biotech, Mandlova 7,10000 Zagreb, Croatia}
\affiliation{$\,^3$ Helmholtz-Institut f\"{u}r Strahlen- und Kernphysik der Universit\"{a}t Bonn,  53115 Bonn, Germany}
\affiliation{$\,^4$ Institut f\"{u}r Kernphysik, Universit\"{a}t Mainz, D-55099 Mainz, Germany}


\date{\today }

\begin{abstract}
\vspace*{0.5cm}
Amplitude- and truncated partial-wave analyses are combined into a single procedure and a novel, almost theory-independent single-channel method for extracting multipoles directly from measured data is developed. In practice, we have created a two-step procedure which is fitted to the same data base: in the first step we perform an energy independent amplitude analysis where continuity is achieved by constraining the amplitude phase, and the result of this first step is then taken as a constraint for the second step where a constrained, energy independent, truncated partial-wave analysis is done. The  method is tested on the world collection of data for $\eta$ photoproduction, and the obtained fit-results are very good. The sensitivity to different possible choices of amplitude phase is investigated and it is demonstrated that the present data base is insensitive to notable phase changes, due to an incomplete database. New measurements are recommended to remedy the problem.
\end{abstract}

\maketitle

\section{Introduction} \label{sec:Introduction}
Finding a connection between QCD and experiment is a conditio sine qua non for establishing whether a particular {description of the effects of non-perturbative QCD} is close to being correct or not, and a lot of effort has in last decades been put into doing it via comparing resonance spectra. While on {the} QCD side, a resonance spectrum is standardly predicted by lattice QCD and various {QCD-inspired phenomenological models}, on the experimental side it is standardly extracted by identifying poles of the scattering matrix~\cite{pdg}.  However, as resonances/poles must have definite quantum numbers, finding pole structure of experimental data must necessarily go through {a} partial wave decomposition where  {the} angular dependence at a fixed energy is represented by a decomposition over the complete set of Legendre polynomials which then define proper  {eigen}values of  {the} angular momentum operator. Combining good quantum numbers of angular momenta with  {the} known  {spins} of the reacting particles, resonance quantum numbers are fully defined.  However,  {one should} be aware that observables  {which are measured} are most generally given in terms of amplitudes, and not partial waves, and to obtain partial waves {one has} to invest some extra work.  Unfortunately, in that process  {the} single-channel partial wave decomposition turned out to be rather non-unique. For decades, it has been known that in  {the} single-channel case, even a complete set of observables is invariant with respect to the phase rotation of  {all} reaction amplitudes  {by the same} arbitrary real function of energy and angle (continuum ambiguity) \cite{Atk73,Bow75,Atk85}, and this free rotation either causes  {a} rearrangement of strength between real and imaginary parts of amplitudes and partial waves for energy dependent  {phase-rotation} functions, or  {it} even mixes  {partial waves} for angular dependent {phase-rotation functions}. These effects lead to unacceptable discontinuities in amplitudes and partial waves, and have been extensively discussed in  {refs~\cite{Svarc2018,Wunderlich:2017dby}}. The main conclusion is that at least one of the reaction amplitude phases must be forced to be continuous in energy and angle in order to restore  {a} continuous, unique solution. The open question is how to {accomplish this task} with minimal model dependence.  With all these issues at hand, finding an optimal method for extracting partial waves with minimal reference to a particular theoretical model for fixing the phase  {turns} out to be of utmost importance.

 {A} direct consequence of  {the} continuum ambiguity is that  {an} unconstrained single-channel, single energy partial wave analysis (SE PWA), in the sense that there is absolutely no correlation among SE PWA solutions at neighboring energies, must be discontinuous. This is the consequence of the fact that if a phase is unspecified at an isolated energy, then the free fit chooses a random phase value  as there is  {an} infinite number of phases which give {an} absolutely identical set of observables. So, the variation of  {the} phase between neighboring energies may be random and discontinuous. If the variation of  {the} phase between neighboring energies is discontinuous,  {the} redistribution of strength between real and imaginary part at each energy will be random, so the partial wave must be discontinuous too. The standard way of achieving the continuity was to  {implement} it on the level of partial waves, so one {resolved to constrain} partial waves directly to values originating from some particular theoretical model. In this case {, the} model dependence is strong. First ideas to use more general principles of analyticity for imposing the continuity instead of referring to a particular model were introduced in  {the mid 1980s} by  {the} Karlsruhe-Helsinki group for {Pion-Nucleon ($\pi N$)} elastic scattering in the form of fixed-$t$ analyticity~\cite{Hoehler84}. In this case instead of demanding the proximity of fitted partial waves to some model values, the continuity is imposed on the level of reaction amplitudes by requiring fixed-$t$ analyticity. In other words, the group was fitting the world collection of data requiring that the reaction amplitude for a fixed-$t$ have  {a} certain analytic, hence continuous form. In this way{, any} specific dependence on a particular model was reduced to the level of discussing what is the correct analytic structure of {the} reaction amplitudes, and this form is fairly well defined by the branch-points of  {the} analyzed reaction. Unfortunately, imposing analyticity in  {the Mandelstam-$t$ variable} opened quite some additional issues, and this is extensively discussed in~\cite{Osmanovic2018}.

The aim of this  {work} is to show that invoking  {analyticity in the} Mandelstam-$t$ variable is not really needed and raises unwanted complications; the required continuity can be obtained by amplitude analysis (AA) in the Mandelstam-$s$ variable, and the continuity is imposed by requiring the proper analyticity of amplitude phases only. So, everything is done in the Mandelstam-s variable, and by this the analysis becomes much simpler. To our knowledge, this is the first time that amplitude analysis and truncated partial wave analysis are joined into one compact, self-sustained {analysis-scheme}.

Let us stress that the continuum ambiguity problem, and all problems of  {the} continuity of  {the} phase related to it are typical and inherent for single-channel analyses where unitarity is violated in the sense that there exists the loss of {probability-}flux into other channels. Unitarity  {equations} become inequalities, and the free phase arises. However, in full coupled-channel formalisms where unitarity is at the end restored by summing up the flux in all channels, the invariance to phase rotations disappears as the phase is fixed, and uniqueness is automatically restored. {However}, this {work} analyzes only  {the} single-channel case.

Let us also warn the reader about another aspect of PWA: the number of partial waves involved. As  {the} partial wave decomposition is an expansion over the complete set of Legendre polynomials, it is inherently infinite, but in practice it must be finite, so all we can talk about is a truncated PWA (TPWA). A lot of effort has recently been put into analyzing the features of  {a} TPWA  {for pseudoscalar meson photoproduction}~ {\cite{TPWAall,YannickPhD}}.  {A} theoretical model was chosen, all observables were generated from this model with a fixed  {angular-momentum cutoff} $\ell_{\text{max}}$, and a complete set of observables generated this way was taken as intput to TPWA. In that way the outcome of TPWA is known in advance, and a lot of conclusion on the symmetries and inter-relation among  {the thus} formed pseudo-observables have been drawn. Unfortunately, as this is  {an idealized} case, most of {these conclusions are not applicable for our practical purposes}. Those pseudo-observables generated {from a} model by default possess explicit properties like unrealistically high precision, continuity in energy and angle, various inter-dependence among observables due to finite truncation order,  etc., which our real data do not necessarily have. This is in particular pronounced if the truncation order is too low. Therefore, we have to be very careful in our analysis of real data to take the truncation order high enough to avoid introducing additional, nonexisting symmetries into the analysis which may raise quite some problems. If we are careful enough, our obtained partial waves are not exact, but indeed are a good representation of  {the} amplitude analysis representing the process.

The paper is organized as follows: the main text goes directly into media res by proposing the new fit-method and showing applications to polarization data in $\eta$ photoproduction in a detailed way. Discussions on the necessary background knowledge concerning the photoproduction formalism, as well as a more elaborate mathematical discussion on the motivation of the proposed novel analysis-scheme, have been relegated to the appendices. In this way, we can present our main results quickly  {and} concisely, while the interested reader can read the more elaborate mathematical discussion in parallel.

\section{The novel approach to single-channel PWA and application to $\eta$ photoproduction} \label{sec:ProposalAndResults}
The main intention of  {our proposed scheme} is to obtain a continuous set of partial waves, directly from experimental data, with minimal involvement of theoretical models.

The 0-th step of our procedure is to perform an unconstrained single energy partial wave analysis (SE PWA); namely to fit the available set of measured data with  {a} chosen number of partial waves at each available energy independently (at each energy the fit is independent of the neighboring energy).  We know that such a process, due to  {the} continuum ambiguities on the level of reaction amplitudes, must  {produce a set of partial waves that are discontinuous in energy}, even for {a} complete set of pseudo-data with very high precision. However, this procedure gives the best possible fit to the data with {the} chosen number of partial waves, and directly measures the consistency of  {the} data. So, this gives us {a} benchmark-set of  {values for the goodness-of-fit} parameter chi squared  {- which we call $ \chi^{2}_{\text{unc.}} (W)$ ('unc.' for 'unconstrained') -} and any method of enforcing continuity of partial waves must be as close as possible to this set, but can never be better. Achieving the continuity of partial waves is, however, a demanding task. For the case of very precise pseudo-data, it has been shown that  {the task} is  {still relatively} simple: it is enough to impose the continuity only in one amplitude phase to achieve the goal that the SE PWA becomes continuous~\cite{Svarc2018}. However, for the real data we unfortunately have a serious problem.  The existing set of observables is incomplete, and errors are realistic, so the situation changes drastically.  Simple methods of imposing continuity on one phase only do not work anymore.

The standard way to impose continuity in discontinuous SE PWA is the penalty function  {methodology}. The idea is to require that the solution one obtains by fitting the data at one isolated energy SIMULTANEOUSLY reproduces the data AND is also close to some continuous function. So, out of  {an} in principle infinite number of solutions at isolated energy for {the} unconstrained SE PWA one picks only those which are also close to  {a} predetermined continuous penalization factor. Of course, the solution will depend on the  {size of the} penalization coefficient. The smaller the coefficient, the more the solution will tend to reproduce the fitted data; it will be more discontinuous and it will less satisfy the penalization function. On the other hand, if one increases the penalization coefficient, the more the fit will reproduce the penalization function and be continuous, but it will less describe the fitted data. In the final limit of extremely low penalization coefficient, the fit will ideally describe the data and be discontinuous, and in the final limit of extremely big penalization coefficient the fit will be continuous, perfectly describe the penalization function, and definitely disagree with the fitted data. The optimum lies somewhere in-between.

\begin{figure}[h!]
\bc
\includegraphics[width=0.4\textwidth]{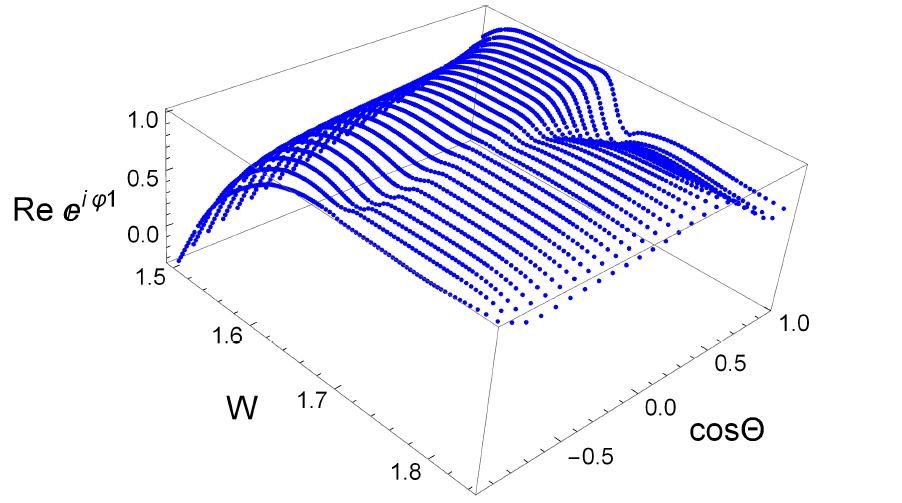}
\includegraphics[width=0.4\textwidth]{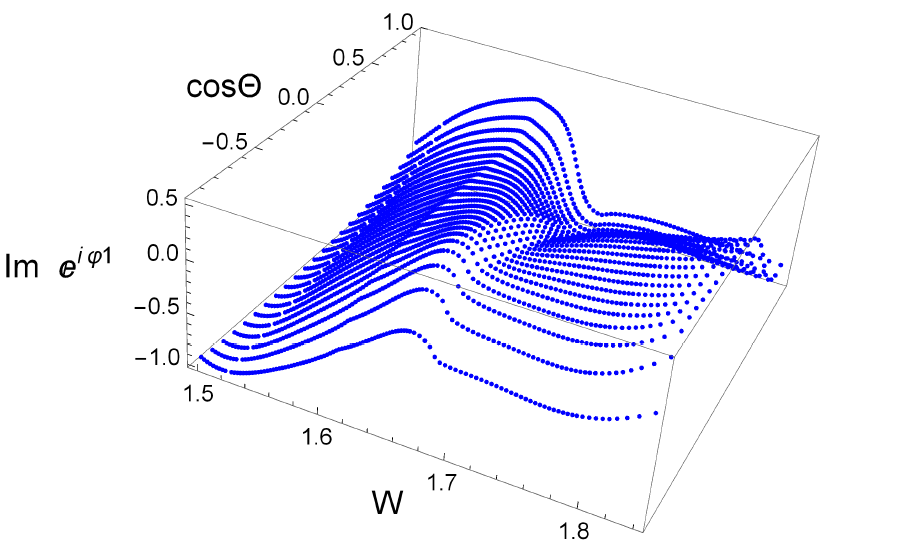} \\
\caption{\label{b1phase}(Color online)  {The normalized transversity amplitude (phase) $e^{i \varphi_{1}} := b_{1} / \left| b_{1} \right|$ } from  {the} BG2014-2 solution  {is shown}.     }
\ec
\end{figure}

The  {first,} most standard approach  {found in the literature} was to penalize partial waves. We require that {the} fitted partial waves reproduce the observable $\Ocal$ and are at the same time close to some partial waves taken from {a} theoretical model. So, for one observable we may at one fixed energy W write  {(for a literature-example of a penalization-scheme which acts on the level of partial waves, though not quite in the same way as in the definition given below, see for instance ref.~ {\cite{Workman:2011hi}})}:
\be
\label{Eq1}
\chi^2(W) & = & \sum_{i=1}^{N_{\text{data}}}w^i \left[ {\cal O}^{\text{exp.}}_i (W,\Theta_i) - {\cal O}^{\text{th.}}_i ({\cal M}^{\text{fit}}(W,\Theta_i)) \right]^2 + \lambda_{\text{pen.}} \sum_{i=1}^{N_{\text{data}}}\left| {\cal M}^{\text{fit}}(W,\Theta_i)- {\cal M}^{\text{th.}}(W,\Theta_i) \right|^2
\ee
where
\be
{\cal M} & \stackrel{\text{def.}}{=} & \left\{ {\cal M}_0, {\cal M}_1, {\cal M}_2, ..., {\cal M}_j \right\} \nonumber
\ee
is the generic notation for the  {set of} all multipoles, $w_i$   is the statistical weight and $j$ is the number of  partial waves (multipoles). Here,  {$\mathcal{M}^{\text{fit}}$}  are fitting parameters and {$\mathcal{M}^{\text{th.}}$} are continuous functions taken from a particular theoretical model.
\\ \\ \noindent
In this case the procedure is strongly model-dependent.
\\ \\
 {A} possibility to make the penalization function independent of a particular model was first formulated in {the} Karlsruhe-Helsinki (KH) $\pi$N{-}elastic PWA  by G. H\"{o}hler  {and collaborators in the mid 1980s}~\cite{Hoehler84}.  Instead of using partial waves which are inherently model dependent, the penalization function was chosen to be constructed from reaction amplitudes which can be in principle directly linked to experimental data with only analyticity requirements  {imposed} in  {the} amplitude reconstruction procedure. So, the equation  {\eqref{Eq1}} was changed to:

\be
\label{Eq2}
\chi^2(W) & = &  \chi^{2}_{\text{data}} (W) + \chi^{2}_{\text{pen}} (W) \nonumber \\
 \chi^{2}_{\text{data}} (W) & = &  \sum_{i=1}^{N_{\text{data}}}w^i \left[ {\cal O}^{\text{exp.}}_i (W,\Theta_i) - {\cal O}^{\text{th.}}_i ({\cal M}^{\text{fit}}(W,\Theta_i)) \right]^2  \nonumber \\
 \chi^{2}_{\text{pen}} (W) & = &  \lambda_{\text{pen.}} \sum_{i=1}^{N_{\text{data}}} \sum_{k=1}^{N_{\text{amp}}}\left| {\cal A}_k({\cal M}^{\text{fit}}(W,\Theta_i))- {\cal A}_k^{\text{pen.}}(W,\Theta_i) \right|^2
\ee

where ${\cal A}_k$  is the generic name for any  {kind} of reaction amplitudes (invariant, helicity, transversity…) {. The amplitudes ${\cal A}_k({\cal M}^{\text{fit}}(W,\Theta_i))$ are} discontinuous ones obtained from fitted multipoles, and {the amplitudes} ${\cal A}_k^{\text{pen.}}(W,\Theta_i)$   {are} continuous ones obtained in  {the} penalization procedure. In this way, one is now responding to two challenges: to get reaction amplitudes which fit the data, and also to make them continuous. In Karlsruhe-Helsinki case{, this} was accomplished by implementing fixed-$t$ analyticity and fitting the data base for fixed-$t$ with reaction amplitudes whose analyticity is achieved by using  {the} Pietarinen expansion, and using the obtained, continuous reaction amplitudes as penalization functions  ${\cal A}_k^{\text{pen.}}(W,\Theta_i)$.\footnote{In their case they have chosen to use invariant amplitudes.}  So, the first step of the KH fixed-$t$ approach was to create the data base ${\cal O}(W)|_{t=\text{fixed}}$  using the measured  {data} base $ {\cal O}(\cos \, \theta)|_{W=\text{fixed}}$, and then to fit them with  {a} manifestly analytic representation of reaction amplitudes for a fixed-$t$. Then, the second step was to perform a penalized PWA defined by Eq.~ {\eqref{Eq2}} in a fixed-$W$ representation where the penalizing function  ${\cal A}_k^{\text{pen.}}(W,\Theta_i) $  was obtained in the first step in a fixed-$t$ representation.  In that way a stabilized SE PWA was performed.
\\ \\ \indent
This approach  was revived very recently for SE PWA of $\eta$ photoproduction by  {the} Main-Tuzla-Zagreb collaboration~\cite{Osmanovic2018}, and analyzed in details. The  {basic result} of that paper is that this fixed-$t$ method works very reliably, but is rather complicated. First it required the creation of a completely new data base   ${\cal O}(W)|_{t=\text{fixed}}$  from the measured data base $ {\cal O}(\cos \, \theta)|_{W=\text{fixed}}$, which introduced  {a} certain model dependence connected with  {the} interpolation, and second it involved quite some problems with the importance of  {the} unphysical regions.
\\ \\ \noindent
Therefore, we propose  {an} alternative.
\\ \\ \indent
We also use Eq.~ {\eqref{Eq2}}, but the penalizing  {function} ${\cal A}_k^{\text{pen.}}(W,\Theta_i) $ is  generated by the amplitude analysis in the same, fixed-$W$ representation, and not in the fixed-$t$ one. This simplifies the procedure significantly, and avoids quite some theoretical assumptions on the behaviour in the fixed-$t$ representation.
\\ \\ \noindent
We also propose a 2-step process:
\begin{itemize}[leftmargin=2.cm]
  \item[\emph{Step 1:}] \hspace*{1.cm} \\
    {Complete experiment analysis/amplitude analysis (CEA/AA)} of experimental data in  {a} fixed- {$W$} {representation} to generate  {the} penalizing  {function}  ${\cal A}_k^{\text{pen.}}(W,\Theta_i) $
  \item[\emph{Step 2:}] \hspace*{1.cm} \\
   Penalized TPWA{, using} Eq.(2) with the penalty function from
         \emph{Step 1}.
\end{itemize}
 {One has to observe} one very important fact: \\ \\ \indent
The "main event" happens in \emph{Step 2};  \emph{Step 1} serves only to impose continuity of \emph{Step 2}.  Therefore, the reaction amplitudes obtained in \emph{Step 1} need not absolutely reproduce the data, it is important that they are close to the experiment, and that they are continuous. The best agreement is then achieved in \emph{Step 2}. Of course, finding the optimal value of  {the} penalization coefficient $\lambda_{\text{pen.}}$ is of utter importance.
\\ \\
\noindent
\underline{\emph{\textbf{Step 1:  CEA/AA}}} \\ \\ \indent
In Appendix~\ref{sec:PhotoFormalism}, we give the formalism of  {pseudoscalar meson} photoproduction, and in Appendix~\ref{sec:CEAandTPWAandNewMethod} we discuss the details of CEA/AA. Out of detailed presentation of the problem we stress the most important  {fact}: that  {the} unconstrained CEA/AA is non-unique and discontinuous because of {the} continuum ambiguity. In this step, therefore, we have to achieve two  {goals}: to find the amplitudes which achieve the best possible agreement with the data and are continuous at the same time. As we do not have a complete set of data of infinite precision at our disposal, it is by definition impossible to obtain the unique solution. We can only get the solution with errors generated by experiment; in other words these errors originate only in the uncertainty of data, and not in continuum ambiguity effects.  $Step \, 1$ is also a two-part procedure.  The best agreement with the data is achieved in  {the} first part, and the continuity is imposed in the second part.
 \\ \\ \noindent
 We adopt the following strategy:
\begin{itemize}[leftmargin=2.cm]
\item[a.] The best agreement with experiment is achieved in the unconstrained fit of the absolute values of all reaction amplitudes to the data base  (observe that in fitting absolute values we do not have any phases involved)
\item[b.] The continuity of  {the} solution is achieved by fixing the phases of all reaction amplitudes to the analytic phase of our choice.
\end{itemize}
Observe that \emph{Step 1} is because of part  {'b.'} a model dependent step, but this model dependence will be additionally reduced in \emph{Step 2}. Namely, in \emph{Step 2} we fit the data with partial waves directly, so {the} phases of all reaction amplitudes are changed correcting the fact that the penalization which gives continuity is model dependent. We may safely say that the approach which is proposed here forces the phase of the final solution to be in-between  {the} exact and  {the} penalizing solution and  {to} be continuous at the same time. The situation may further improve by iteration, i.e. to repeat \emph{Step 1} with input from \emph{Step 2} as it was done in KH approach for fixed-$t$, but the final result so far does not require that.
\\ \\ \noindent
\textbf{a. Obtaining absolute values}
\\ \\ \indent
For obtaining absolute values it is extremely useful to use  {the} TRANSVERSITY REPRESENTATION. Namely, in {the} transversity representation for  {$\eta$ photoproduction,} all four absolute values  {$\left|b_{i}\right|$} are determined by  {a set of} four observables  {given by the unpolarized differential cross section} $d \sigma/d \Omega \stackrel{def}{=}\sigma_0$,  {the beam asymmetry} $\Sigma$,  {the target-asymmetry} $T$, and  {the recoil-polarization asymmetry} $P$  {(cf. Table~\ref{tab:PhotoproductionObservables} in appendix~\ref{sec:PhotoFormalism})}:
\be
\sigma_0 & = & \frac{\rho}{2}\left( |b_1|^2+ |b_2|^2+|b_3|^2+|b_4|^2 ) \right) \nonumber \\
\hat{\Sigma}  & = & \frac{ {\rho}}{2}\left( -|b_1|^2 - |b_2|^2+|b_3|^2+|b_4|^2 ) \right) \nonumber \\
\hat{T}  & = & \frac{ {\rho}}{2}\left( |b_1|^2 - |b_2|^2-|b_3|^2+|b_4|^2 ) \right) \nonumber \\
\hat{P}  & = & \frac{ {\rho}}{2}\left( -|b_1|^2 + |b_2|^2-|b_3|^2+|b_4|^2 ) \right)
\ee
where  {${\cal O} \, \sigma_0 \stackrel{def}{=} \hat{\cal O}$} and  $\rho$ is defined in appendices. Therefore, having all four observables with sufficient precision and in adequate number of angular points would enable us, up to discrete ambiguities, the unique extraction of  {the} absolute values  {$\left| b_{i} \right|$} in SE PWA. By adequate programming (taking into account similarity of solutions at neighboring energies, one can eliminate discontinuities due to discrete ambiguities. All remaining discontinuities will be of experimental origin.
\\ \\ \noindent
\textbf{b. Determining phases}
\\ \\ \indent
Up to this moment, our model is completely energy and angle independent, and depends only on experimental data. However, results are still not continuous. Introducing analytic phases in this step produces continuity. For  {a} single pole amplitude, the phase is smooth (in the vicinity of the pole the phase just quickly transverses through $\pi/2$ without producing any pronounced structure), but already two poles in the analyzed range produce phase interference, so some structures may be formed. In addition, we know that in our energy range of interest, some threshold openings are present, and this will also produce  {a rather un-smooth} phase behaviour. So, at first it looks  {as if} the phase may behave rather violently. On the other hand, we are not discussing partial waves where poles are directly visible, but reaction amplitudes which are a sum over all known resonances and thresholds, so the effect is smeared out. So, to the best of our knowledge, we can only say that phases are smooth and analytic. At this moment there is very little choice but to take the phase from a theoretical model. We have chosen  {the Bonn-Gatchina model solution}  BG2014-02~\cite{BoGa,BoGaweb}).
\\ \\ \indent
These phases are smooth, and do contain some structures which are introduced by the model. Just as an illustration, {we show in Fig.~\ref{b1phase}} the phase of  {the} $b_{1}$ amplitude. Other phases are very similar.
\\ \\ \noindent
So, our first solution $Sol \, 1$ is obtained by using  {the theoretical BG2014-02 phases} directly.
\\ \\ \noindent
Unfortunately, we do not know how strong and model dependent this assumption is.
\\ \\ \noindent
As  {the} single-spin observables $d\sigma/d\Omega$, $\Sigma$, $T$ and $P$ are phase independent, and only {the double-polarization observables of type} beam-target- ($\mathcal{BT}$), beam-recoil- ($\mathcal{BR}$) and target-recoil- ($\mathcal{TR}$) are, we may hope that this dependence is weak. Using the fact that phases are analytic function offers us the possibility to test the size of this dependence.
\\ \\ \noindent
First, we confirm that all transversity amplitude phases indeed are analytic functions. To do  {this}, we fit all four phases with  {a $2$-dimensional} Pietarinen expansion,  {a} method which has not been formulated up to now. The method is {based} on Pietarinen expansion technique \cite{Svarc2013}, but extended to two variables: energy and angle (more precisely cosine of the angle  {$x := \cos \theta$}). Namely, in energy dimension we use standard Pietarinen expansion, but each of {the} coefficients  {also} depends on  {the} angle. Similar as for  {the} energy part, for  {the} angular dependence we also assume the expansion over a complete set of functions, in this case we use Legendre polynomials. So, one gets an analytic function which  {is} analytic in energy and angle  {and} with the analyticity we control.
\be \label{Eq:2D-Pietarinen}
PT(W,\theta) & = & \frac{\sum_{k=0}^{N}c_k(x)Z(W)^k}{\left|\sum_{k=0}^{N}c_k(x)Z(W)^k \right| } \nonumber \\
Z(W) & = & \frac{\alpha - \sqrt{W_0-W}}{\alpha + \sqrt{W_0-W}} \nonumber \\
c_k(x) & = & \sum_{l=0}^{M} c_{k,l}P_l(x) \nonumber \\
x & = & \cos \, \theta
\ee
 {$M$} and  {$N$} are small numbers (angular momentum index M is always around 3, and energy index can vary from 4 for very simple energy analyticity to 20 for a fairly complicated one), $P_l(x)$ are Legendre polynomials and $\alpha$ is {the} Pietarinen range parameter.  Then we make a  {2-dimensional} fit to the four normalized transversity amplitudes and get {the} coefficients $\alpha$, $W_0$, and  $c_{k,l}$  for all four absolute values.
\\ \\ \noindent
If we are able to fit the phases with such an expansion,  {the} phases  {have to be} analytic functions.
\\ \\ \noindent
We use  {a} very simple Pietarinen expansion, with only one branch-point at  {the} $\eta$ photoproduction {threshold} and only 4 terms in  {the} angular expansion. However, we see that the analytic structure of the fitted phase is rather complicated in energy, and we need as much as $N=20$ energy terms to obtain a decent fit. The result is {shown} in Fig.~\ref{b1fitted}
\\ \\ \noindent
We again show the result only for $b_{1}$ amplitude:
\begin{figure}[h!]
\bc
\includegraphics[width=0.4\textwidth]{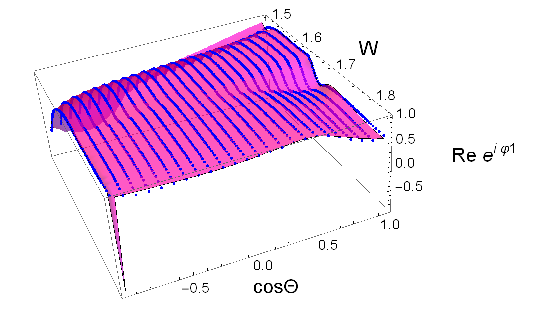}
\includegraphics[width=0.4\textwidth]{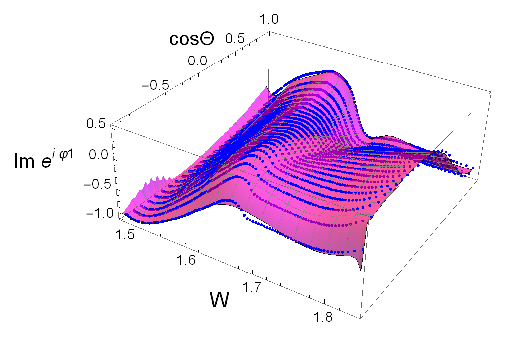}  \\
\caption{\label{b1fitted}(Color online)  {The normalized transversity amplitude (phase) $e^{i \varphi_{1}}$} from {the BG2014-02} solution (Discrete symbols) and  {a 2-dimensional} Pietarinen fit (2D plane)  {are shown}.     }
\ec
\end{figure}
\\ \noindent
The analyticity of  {the phases of} $b_{1}-b_{4}$ offers us the possibility of testing the sensitivity of our method to the phase. Instead of  {the} very physical phases $b_{1}-b_{4}$ of  {the} BG2014-2 model and illustrated for $b_{1}$ in Figs.(\ref{b1phase},\ref{b1fitted}), we shall use a phase with much simpler analyticity, and which is generated by a 2D fit to  {the BG2014-02} phases with  {$N=4$} terms only! This phase is again  {shown} as an illustration for {the} $b_{1}$ amplitude in Fig.~\ref{b1smoothphase}.
\begin{figure}[h!]
\bc
\includegraphics[width=0.4\textwidth]{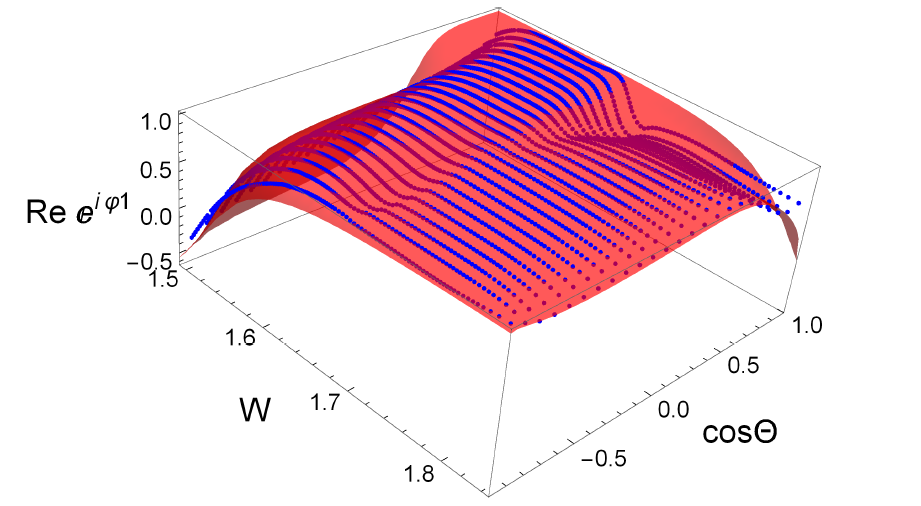}
\includegraphics[width=0.4\textwidth]{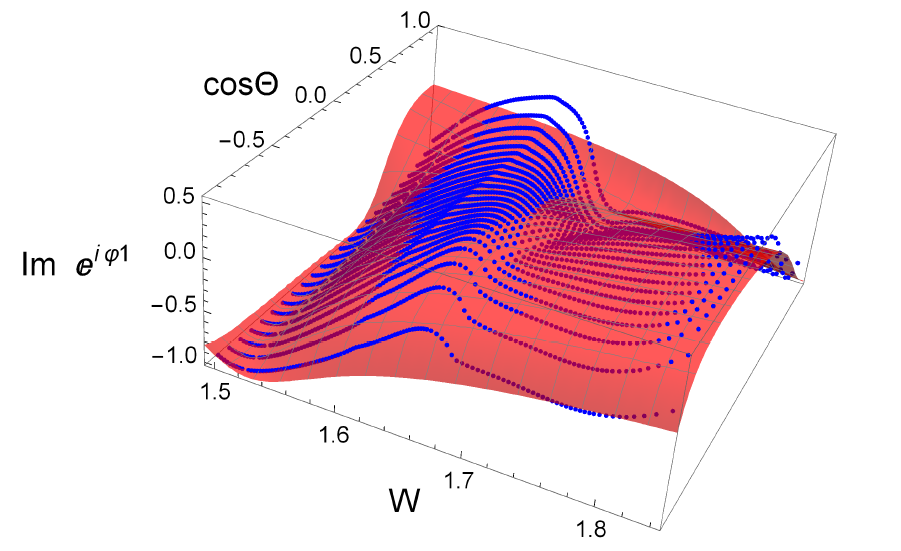}  \\
\caption{\label{b1smoothphase}(Color online)  {The normalized transversity amplitude (phase) $e^{i \varphi_{1}}$ } from {the} smoothed BG2014-02 solution  {is shown}.     }
\ec
\end{figure}
\\ \\  \noindent
So, our second solution $Sol \, 2$ is obtained by using  {the smoothed theoretical BG2014-02} phases.
\\ \\ \noindent
We stress that this is a very unphysical phase as despite being anticipated no structures are allowed, so the amount of dissimilarity between $Sol \, 1$ and $Sol \, 2$ will  {give} us the maximum model dependence of our assumption. It is clear that all structures in  {the} phases in $Sol \, 2$ are eliminated, the obtained functions are smooth, and {represent} the best fit to  {the BG2014-02} phases.

In this way, we complete the \emph{Step 1} by using  {the} original and smoothed analytic phases which are generated by the phases from  {the theoretical} Bonn-Gatchina model (used phases are the best fit of  {BG2014-02} input with 2-D Pietarinen expansion given in Eq.~\ref{Eq:2D-Pietarinen}).
\\ \\ \\ \noindent
\underline{\emph{\textbf{Step 2 : TPWA}}}
\\ \\ \indent
We perform  {a} standard penalized TPWA defined by Eq.~ {\eqref{Eq2}} with $\ell_{\text{max}}=5$. The only issue is finding {an} optimal  {value for the} penalty-function coefficient  {$\lambda_{\text{pen.}}$}.  {This issue will be discussed further below.}
\\ \\ \noindent
We, however, have to discuss two features of TPWA:  {the} threshold behaviour and  {the} data base.
\\ \\ \noindent
\textbf{\emph{\underline{Threshold behaviour:}}} \\ \\ \indent
We know that in the vicinity of  {a} threshold, partial waves have to behave like $q(W)^{L}$  where $q(W)$ is the {absolute value of the meson's} cm momentum, and in our procedure that has not been enforced up to now in any way. $Step \, 1$ is an unconstrained fit  {as far as the} absolute values  {$\left| b_{i} \right|$} are concerned, so no restrictions are coming through the penalty function. The  TPWA itself also does not require that our result obeys that rule. So, we have to impose that  {threshold-}behaviour somehow.
\\ \\ \noindent
 {A} very natural way to do it is, again, via penalty function technique, and we follow the method recommended by {the} KH group in ref.~\cite{Hoehler84}.
\\ \\ \noindent
The logic is the following: we add {another} penalty function to our total $\chi^2 (W)$, which is to be minimized:
\be \label{Eq:threshold}
\chi^2_{\text{thr.}}(W)    & = & \lambda_{\text{thr.}} \sum_{l=1}^{\ell_{\text{max}}} |{\cal M}_{l \pm}|^2 F_{\text{thr.}}(W,b,l)^{2 \cdot l} \nonumber \\
              F_{\text{thr.}}  (W,b,l) &  =  & \tfrac{  b \cdot l}{ q(W)} \, \cdot \, e^{-\tfrac{ q(W)}{ 0.1 \, b} }  \nonumber \\
              {\cal M}_{l \pm}   & = & \left\{ E_{ l \pm}, M_{ l \pm} \right\}
\ee
where $F_{\text{thr.}}  (W,b,l)$ is a phenomenological function instead of  {the} theoretical function $R_2^{2l}$ of ref.~\cite{Hoehler84}, which is connected with convergence radius of the PWA expansion. We have used this function for the value $b=m_N$ where $m_N$ is  {the} nucleon mass, and $\lambda_{\text{thr.}} = 2$. The function  $F_{\text{thr.}}  (W,b,l)^{2\cdot l}$  behaves like $q(W)^{-2l}$ for small $q(W)$, and vanishes for big $q(W)$, so  {it} scales down all multipoles  {with} low $q(W)$, and leaves  {those} unchanged  {that have a big $q(W)$}. In  this way {, the} $q(W)^{l}$ power law is automatically enforced for low $q(W)$.
\\ \\ \noindent
\textbf{\emph{\underline{Data base:}}} \\ \\ \indent
The data selection is particularly important as we want to be as  {close} as possible to  {a} complete set of observables. At this moment {,} we take all available measured data, and we take them without any renormalization, exactly as they are published. In Table~\ref{tab:expdata} we give our data base.
\begin{table*}[htb]
\begin{center}
\caption{\label{tab:expdata} Experimental data from A2@MAMI,
GRAAL, and  CBELSA/TAPS  used in our PWA. Data from  CBELSA/TAPS are taken at the center of the energy bin.}
\bigskip
\begin{ruledtabular}
\begin{tabular}{ccccccc}
 Obs.        & $N$ & $E_{lab}$~[MeV] & $N_E$  & $\theta_{cm}$~[deg] & $N_\theta$ & Reference    \\
\hline
 $\sigma_0$ & $2400$ & $710 - 1395$ & $120$  & $18 - 162$ & $20$ & A2@MAMI(2010)~\cite{McNicoll:2010qk} \\
 $\Sigma$   & $ 150$ & $724 - 1472$ & $ 15$  & $40 - 160$ & $10$ & GRAAL(2007)~\cite{Bartalini:2007fg} \\
 $T$        & $ 144$ & $725 - 1350$ & $ 12$  & $24 - 156$ & $12$ & A2@MAMI(2016)~\cite{Annand:2016ppc} \\
 $F$        & $ 144$ & $725 - 1350$ & $ 12$  & $24 - 156$ & $12$ & A2@MAMI(2016)~\cite{Annand:2016ppc} \\
 $E$        & $ 64$  & $750 - 1450$ & $ 8 $  & $29 - 151$ & $8$ & CBELSA/TAPS(2020)~\cite{Muller:2020plb} \\
 $P$        & $ 66$  & $725 - 908 $ & $ 6 $  & $41 - 156$ & $11$ &  CBELSA/TAPS(2020)~\cite{Muller:2020plb} \\
 $G$        & $ 48$  & $750 - 1250$ & $ 6 $  & $48 - 153$ & $8 $ &  CBELSA/TAPS(2020)~\cite{Muller:2020plb} \\
 $H$        & $ 66 $ & $725 - 908 $ & $ 6 $  & $41 - 156$ & $11$ &  CBELSA/TAPS(2020)~\cite{Muller:2020plb} \\
\end{tabular}
\end{ruledtabular}
\end{center}
\end{table*}

This  {selected} set of data is somewhat specific, and deserves our special attention.  The set is dominated by {the} very dense and very precise  {$\sigma_0$-data from A2@MAMI}, while other spin observables are measured only  at 6-15 energies, and much less angles. So the question arises how these sparse spin data will be combined with very precise results on $\sigma_0$. We shall solve  {this problem via} interpolation.
\noindent
We generate two sets of interpolated data:
\\ \\ \noindent
\emph{\underline{Set 1}}: \\ \indent
We use all $\sigma_0$  data, and all spin data are interpolated. So the whole minimization is performed on {a} set which consists of $\sigma_0$  data + observables interpolated at energies and angles where $\sigma_0$ is measured. These data are marked light grey.
Observe that all data are very dense, but in practice the only factually measured data are  {the} $\sigma_0$  values, all other data are obtained by interpolation from  {the} measured values. This set of data is somewhat model dependent, and serves only as an indication. This set will be used in \emph{Step 1} .
\\ \\ \noindent
\emph{\underline{Set 2}}: \\ \indent
We use only part of  {the} $\sigma_0$  data at energies where at least one additional spin observables is exactly measured.  This set is not so dense in energy, but  {the} model dependence is reduced.  We denote the results corresponding to this set with red discrete symbols. This set will be used in \emph{Step 2} .

\section{Results and Discussion} \label{sec:Results}
First we made an unconstrained fit to produce the bench-mark  {$\chi^{2}_{\text{unc.}}(W)$} function which represents the lowest possible $\chi^2$ value for any PW fit, and consequently indicates how consistent the data base actually is. However, let us remind the reader that  the resulting partial waves for such a fit are random and discontinuous. {Then, we performed two fits using the 2-step analysis-scheme introduced above, with the results being denoted as $Sol \, 1$ and $Sol \, 2$ according to the corresponding sets of $b_{i}$-phases defined in section~\ref{sec:ProposalAndResults}. For both solutions, we adjusted the following value for the penalty coefficient: $\lambda_{\text{pen.}} = 10$. This value represents the lower boundary of the following roughly estimated optimal window of penalty-coefficient values around $\lambda_{\text{pen.}} \simeq 10, \ldots, 50$. This window of values has been determined by 'sweetspot'-fitting techniques similar in spirit, but not exactly equal, to those proposed in sections~II and~III of reference \cite{Landay:2016cjw}. The coefficient for the threshold-penalty~\eqref{Eq:threshold} was set to $\lambda_{\text{thr.}} = 2$.
 \newpage
  At this moment it is essential to show the difference among  {$\chi_{\text{unc.}}^2/ \text{ndf}$} for {the} unconstrained solution,  {as well as $\chi^{2}_{\text{data}} / \text{ndf}$ for} $Sol \, 1$ and $Sol \, 2$ ($ndf$ is number of degrees of freedom). We show it in~Fig.\ref{Chi:comparison}.
\begin{figure*}[h!]
\bc
\includegraphics[width=0.9\textwidth]{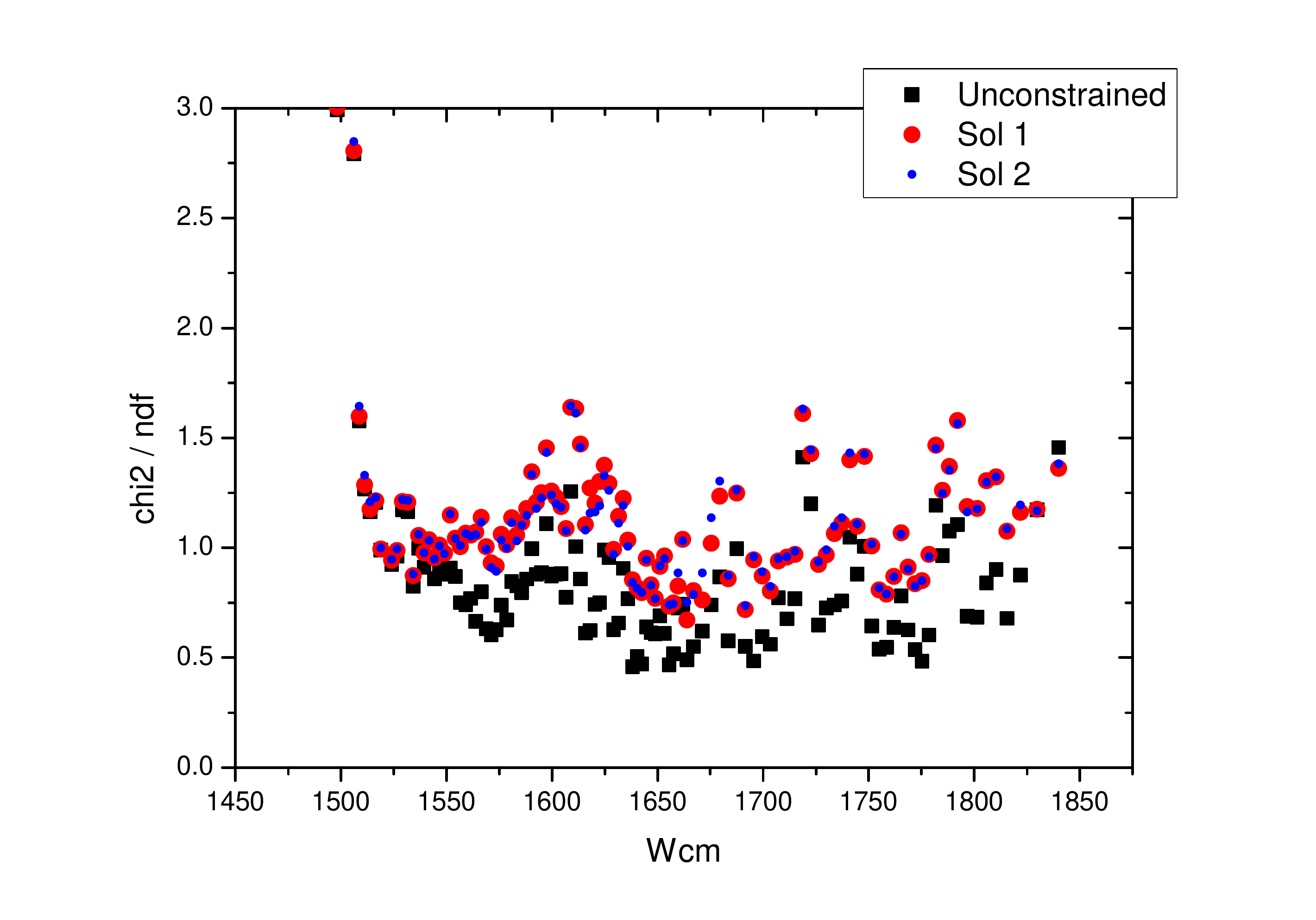}
\caption{\label{Chi:comparison}(Color online)  {A comparison} of  {$\chi_{\text{unc.}}^2/ \text{ndf}$} for {the} unconstrained solution,  {as well as $\chi^{2}_{\text{data}} / \text{ndf}$ for} $Sol \, 1$ and $Sol \, 2$ at all measured energies {, is shown.}  }
\ec
\end{figure*}

The {$\chi_{\text{unc.}}^2/ \text{ndf}$ from the unconstrained solution has by far the smallest values}, however multipoles for this solution are discontinuous. The {$\chi_{\text{data}}^2/ \text{ndf}$} for both solutions $Sol \, 1$ and $Sol \, 2$ is somewhat worse, but still very good.
\\ \\ \noindent
However, {the $\chi_{\text{data}}^2/ \text{ndf}$} for $Sol \, 1$ and $Sol \, 2$, solutions with two different phases,  is barely distinguishable! $\chi^2/ndf$ for  $Sol \, 1$, the solution with phase directly taken over from a very good, multichannel ED model BG2014-2, is systematically better than $\chi^2/ndf$ for  $Sol \, 2$ where the phase is ad hoc smoothed\footnote{There are several points where this is not the case, but this is just the reflection of the fact that the phase from BG2014-2 model is still only a model and not a genuine phase, so there is a possibility that smoothed phase is accidentally better.}. Somewhat more pronounced differences can be seen in the energy range 1600 MeV $\leq$ W $\leq$ 1700 MeV, but this is exactly the area of numerous threshold openings ($K \Lambda$, $K \Sigma$,...) where the phase is expected to have notable structure.
\\ \\ \noindent
Therefore, we in Figs.~\ref{Multipoles:Sol1} and \ref{Multipoles:Sol2} show the lowest multipoles for \emph{Sol 1} and \emph{Sol 2}  and corresponding predictions of BG2014-2 theoretical solution.
\\ \\ \noindent
As it was to be expected, differences are noticeable, but not big. In spite of small differences in $\chi^2/ndf$ for  $Sol \, 1$ and  $Sol \, 2$, the obtained multipoles are not identical.

\clearpage
\underline{\textbf{$Sol \, 1$, solution with BG 2014-2 phase.}}

\begin{figure}[h!]
\bc
\includegraphics[width=0.38\textwidth]{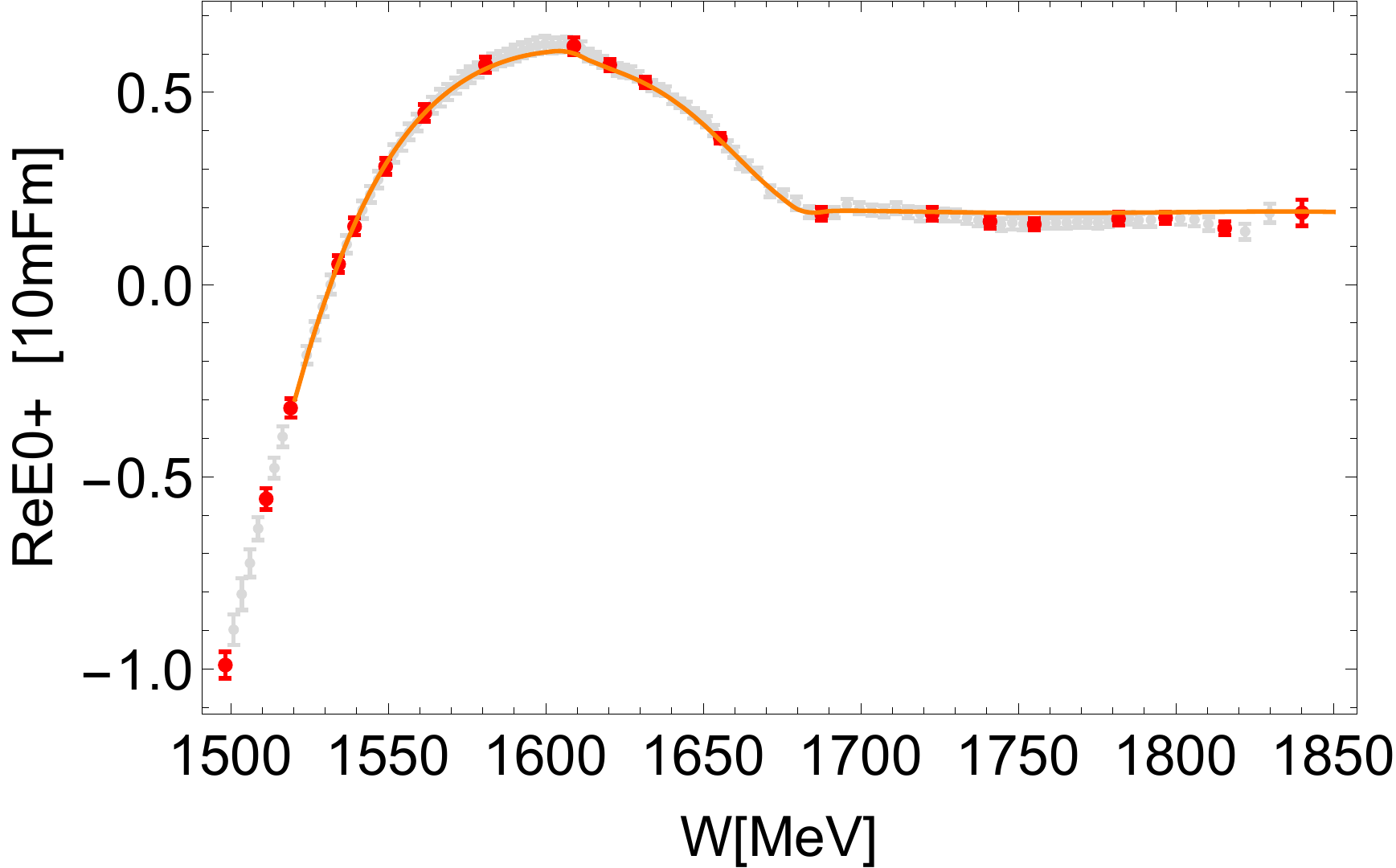} \hspace{0.5cm}
\includegraphics[width=0.38\textwidth]{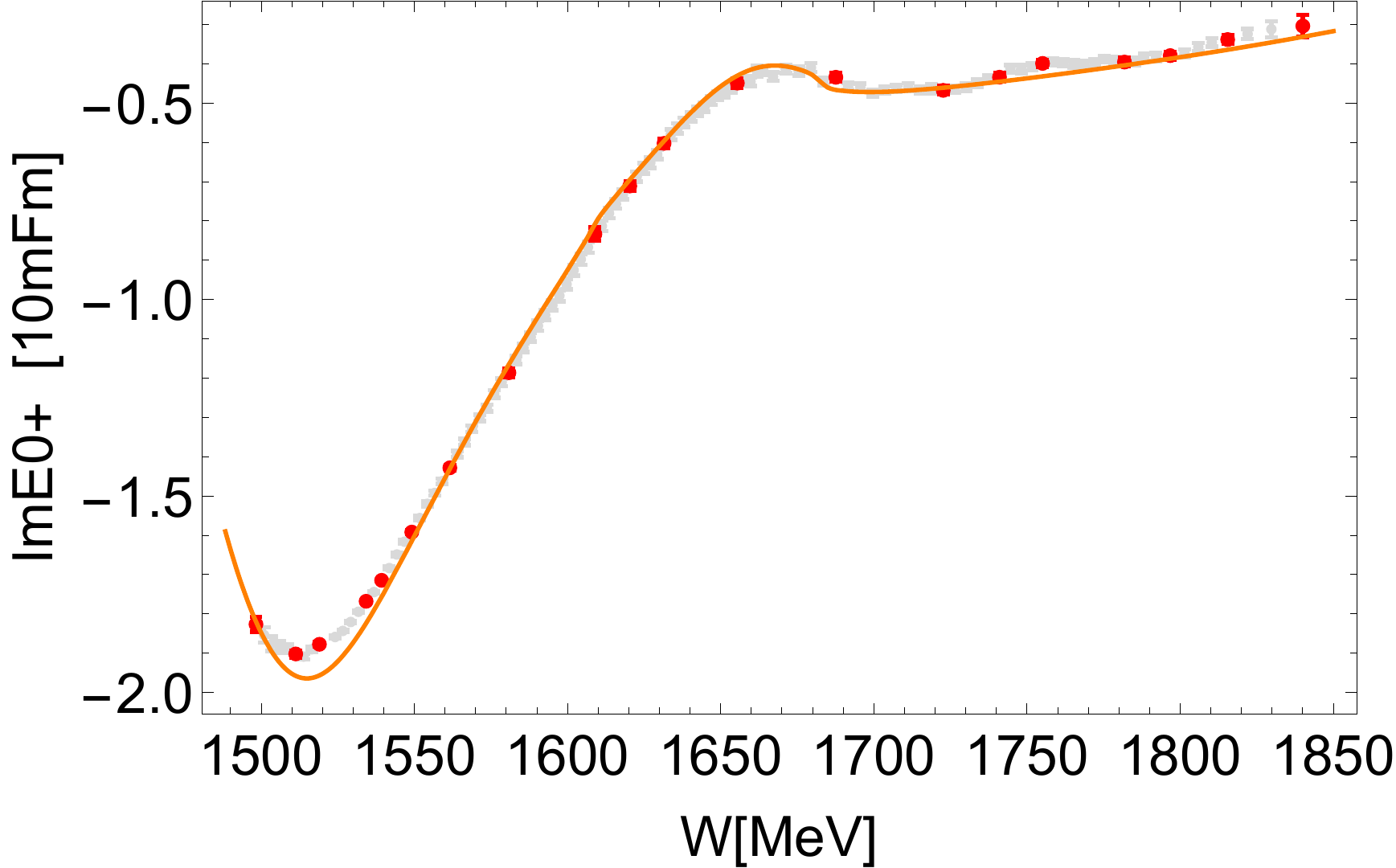}  \\
\includegraphics[width=0.38\textwidth]{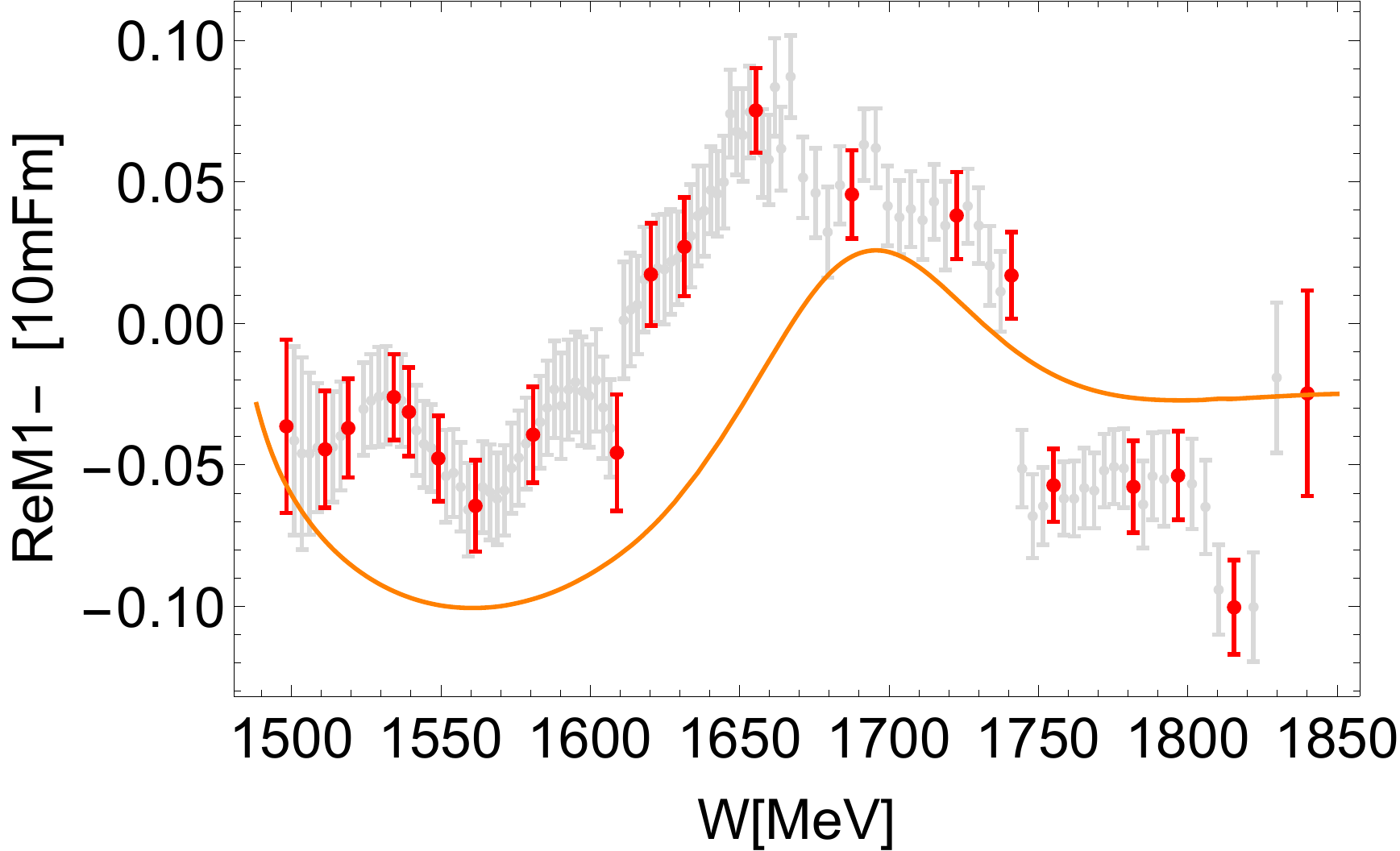} \hspace{0.5cm}
\includegraphics[width=0.38\textwidth]{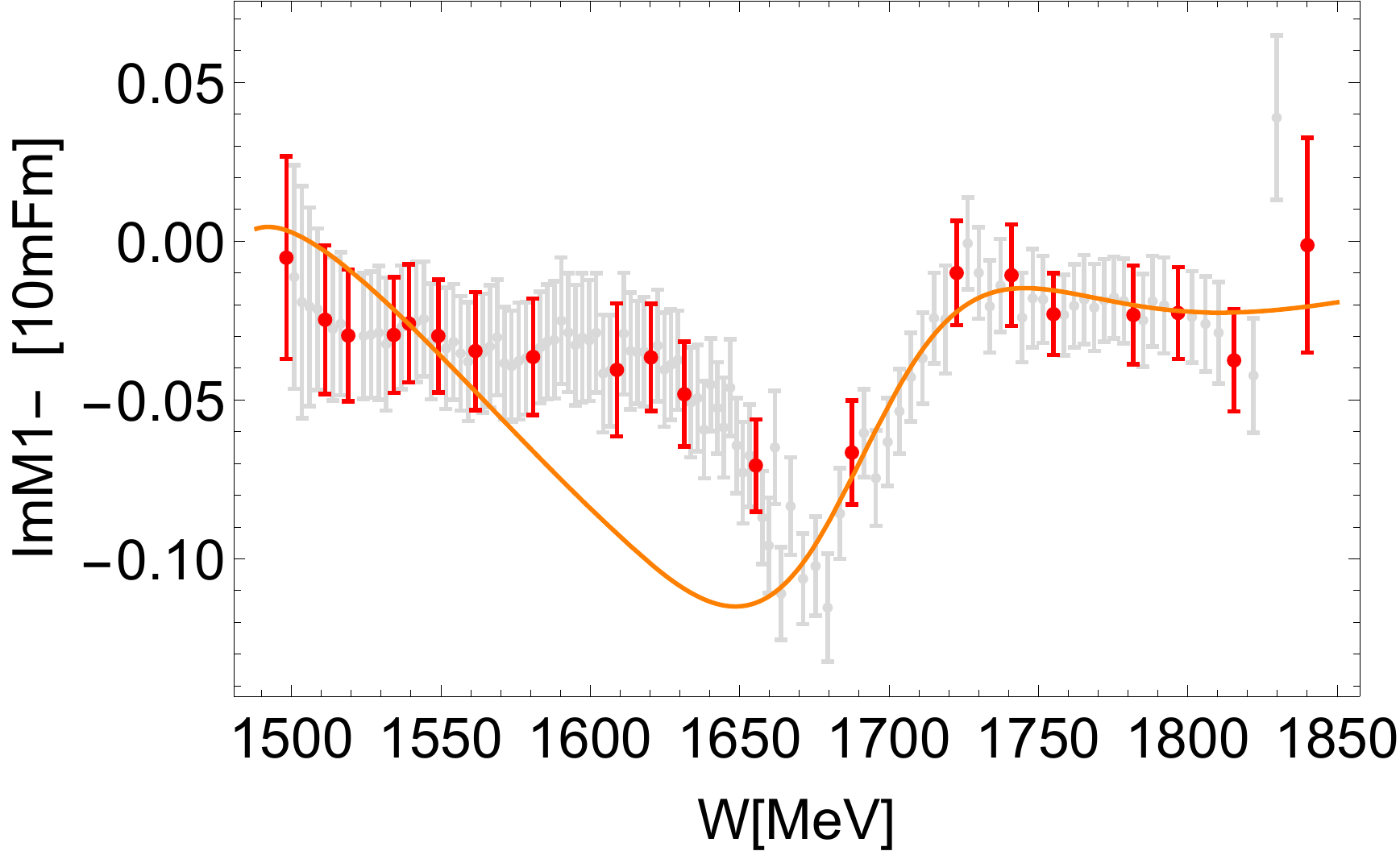}  \\
\includegraphics[width=0.38\textwidth]{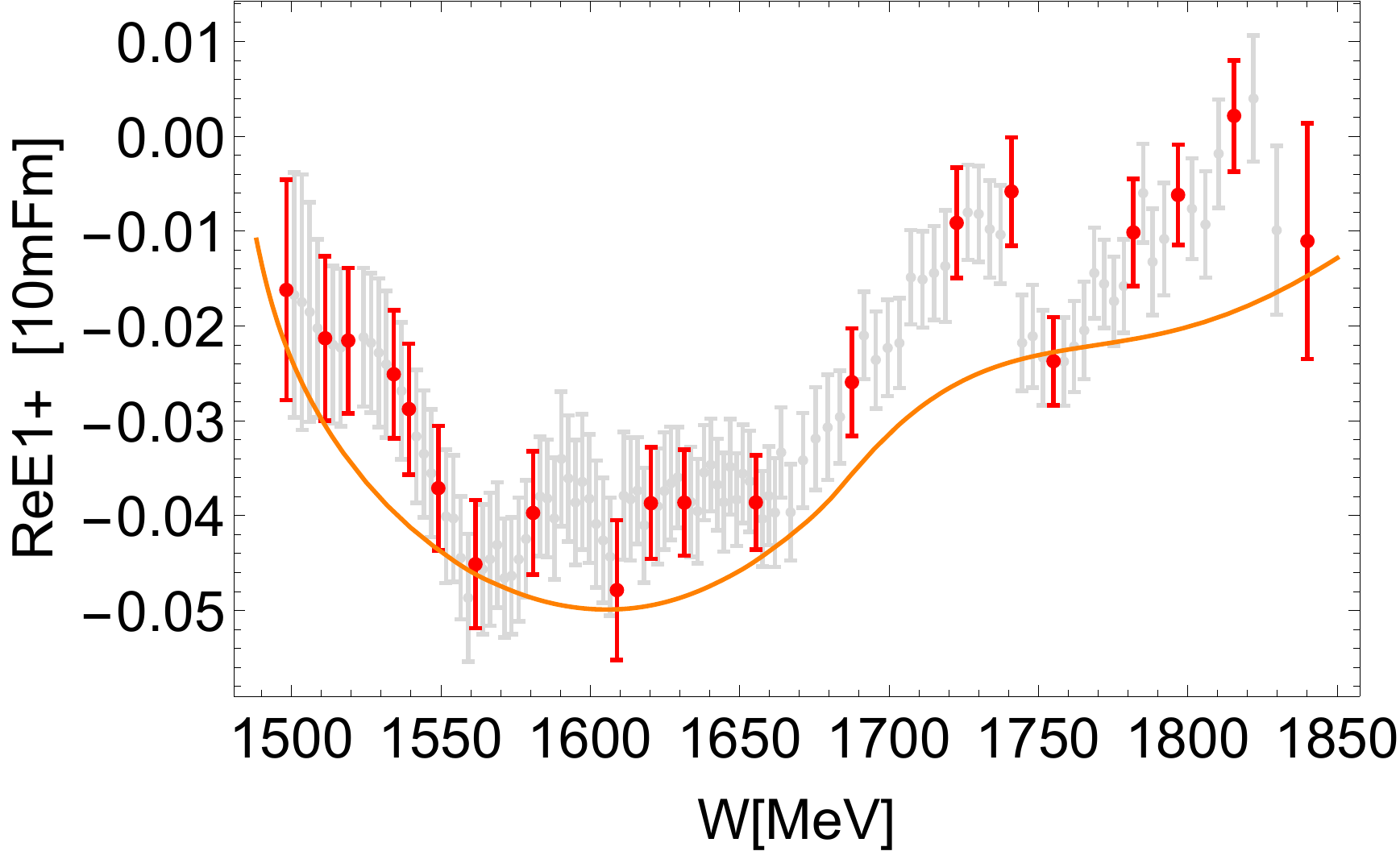} \hspace{0.5cm}
\includegraphics[width=0.38\textwidth]{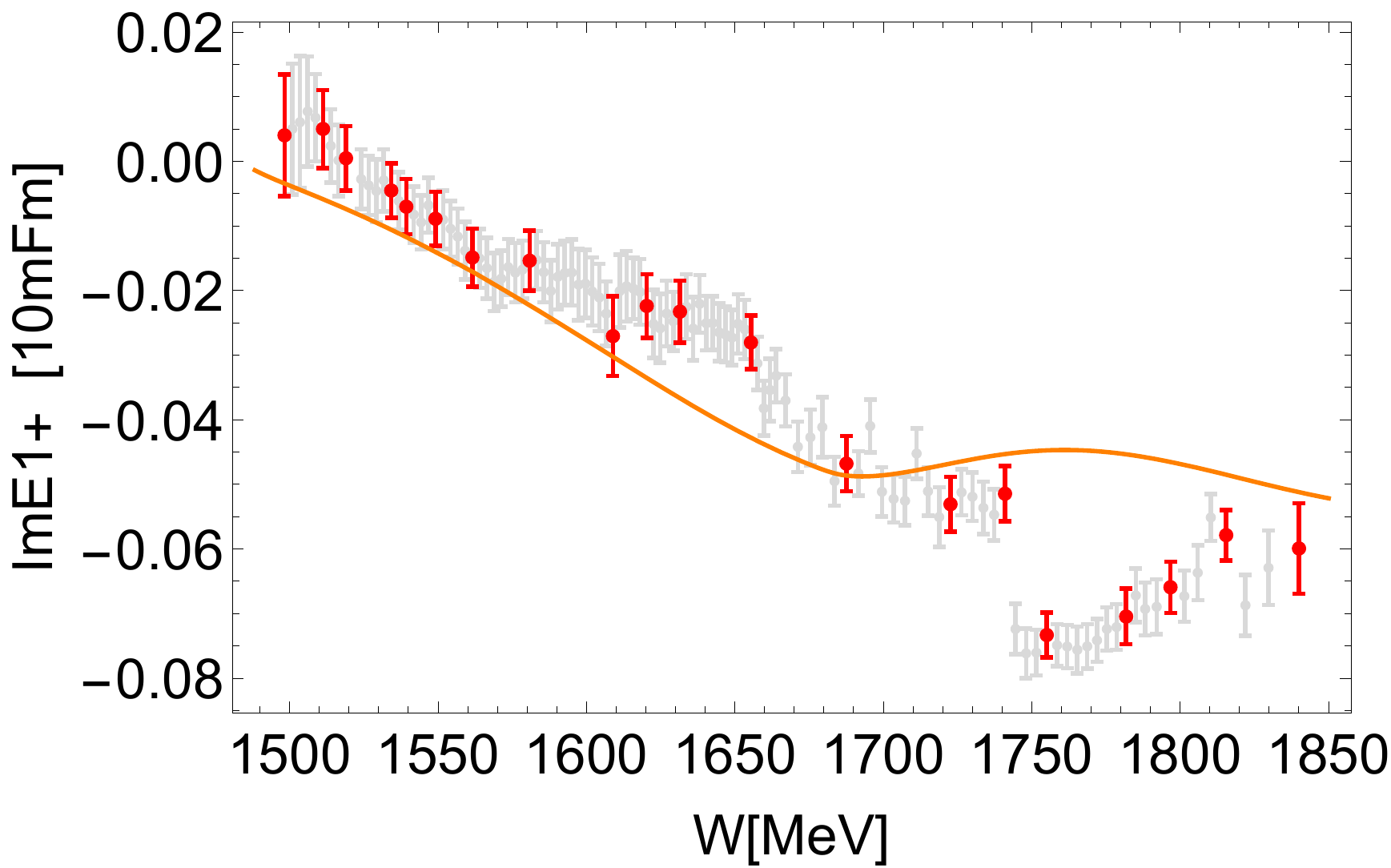}  \\
\includegraphics[width=0.38\textwidth]{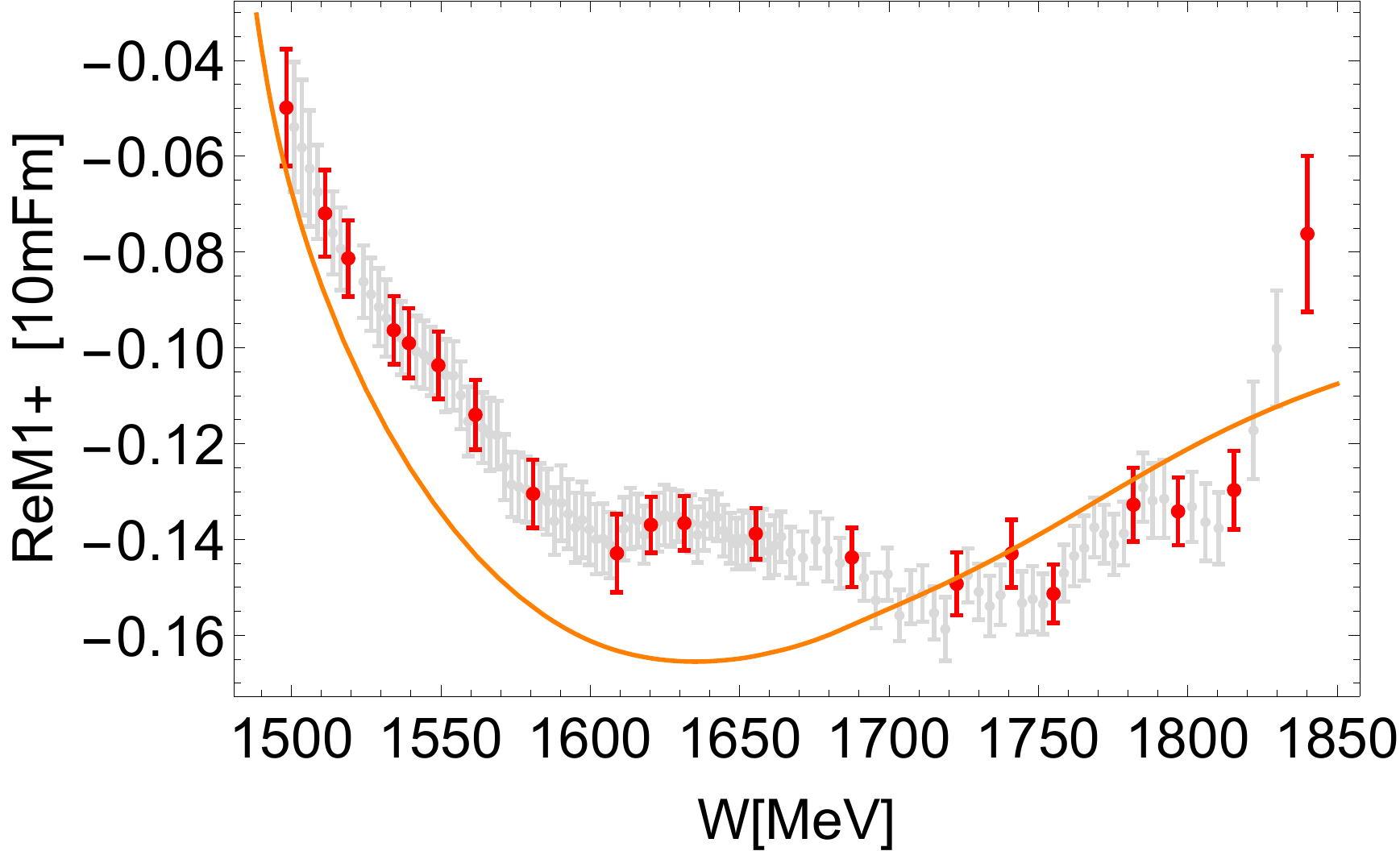} \hspace{0.5cm}
\includegraphics[width=0.38\textwidth]{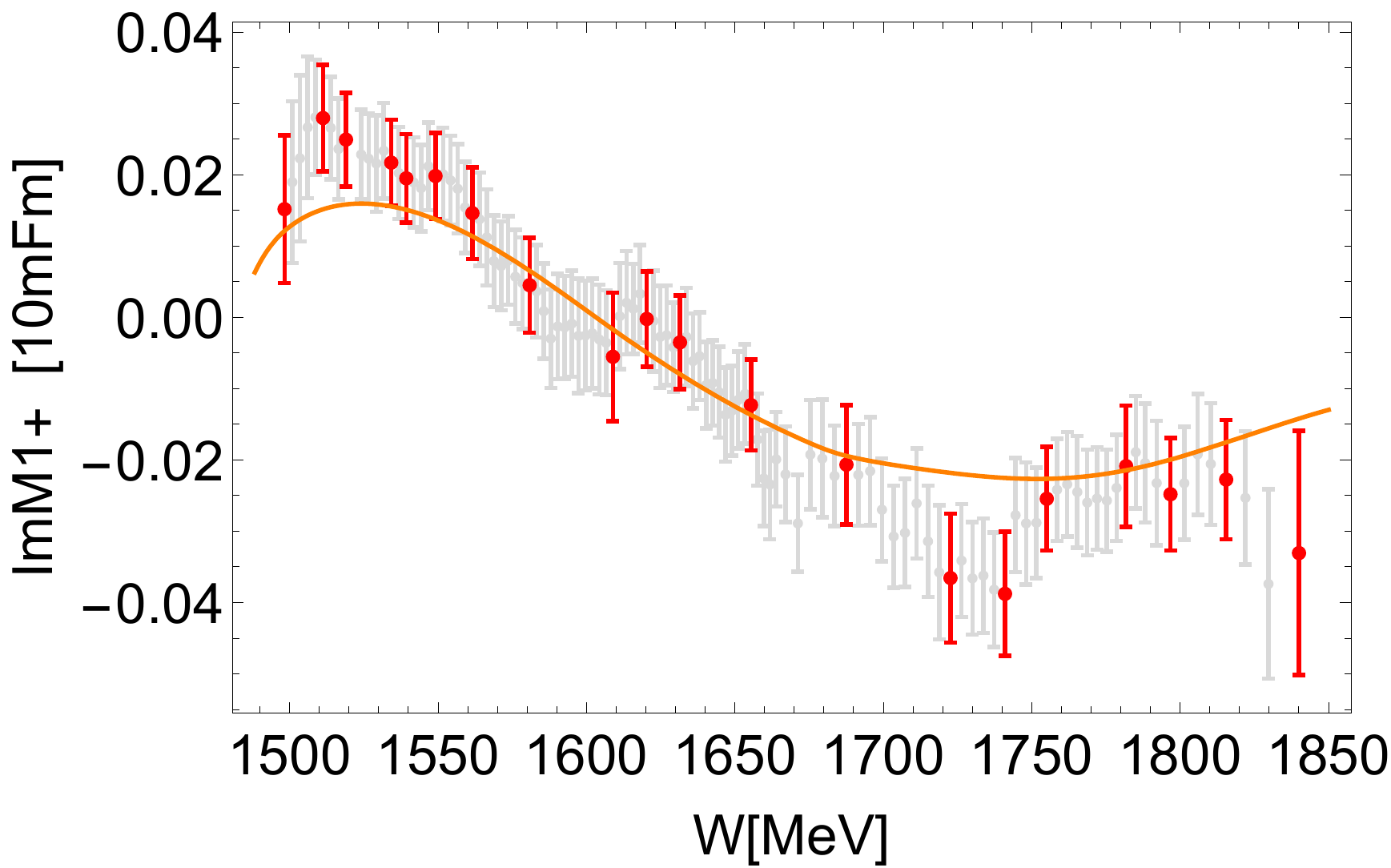}  \\
\includegraphics[width=0.38\textwidth]{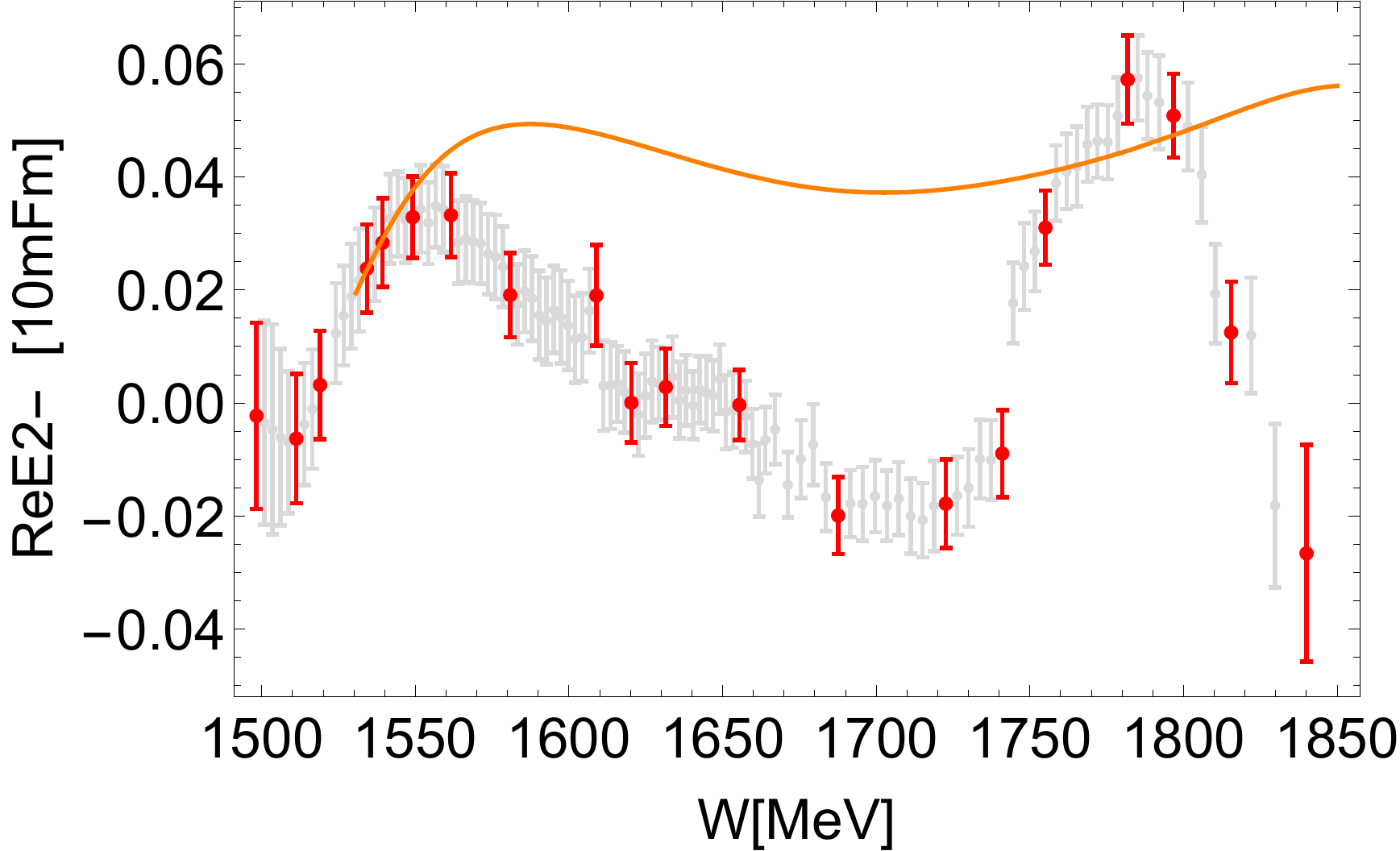} \hspace{0.5cm}
\includegraphics[width=0.38\textwidth]{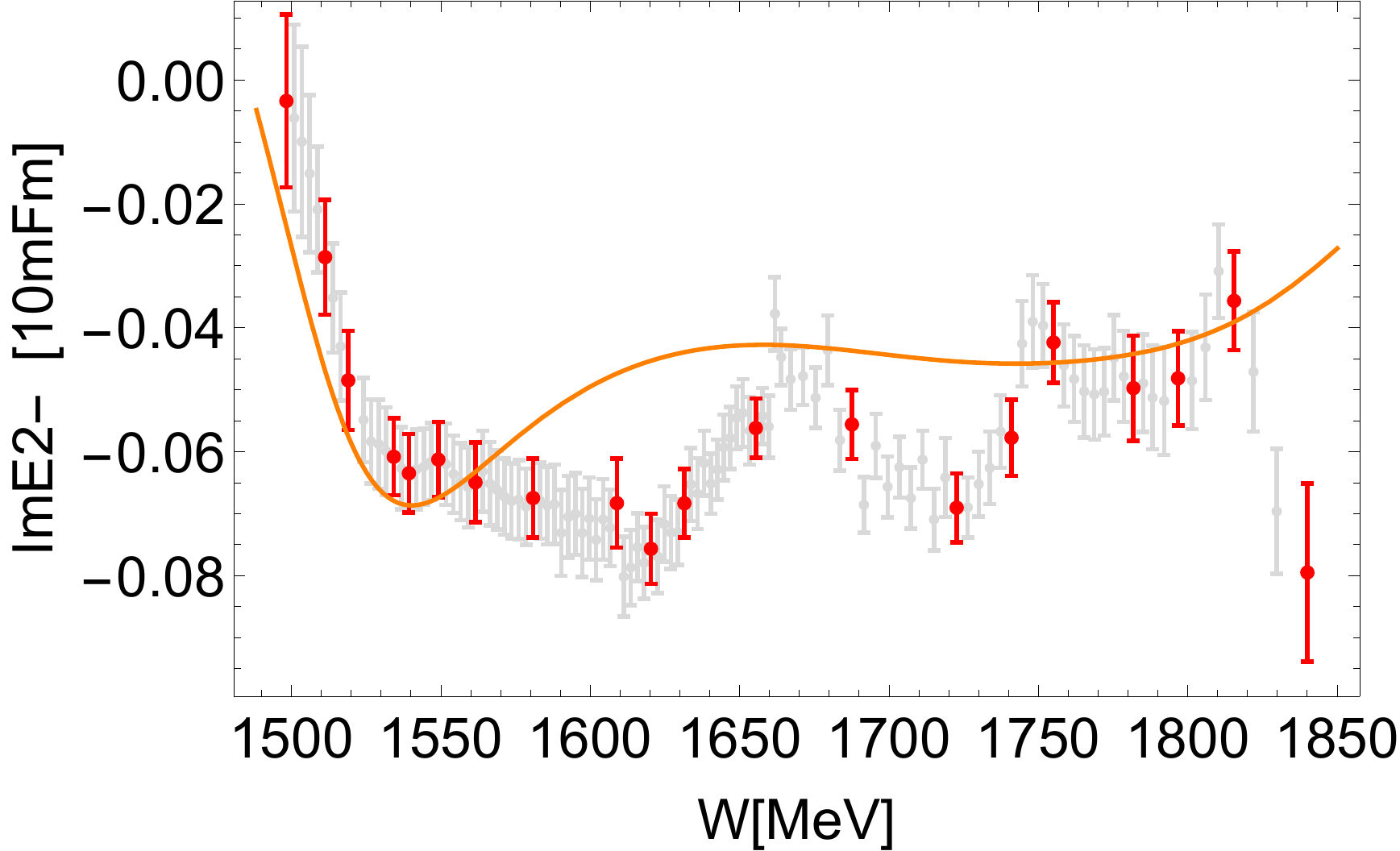}  \\
\ec
\end{figure}

\begin{figure}[h!]
\bc
\includegraphics[width=0.4\textwidth]{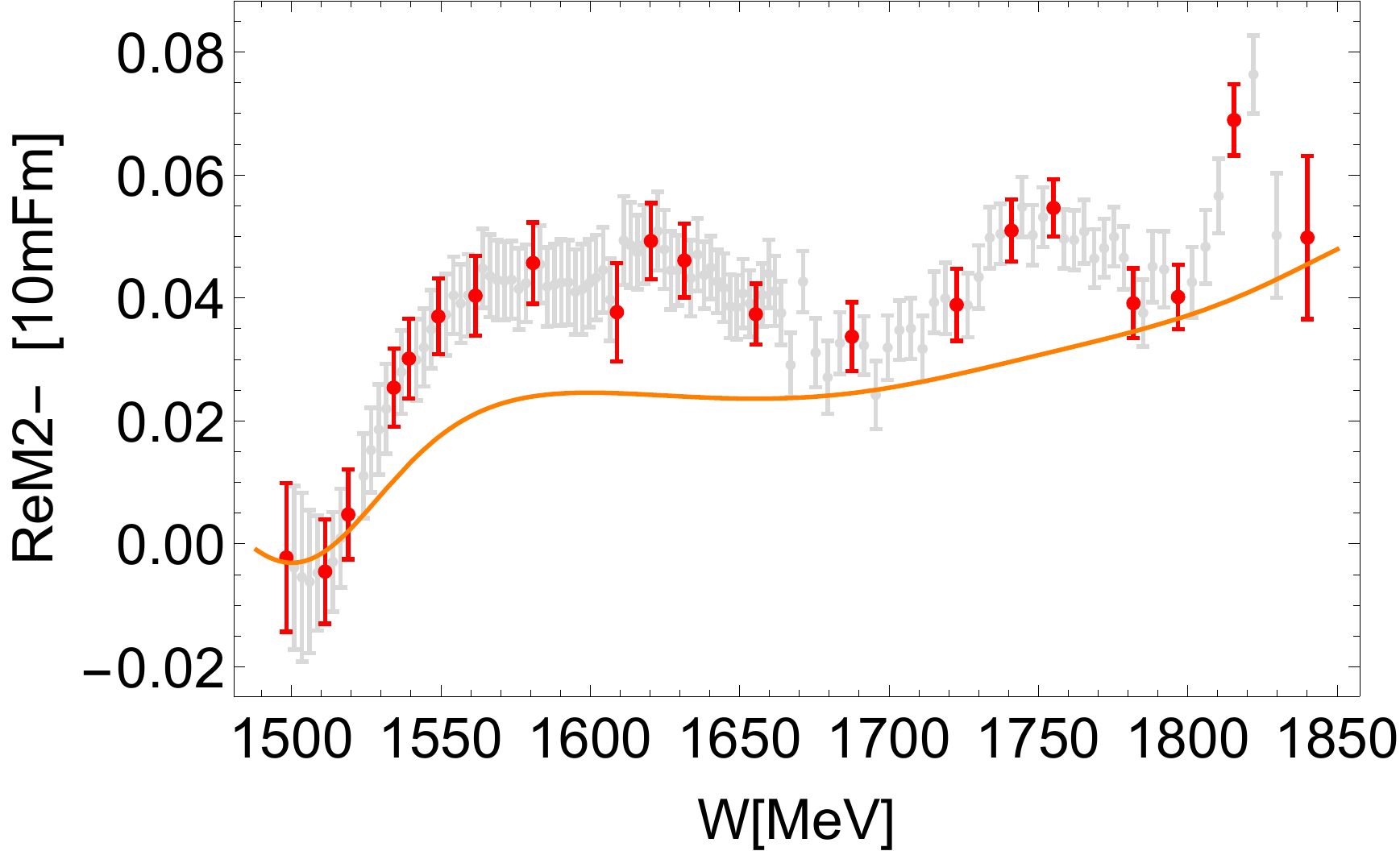} \hspace{0.5cm}
\includegraphics[width=0.4\textwidth]{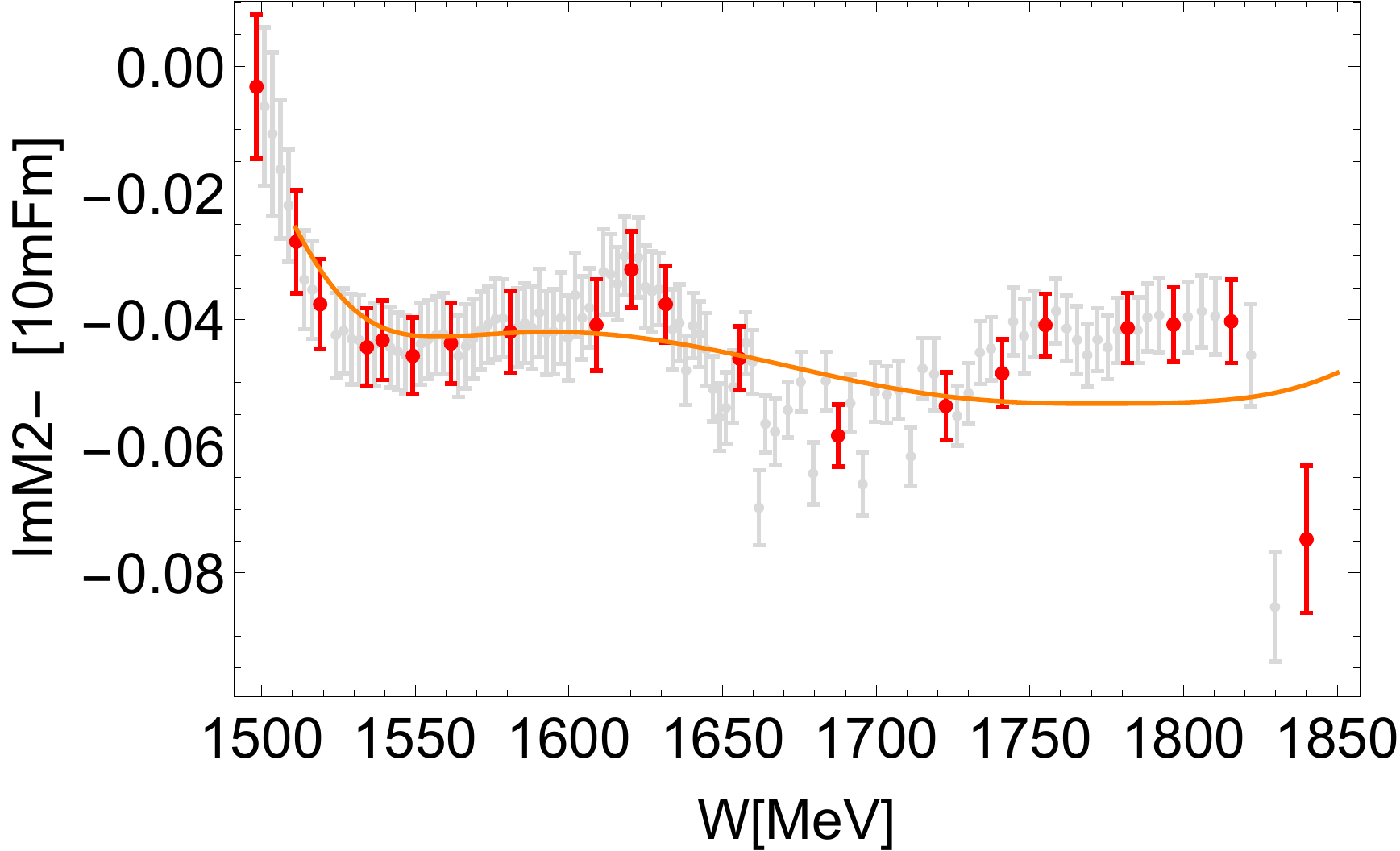}  \\
\includegraphics[width=0.4\textwidth]{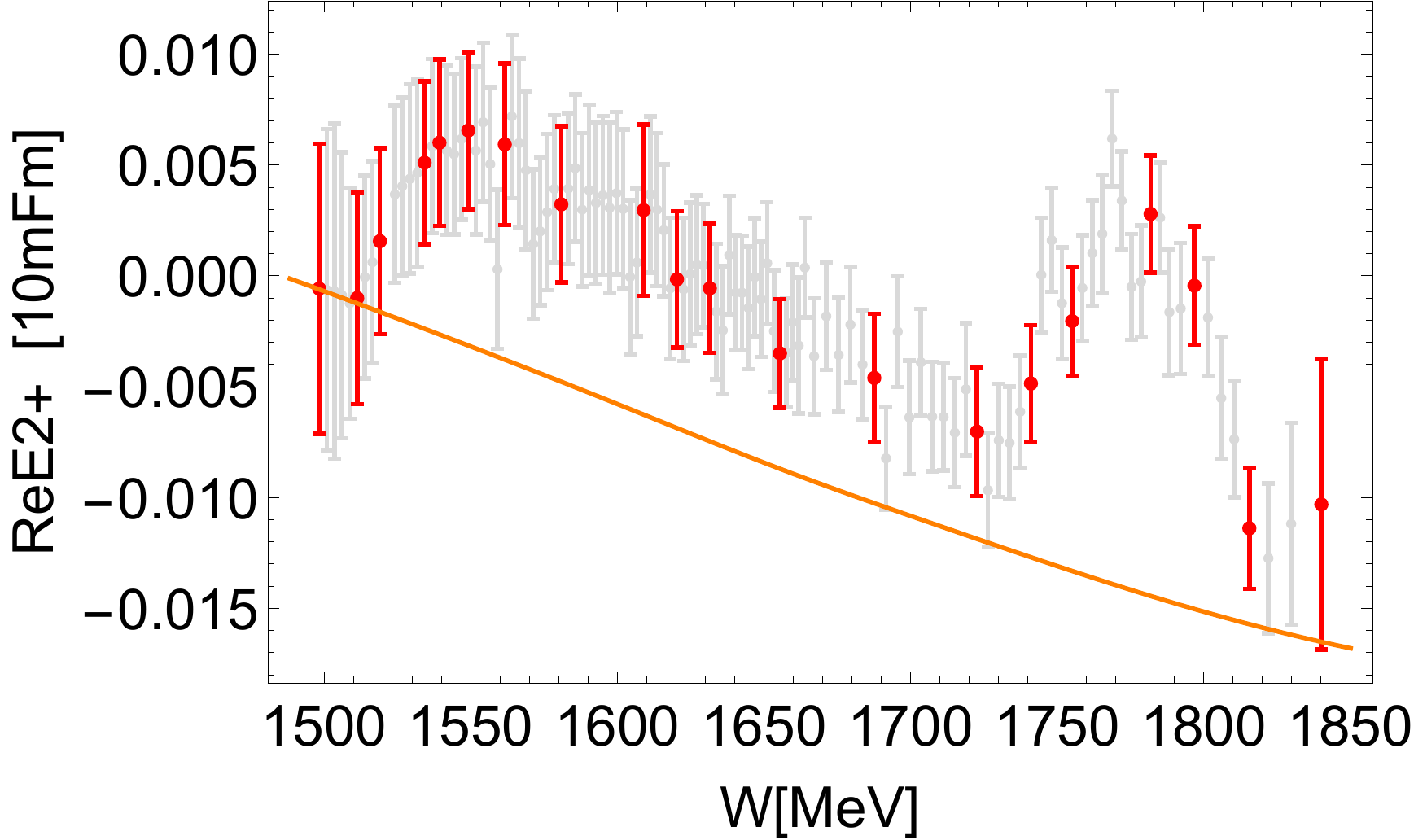} \hspace{0.5cm}
\includegraphics[width=0.4\textwidth]{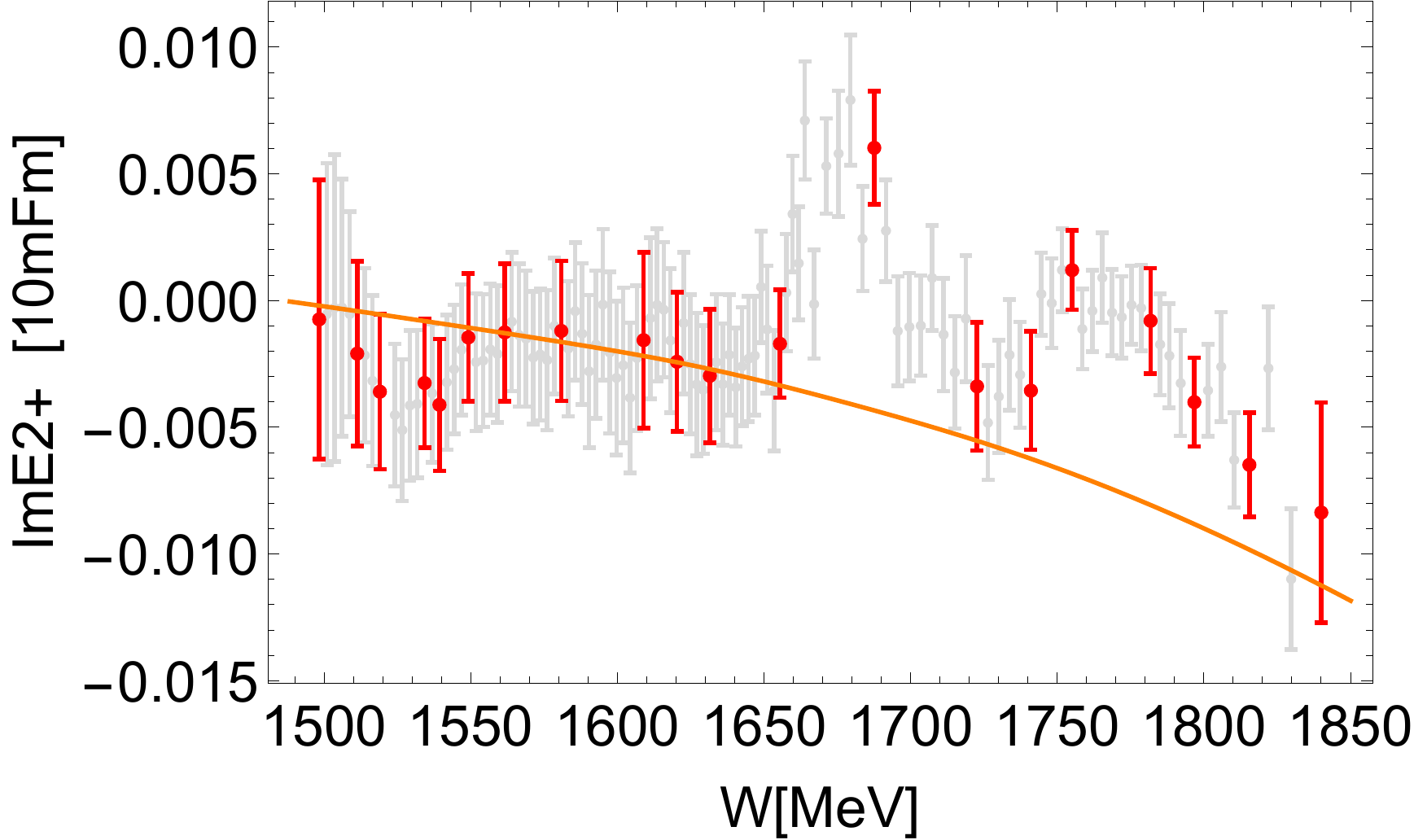}  \\
\includegraphics[width=0.4\textwidth]{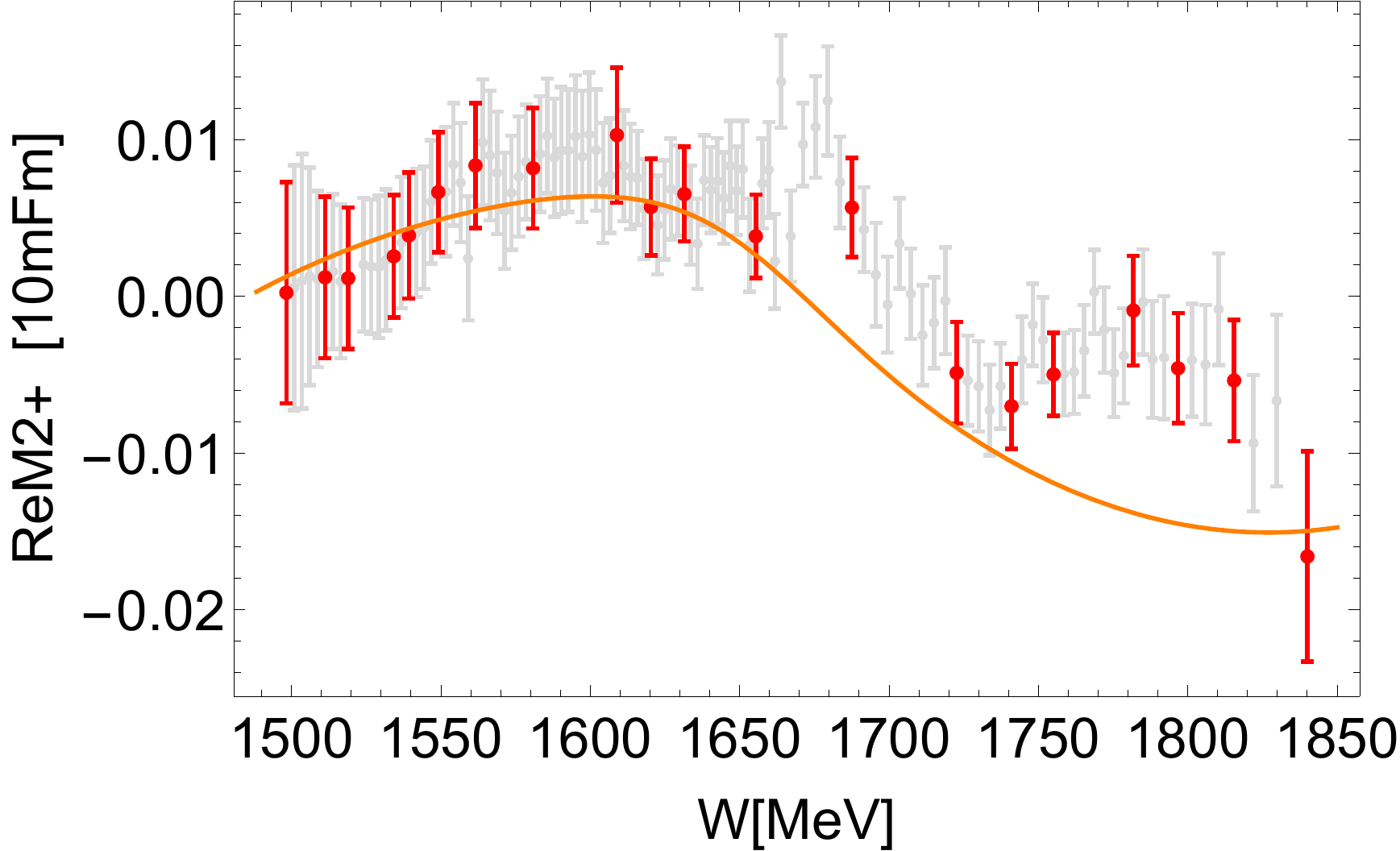} \hspace{0.5cm}
\includegraphics[width=0.4\textwidth]{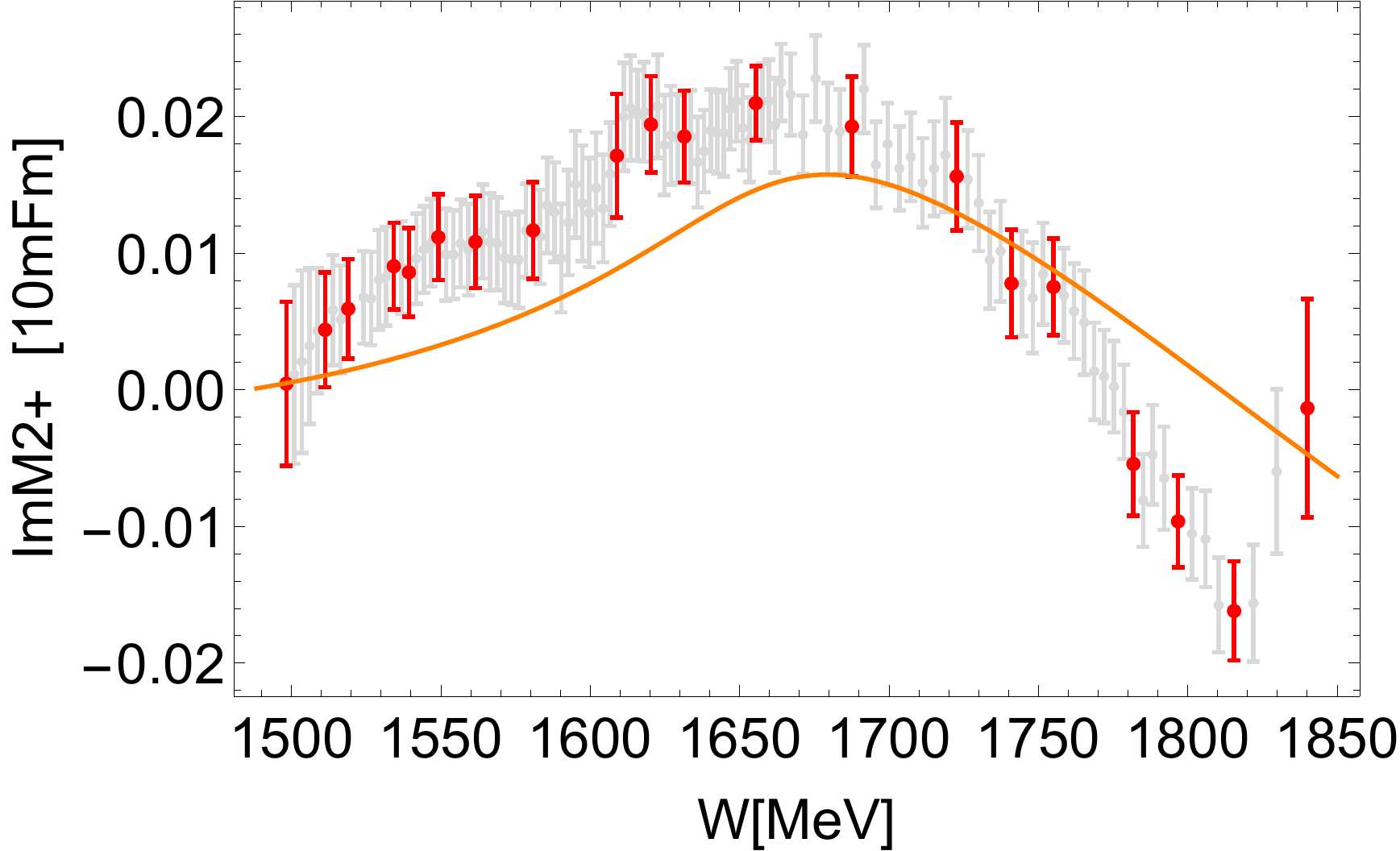}  \\
\includegraphics[width=0.4\textwidth]{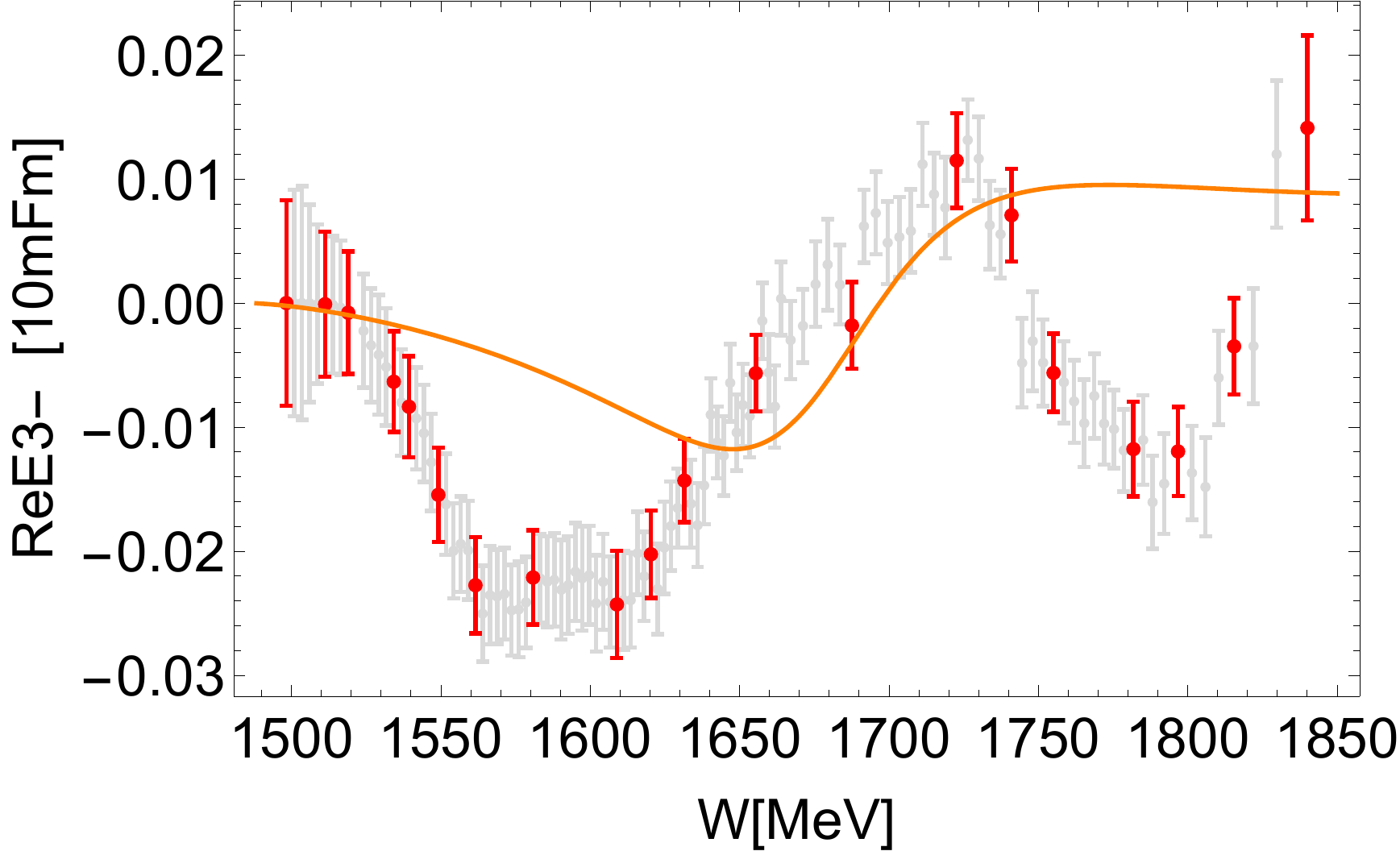} \hspace{0.5cm}
\includegraphics[width=0.4\textwidth]{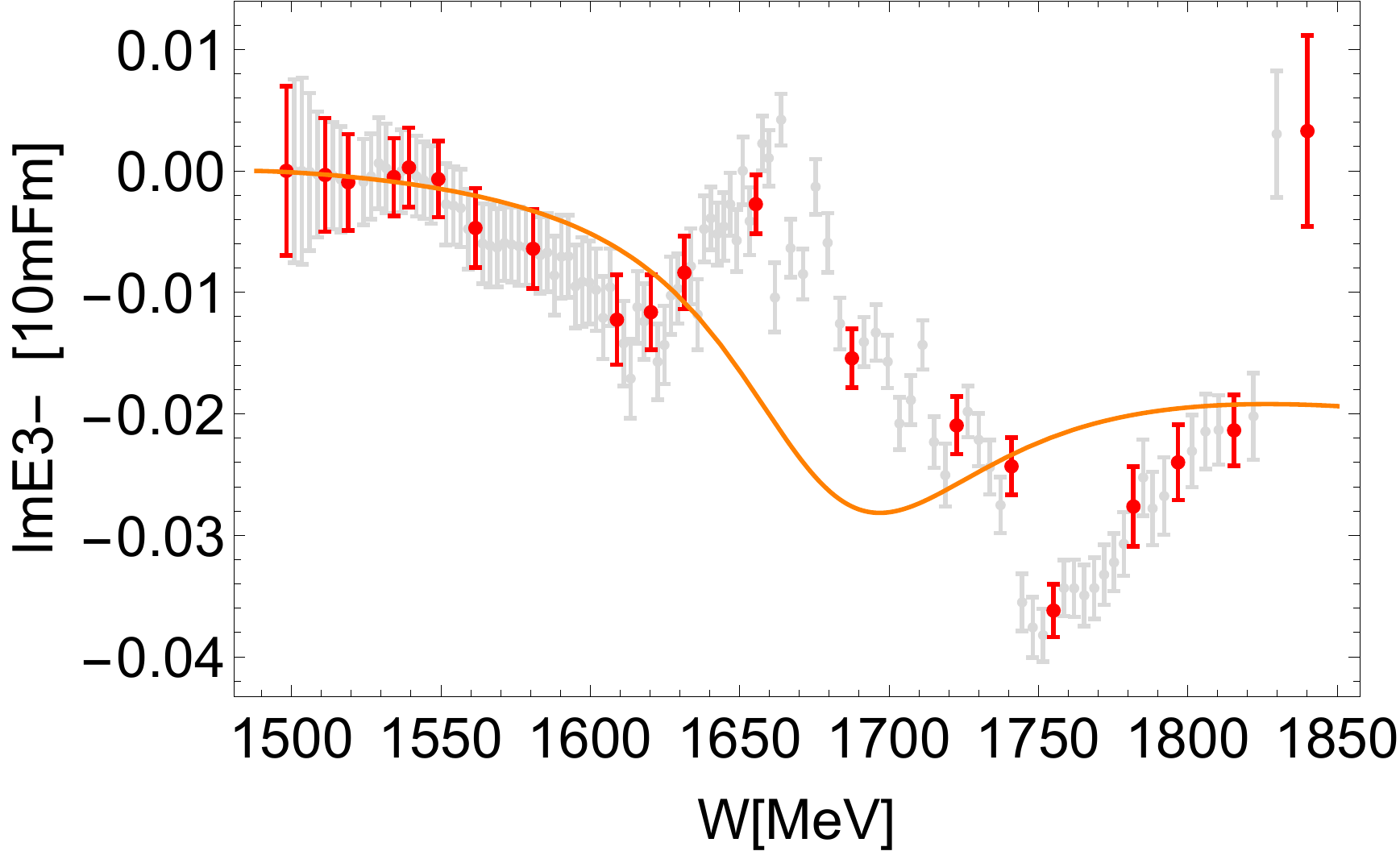}  \\
\includegraphics[width=0.4\textwidth]{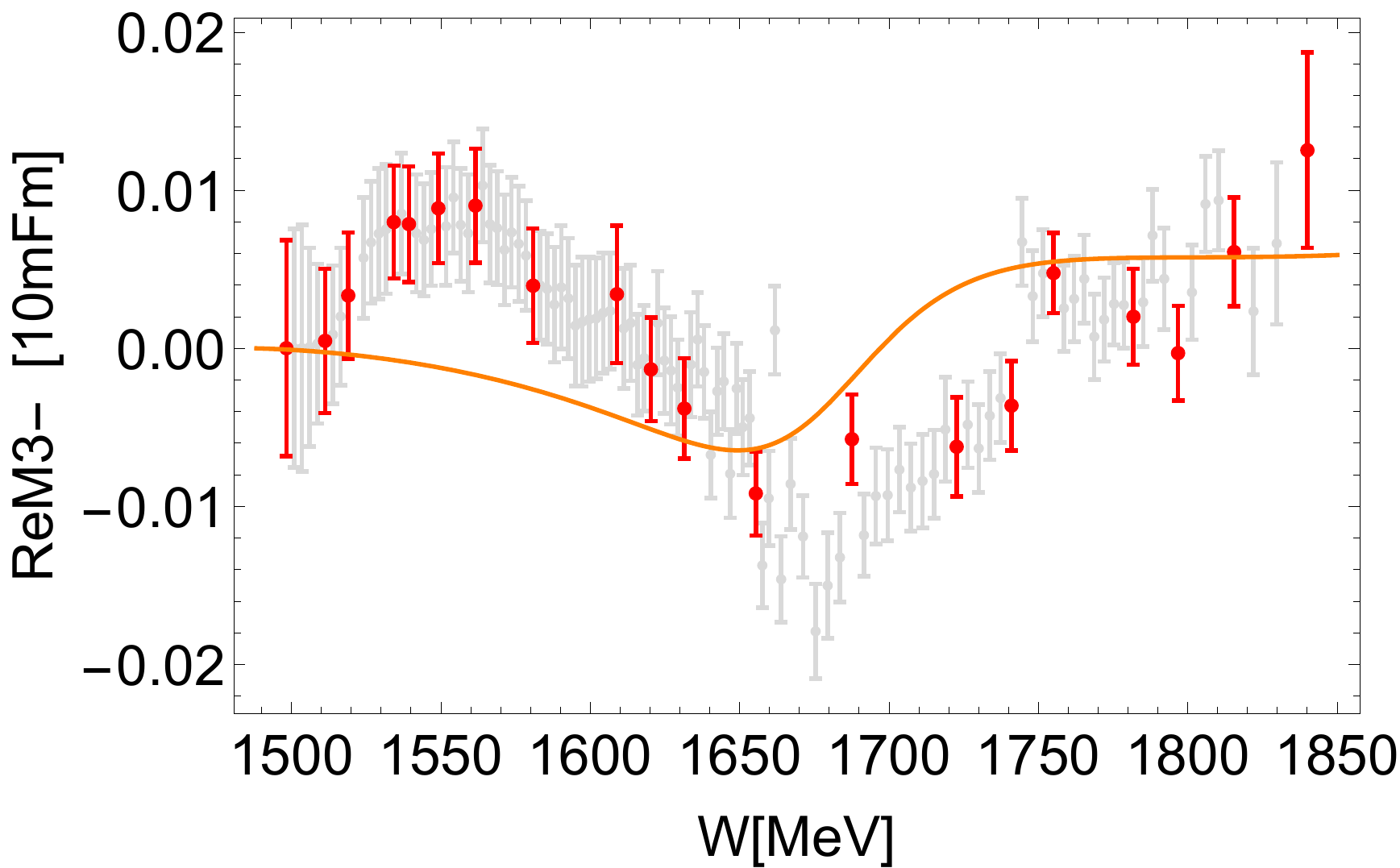} \hspace{0.5cm}
\includegraphics[width=0.4\textwidth]{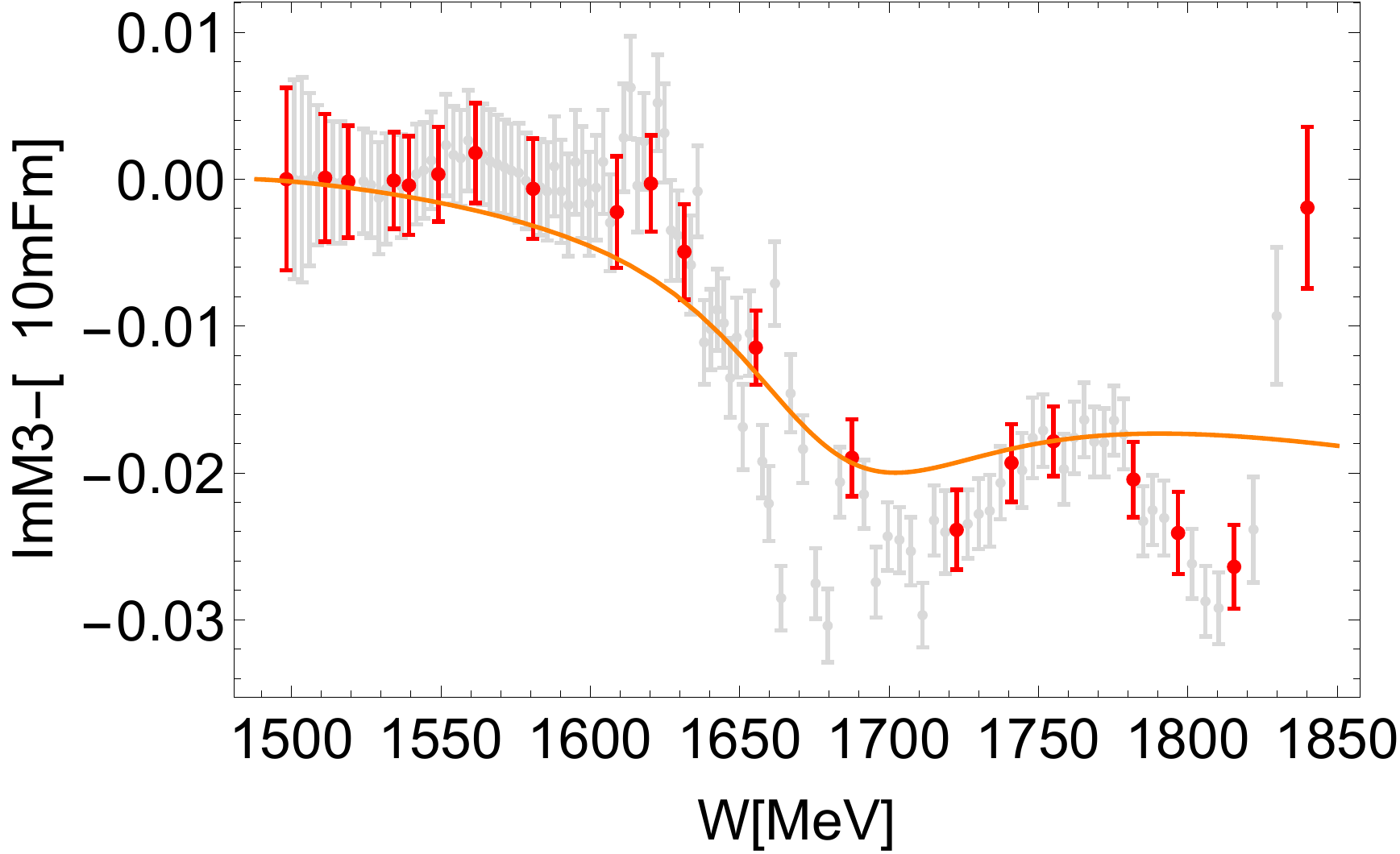}  \\
\caption{\label{Multipoles:Sol1}(Color online) Multipoles for L=0, 1 ,2 and 3 partial waves for Sol 1. Grey discrete symbols correspond to Set 1, and red discrete symbols correspond to Set 2. Orange full line is BG2014-2 solution for comparison.      }
\ec
\end{figure}

\clearpage

\underline{\textbf{$Sol \, 2$, solution with smoothed BG2014-2 phase.}}

\begin{figure}[h!]
\bc
\includegraphics[width=0.38\textwidth]{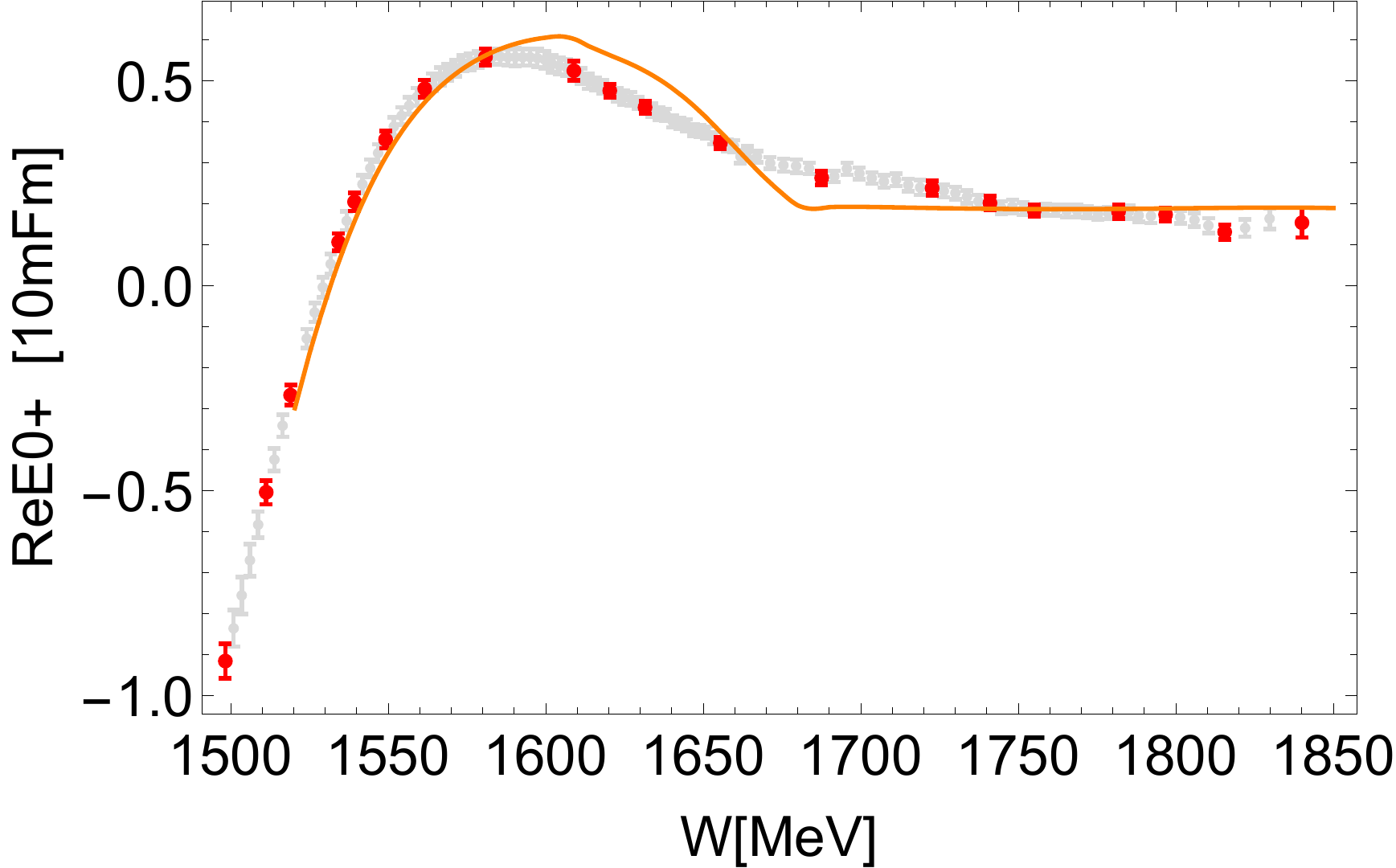} \hspace{0.5cm}
\includegraphics[width=0.38\textwidth]{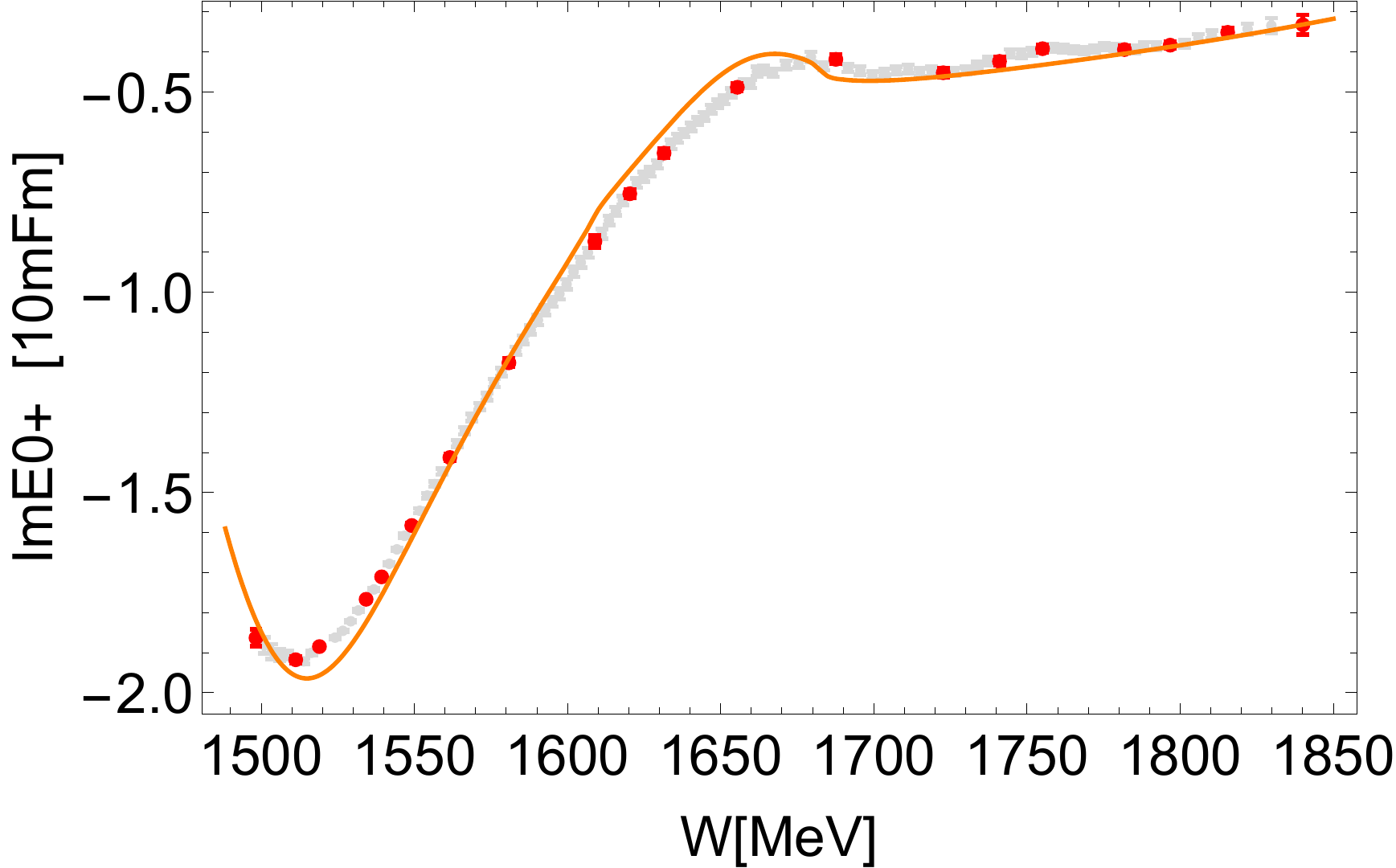}  \\
\includegraphics[width=0.38\textwidth]{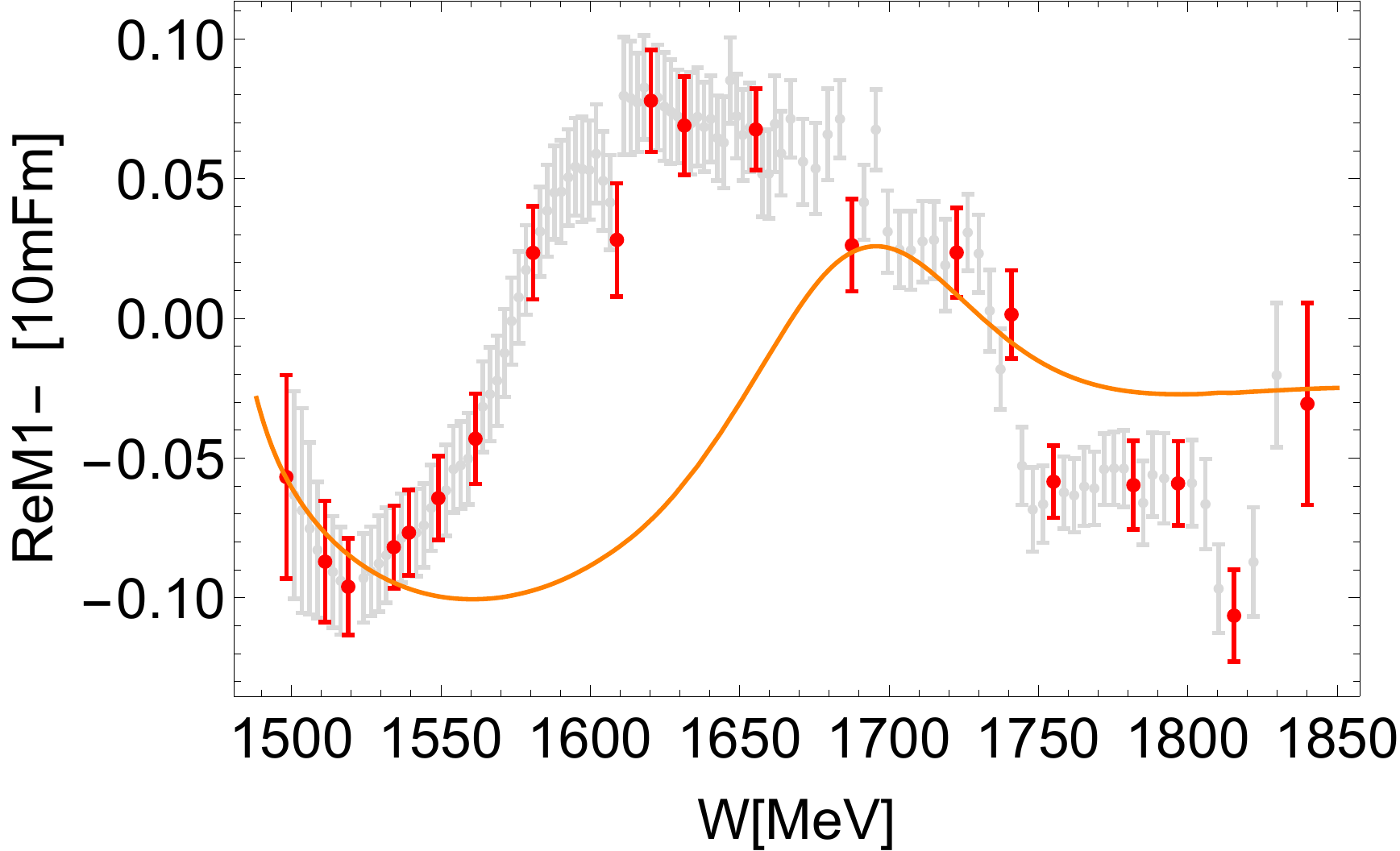} \hspace{0.5cm}
\includegraphics[width=0.38\textwidth]{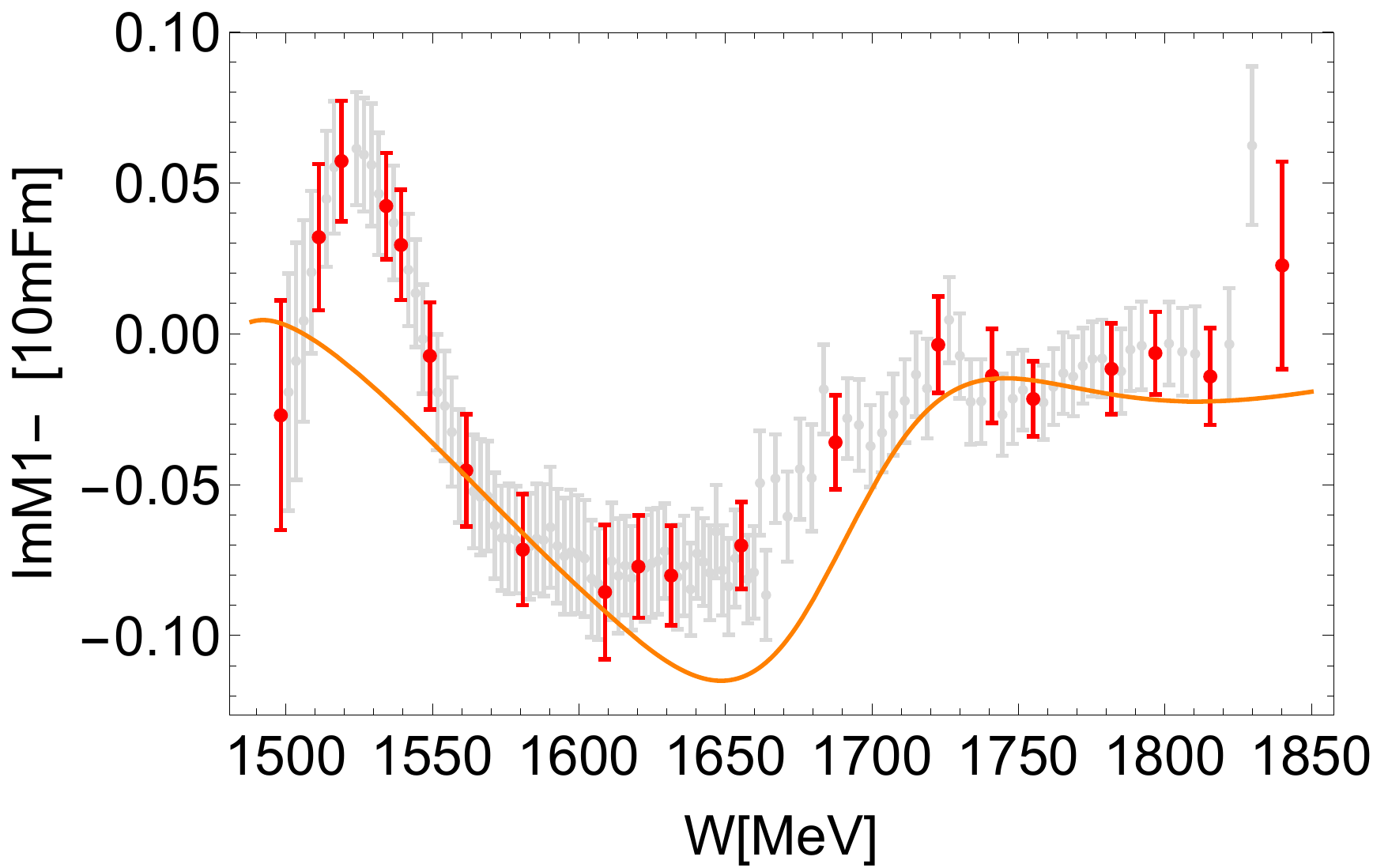}  \\
\includegraphics[width=0.38\textwidth]{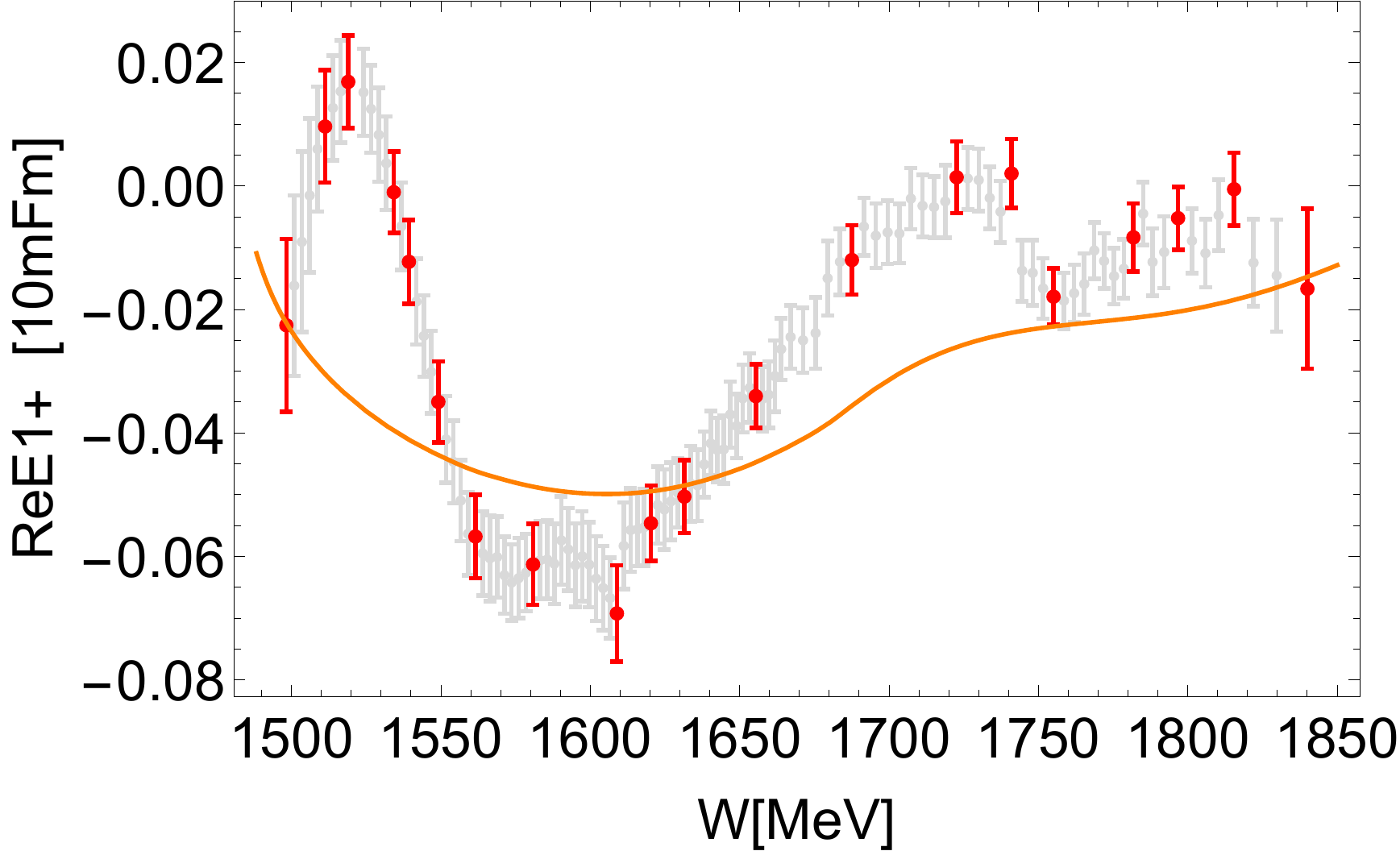} \hspace{0.5cm}
\includegraphics[width=0.38\textwidth]{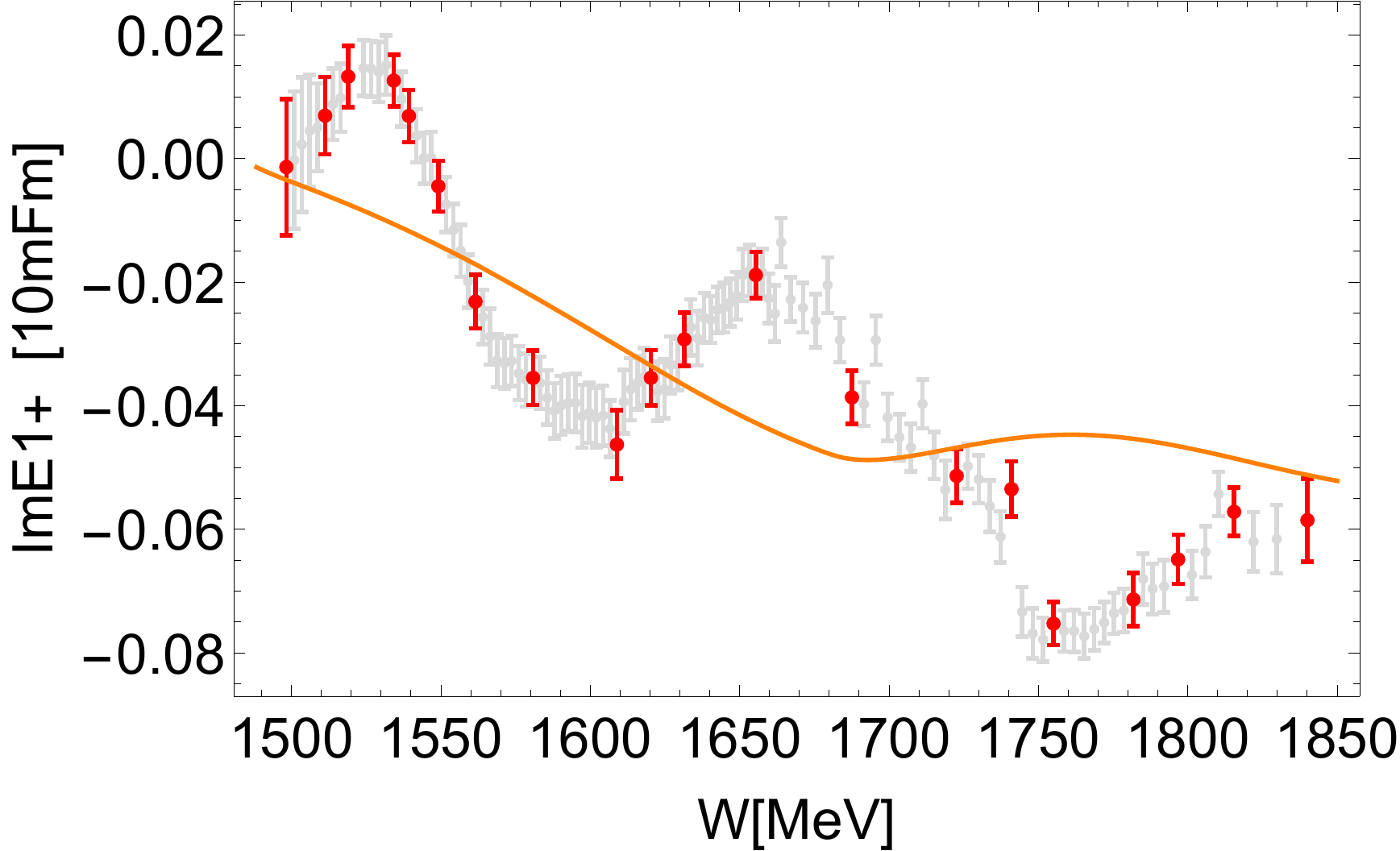}  \\
\includegraphics[width=0.38\textwidth]{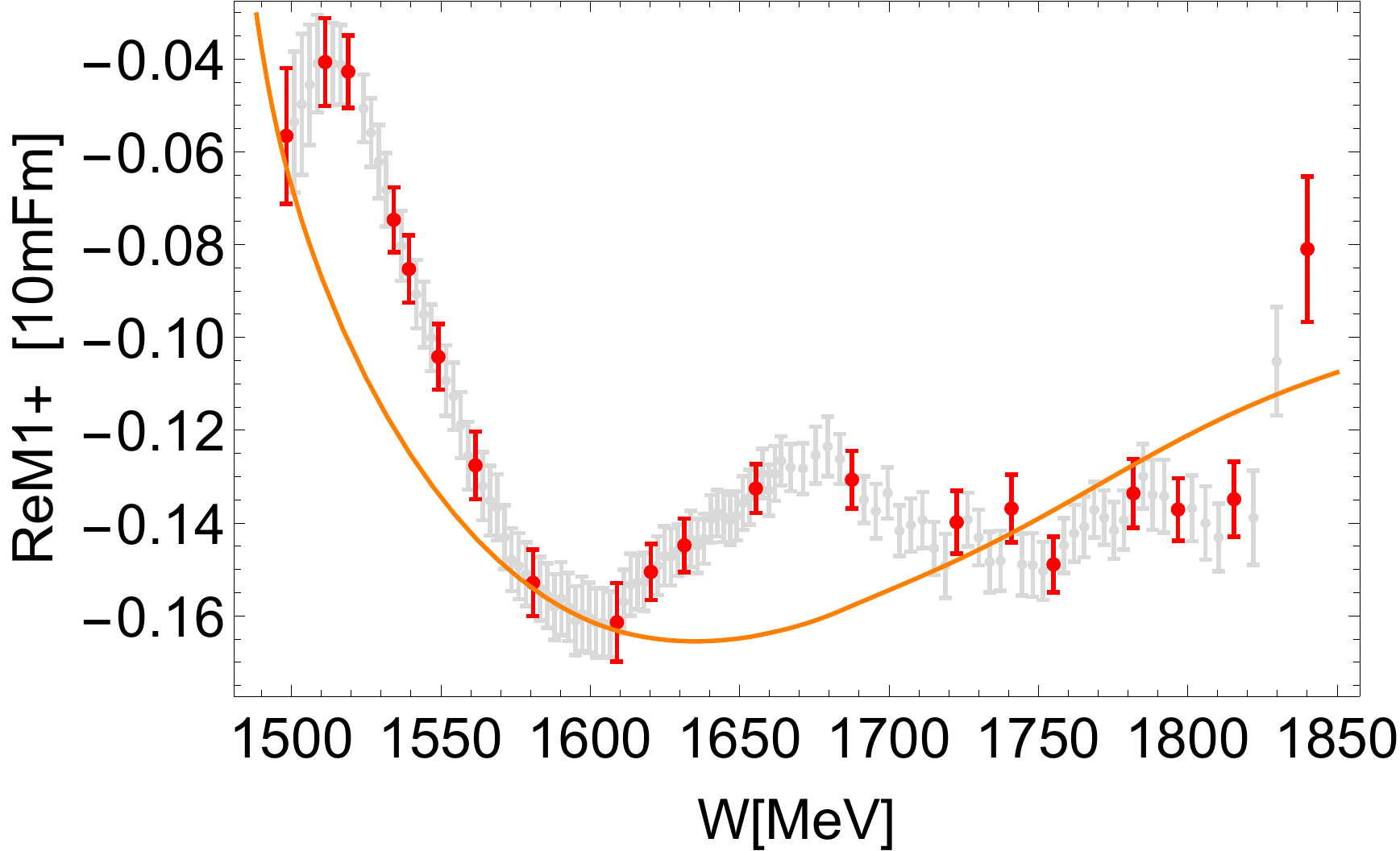} \hspace{0.5cm}
\includegraphics[width=0.38\textwidth]{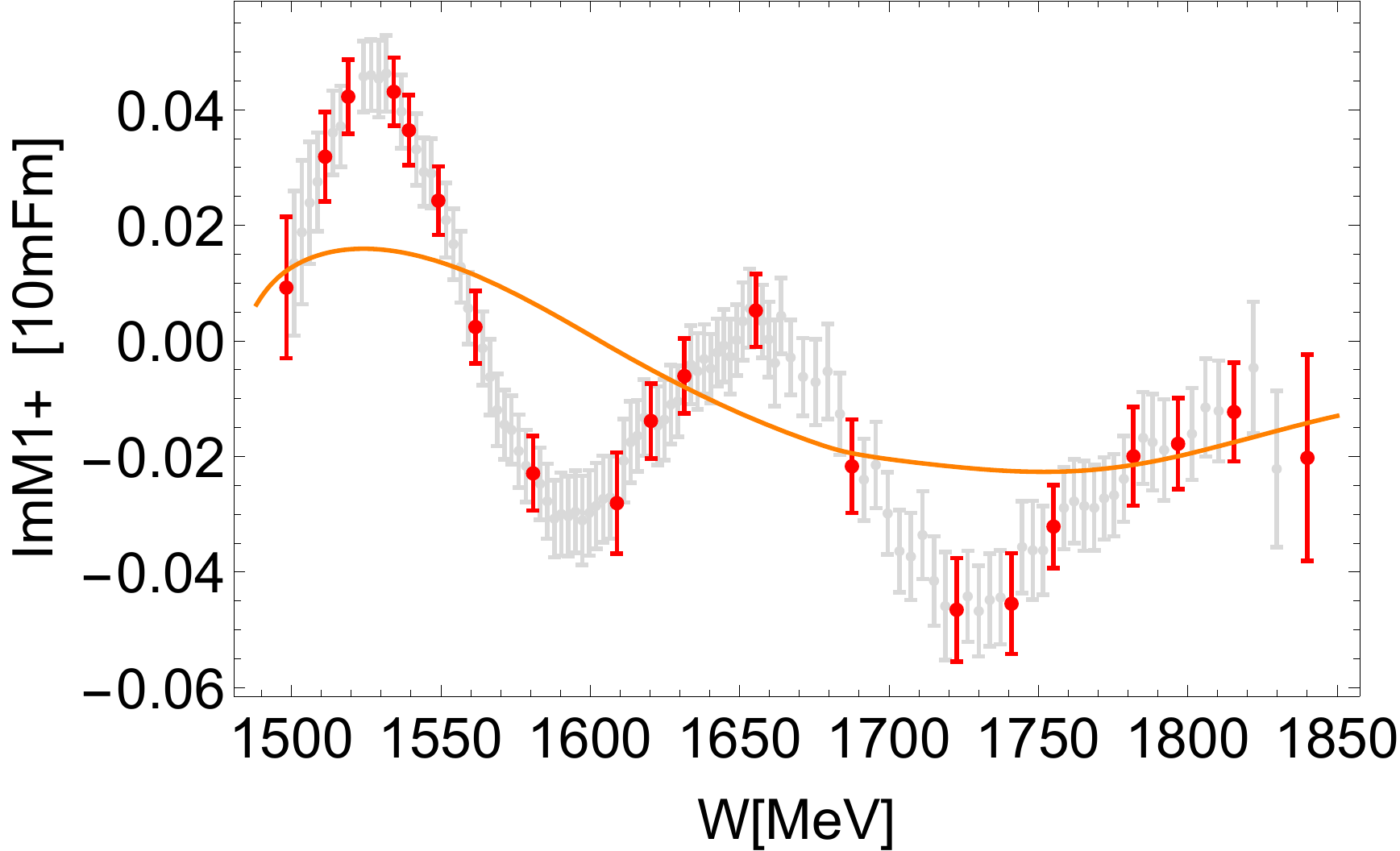}  \\
\includegraphics[width=0.38\textwidth]{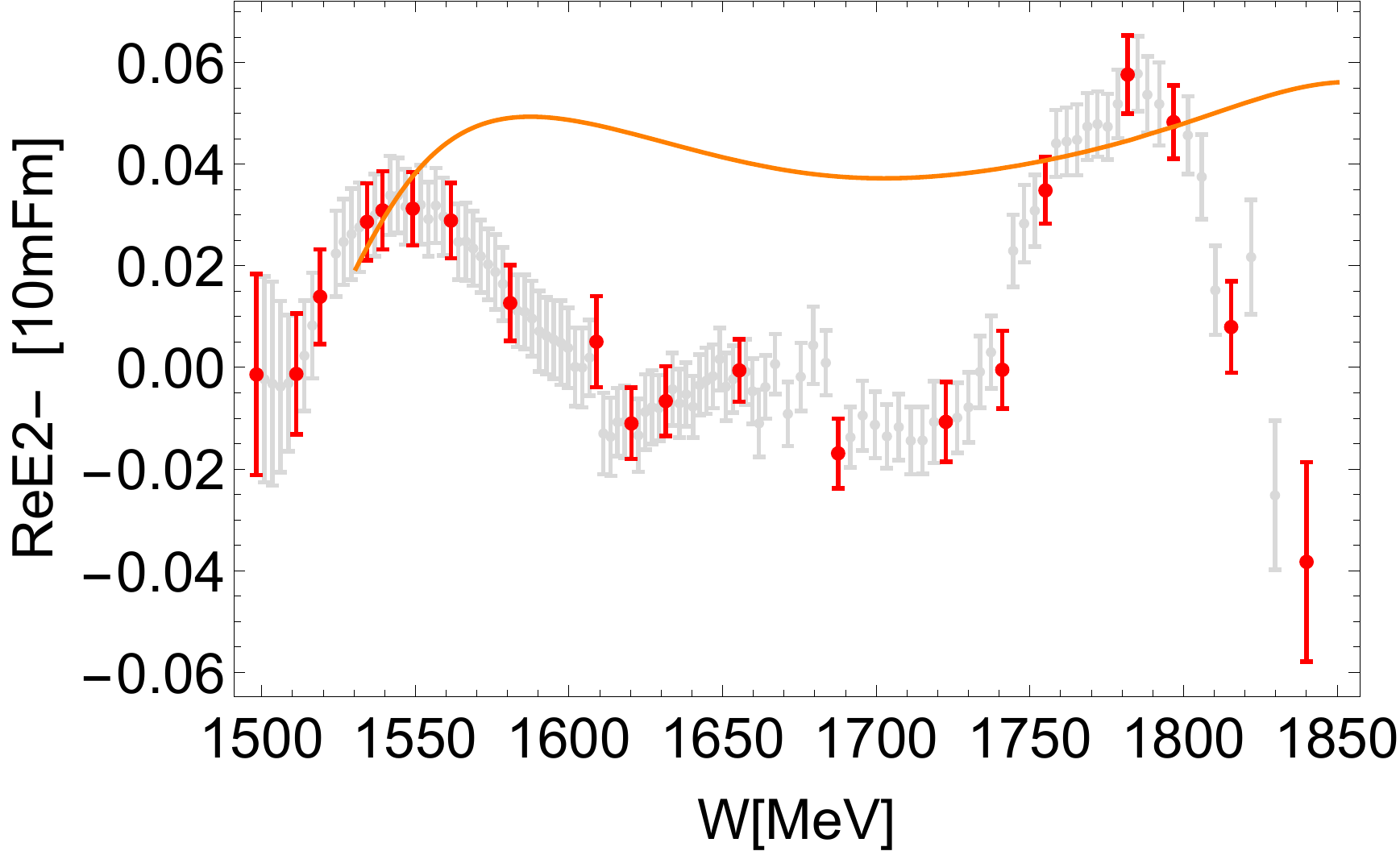} \hspace{0.5cm}
\includegraphics[width=0.38\textwidth]{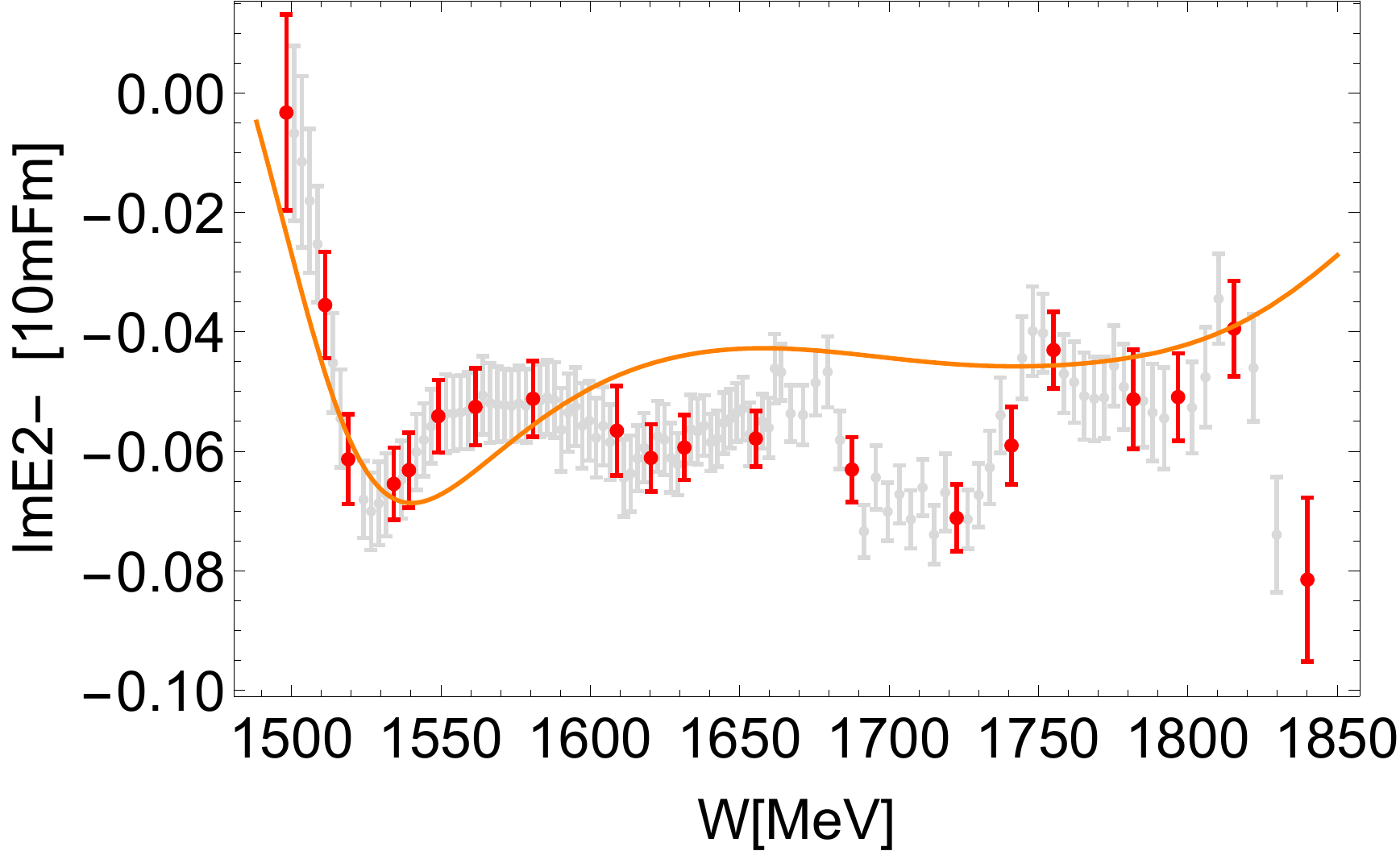}  \\
\ec
\end{figure}

\begin{figure}[h!]
\bc
\includegraphics[width=0.4\textwidth]{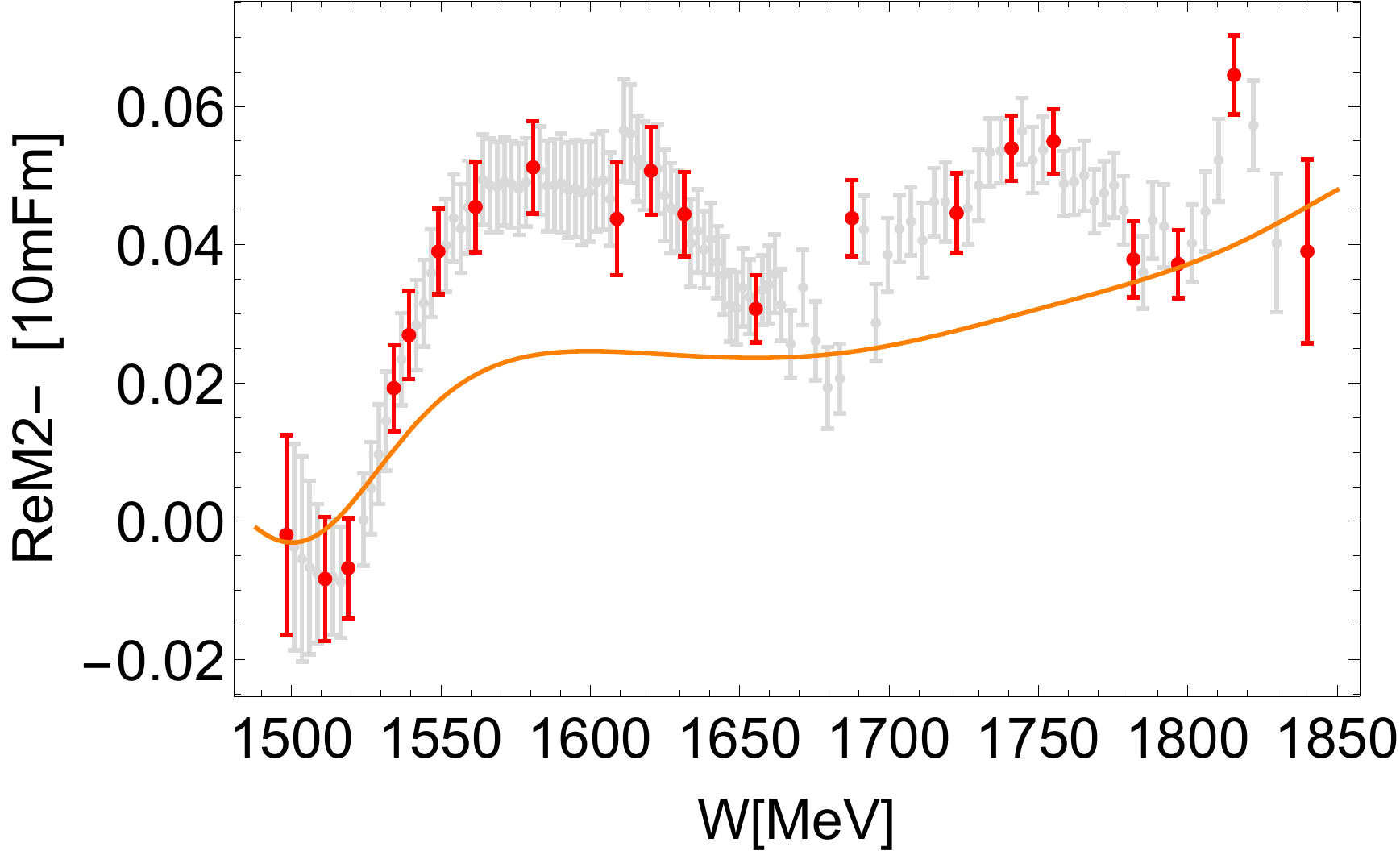} \hspace{0.5cm}
\includegraphics[width=0.4\textwidth]{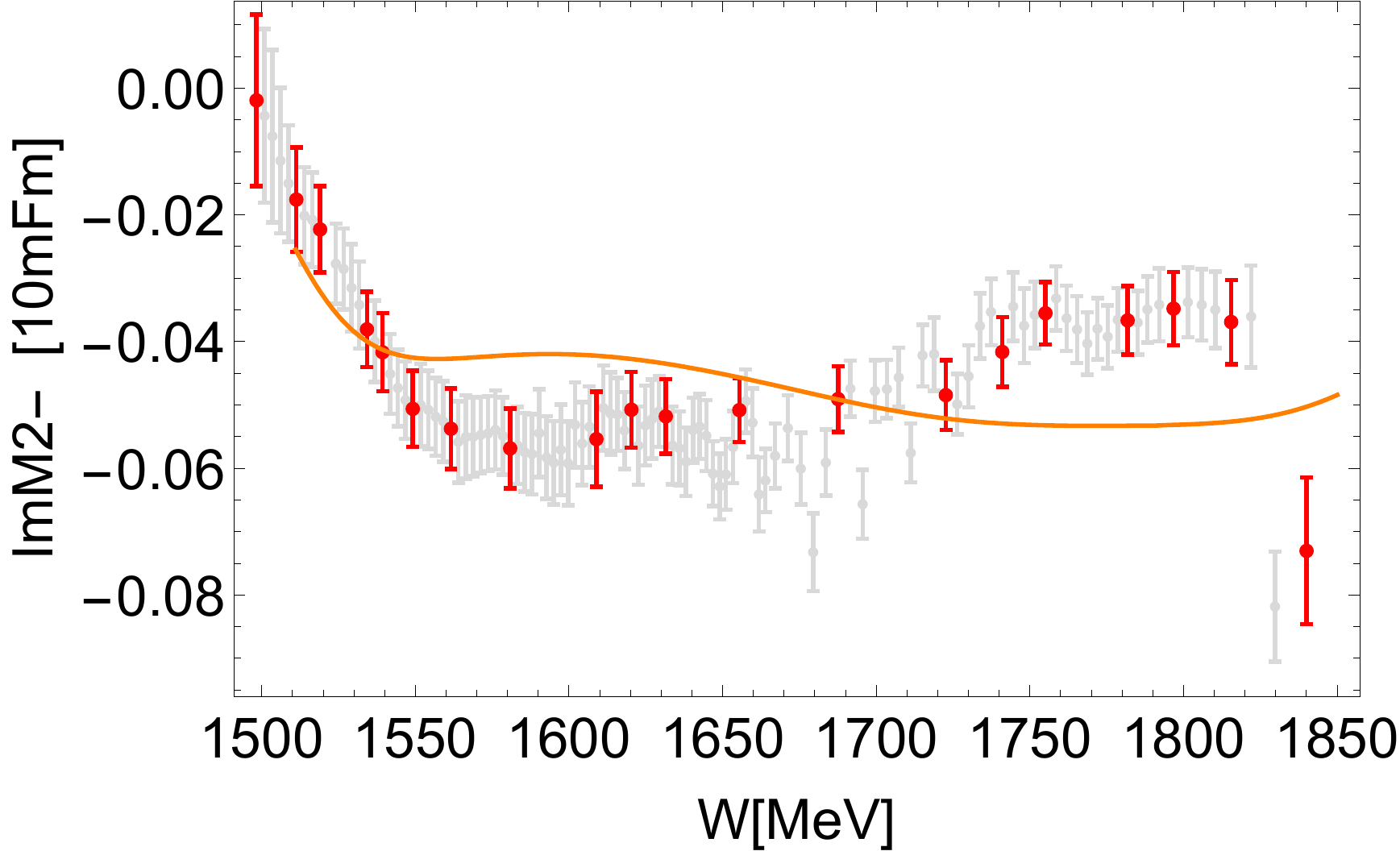}  \\
\includegraphics[width=0.4\textwidth]{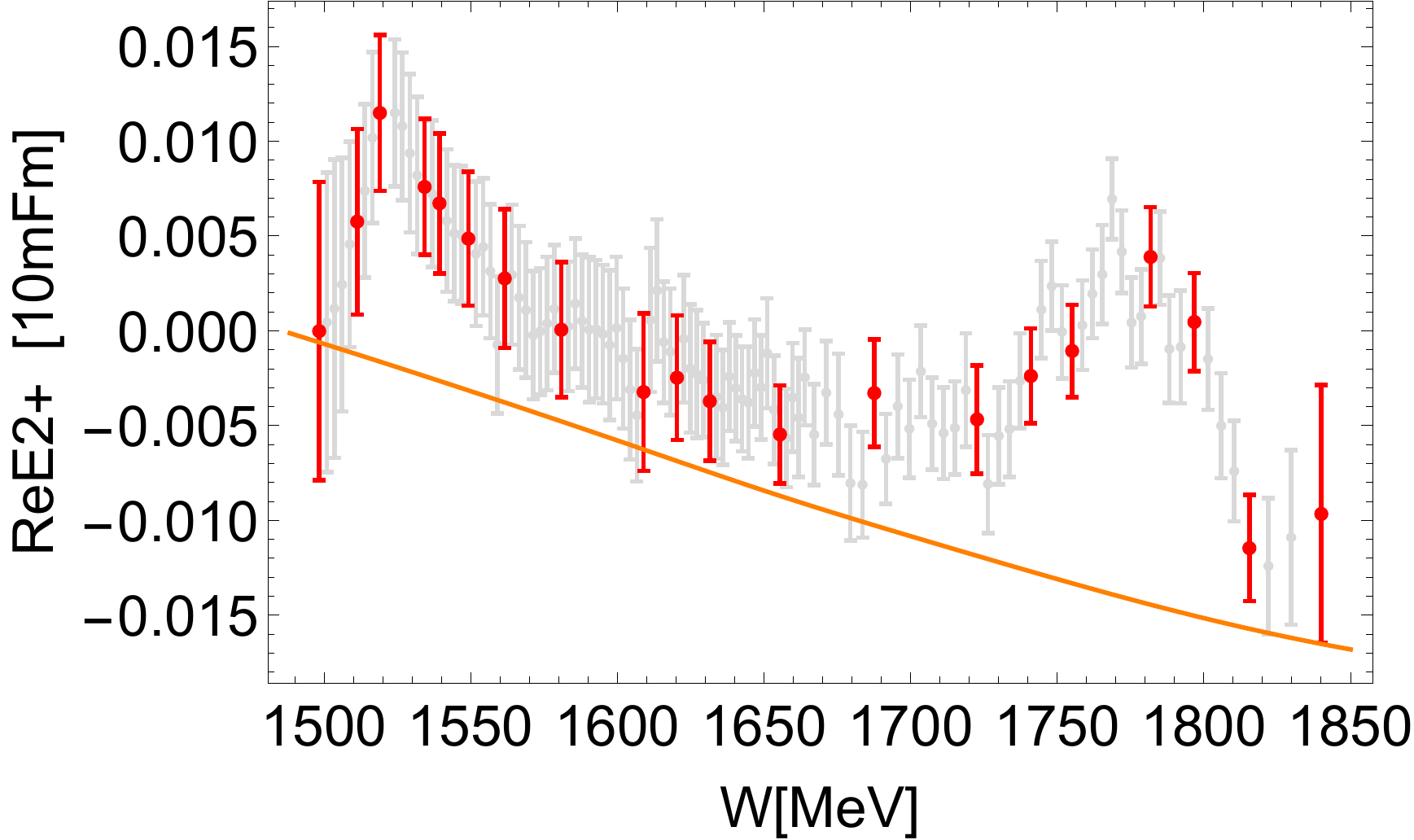} \hspace{0.5cm}
\includegraphics[width=0.4\textwidth]{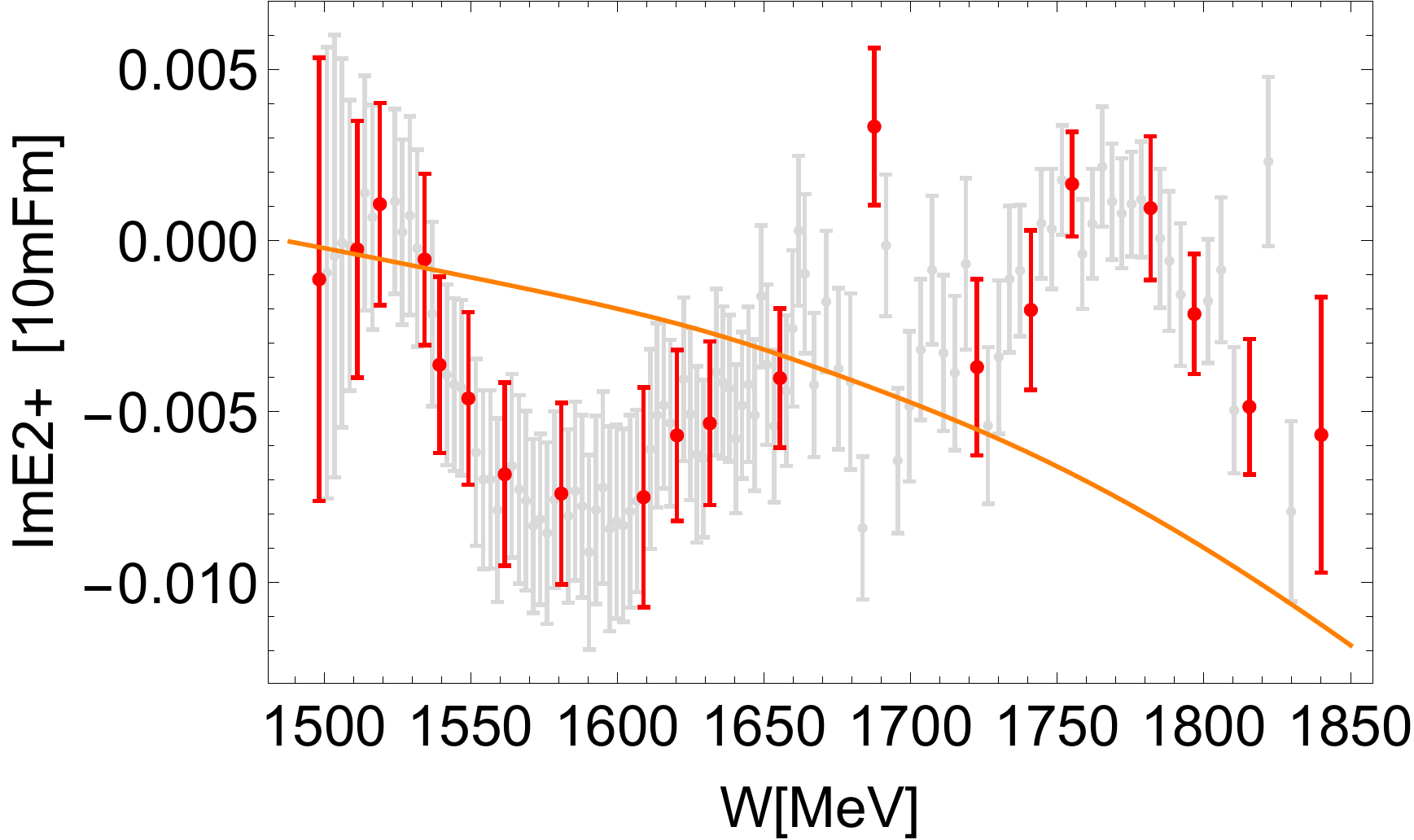}  \\
\includegraphics[width=0.4\textwidth]{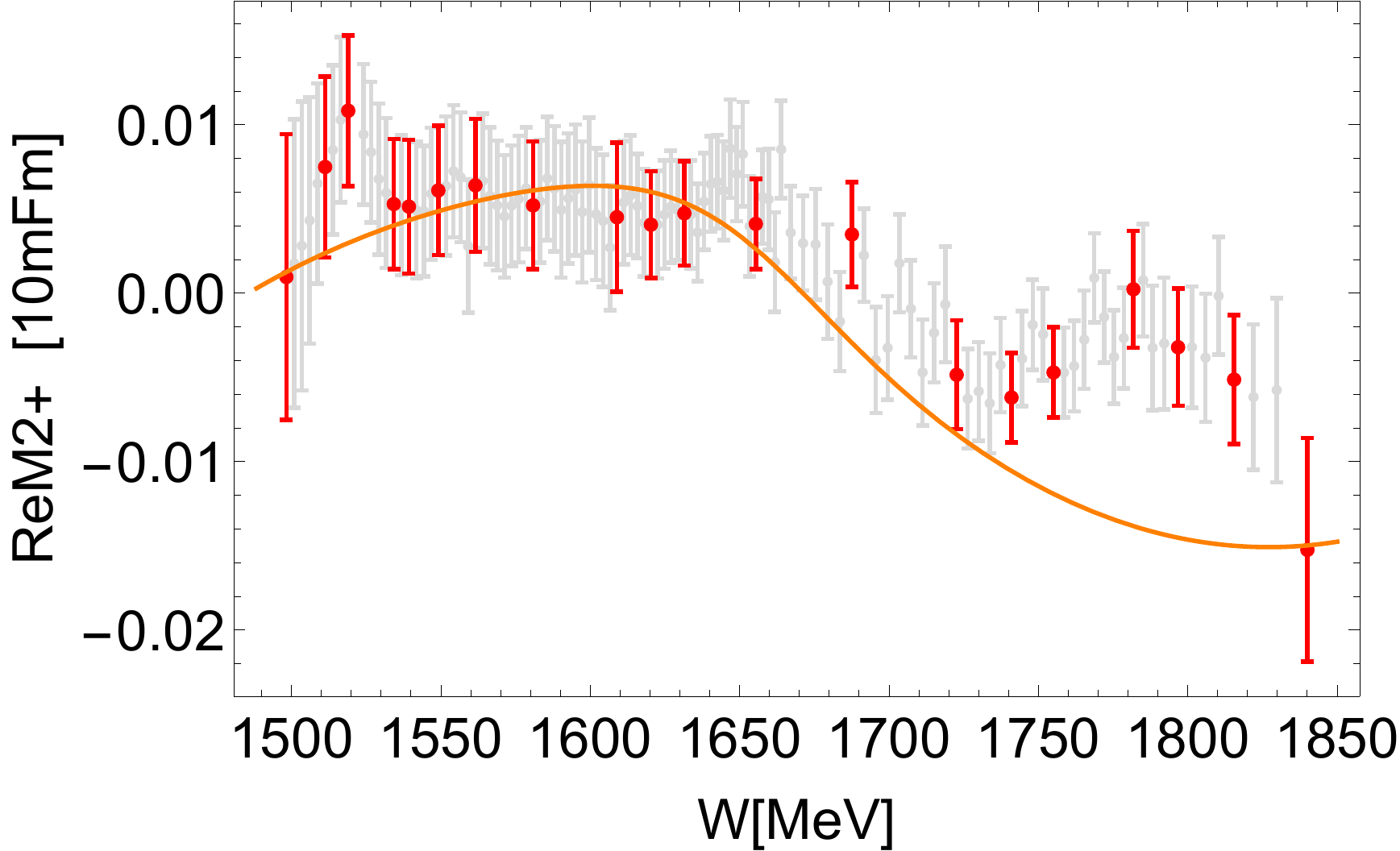} \hspace{0.5cm}
\includegraphics[width=0.4\textwidth]{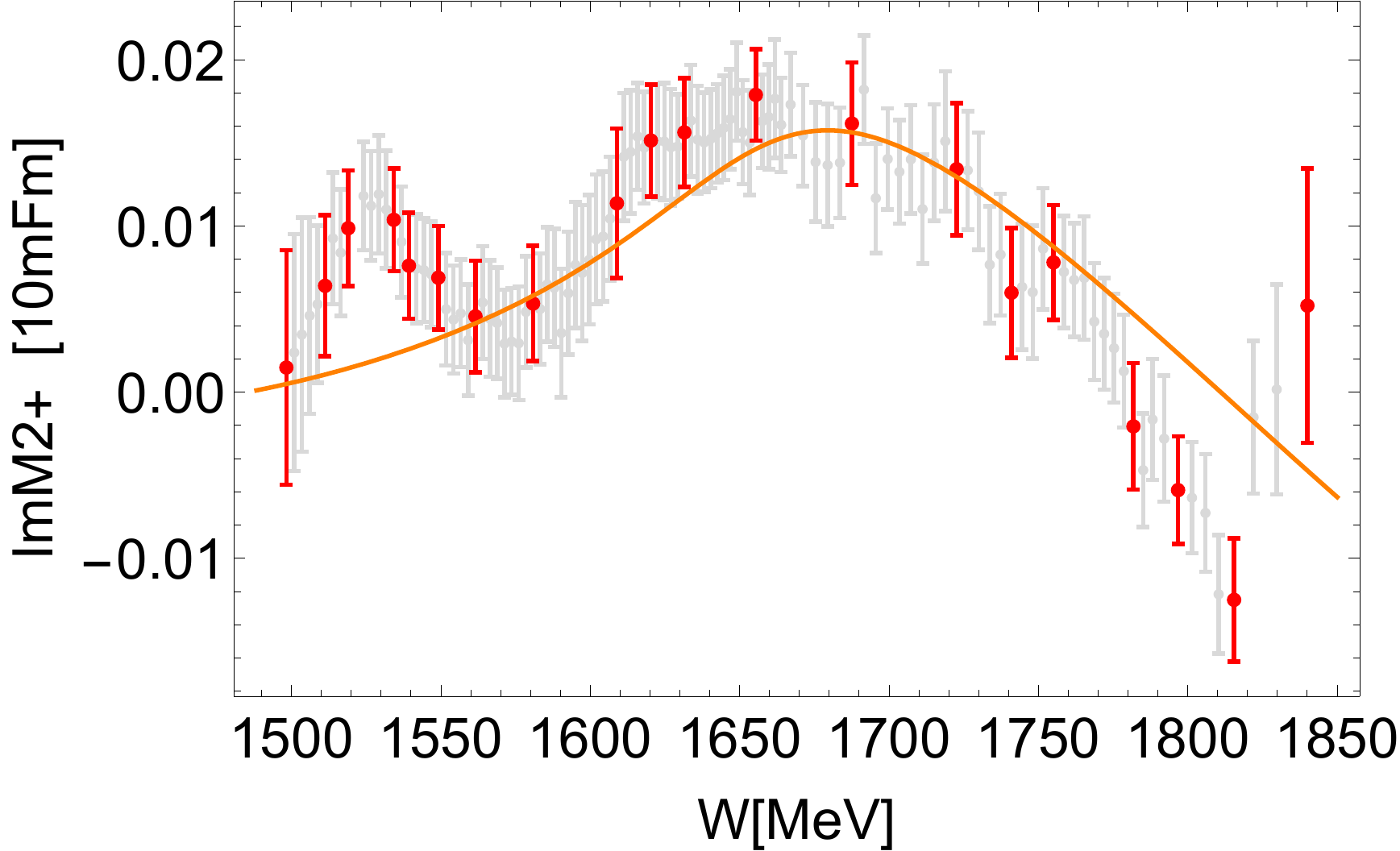}  \\
\includegraphics[width=0.4\textwidth]{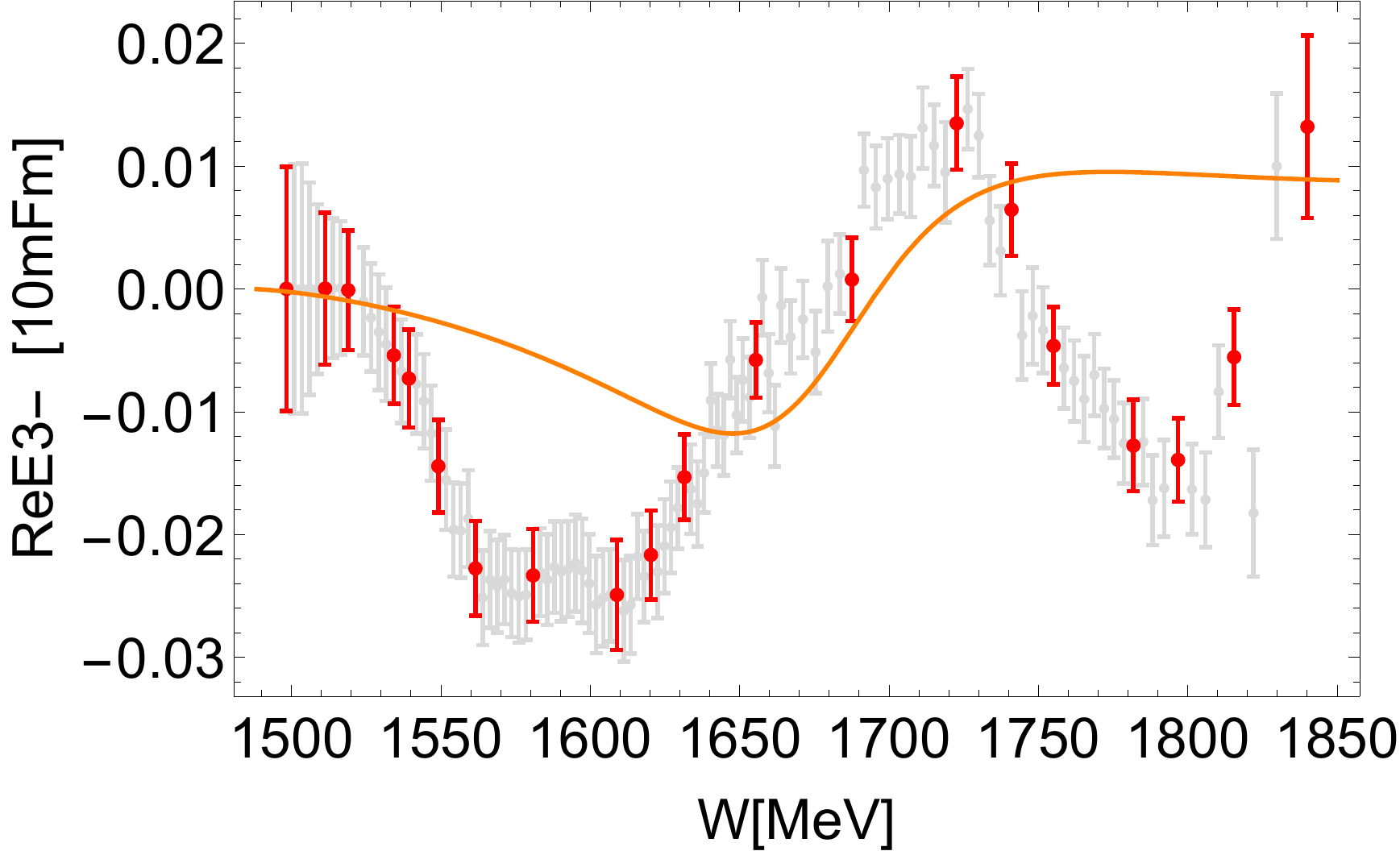} \hspace{0.5cm}
\includegraphics[width=0.4\textwidth]{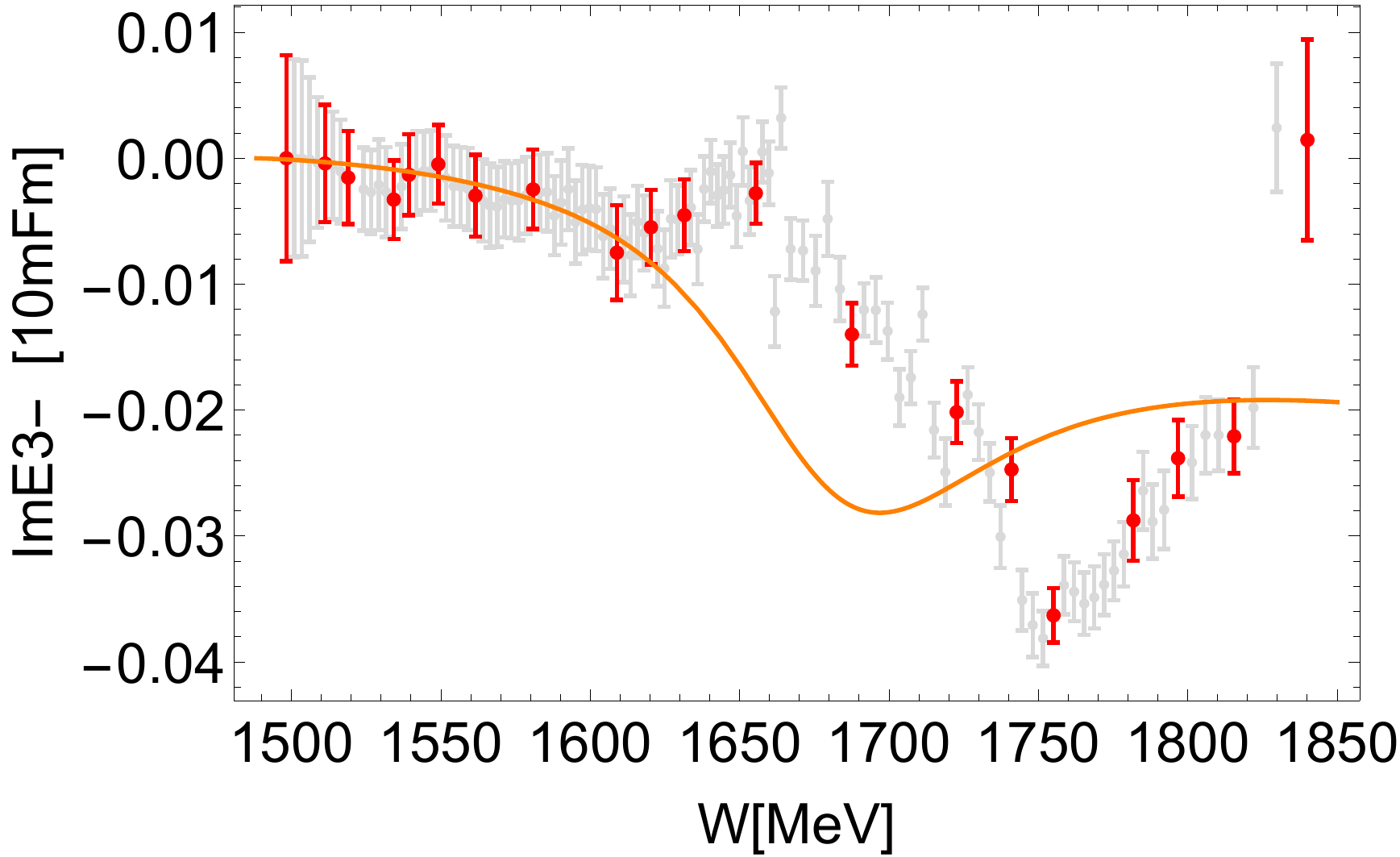}  \\
\includegraphics[width=0.4\textwidth]{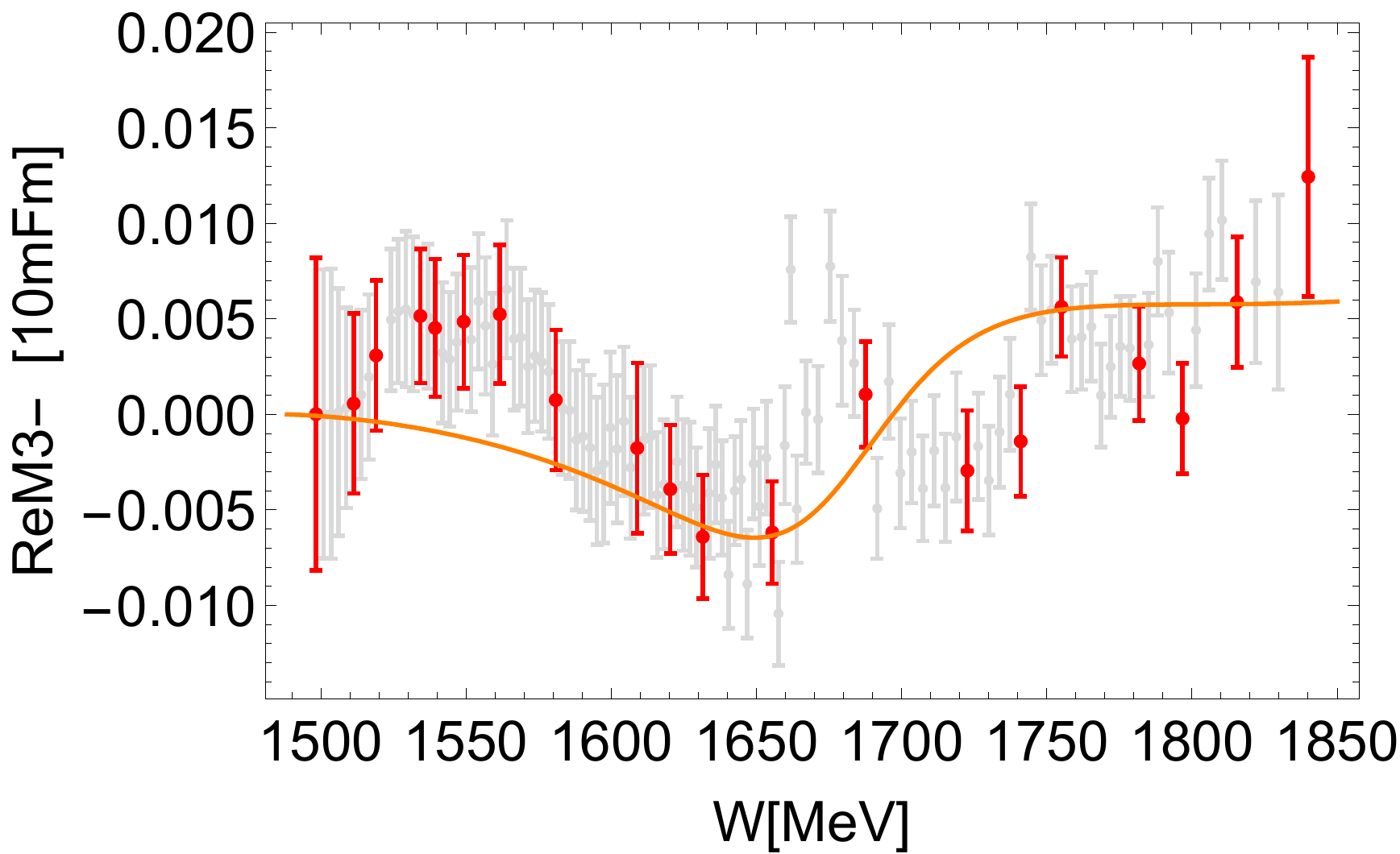} \hspace{0.5cm}
\includegraphics[width=0.4\textwidth]{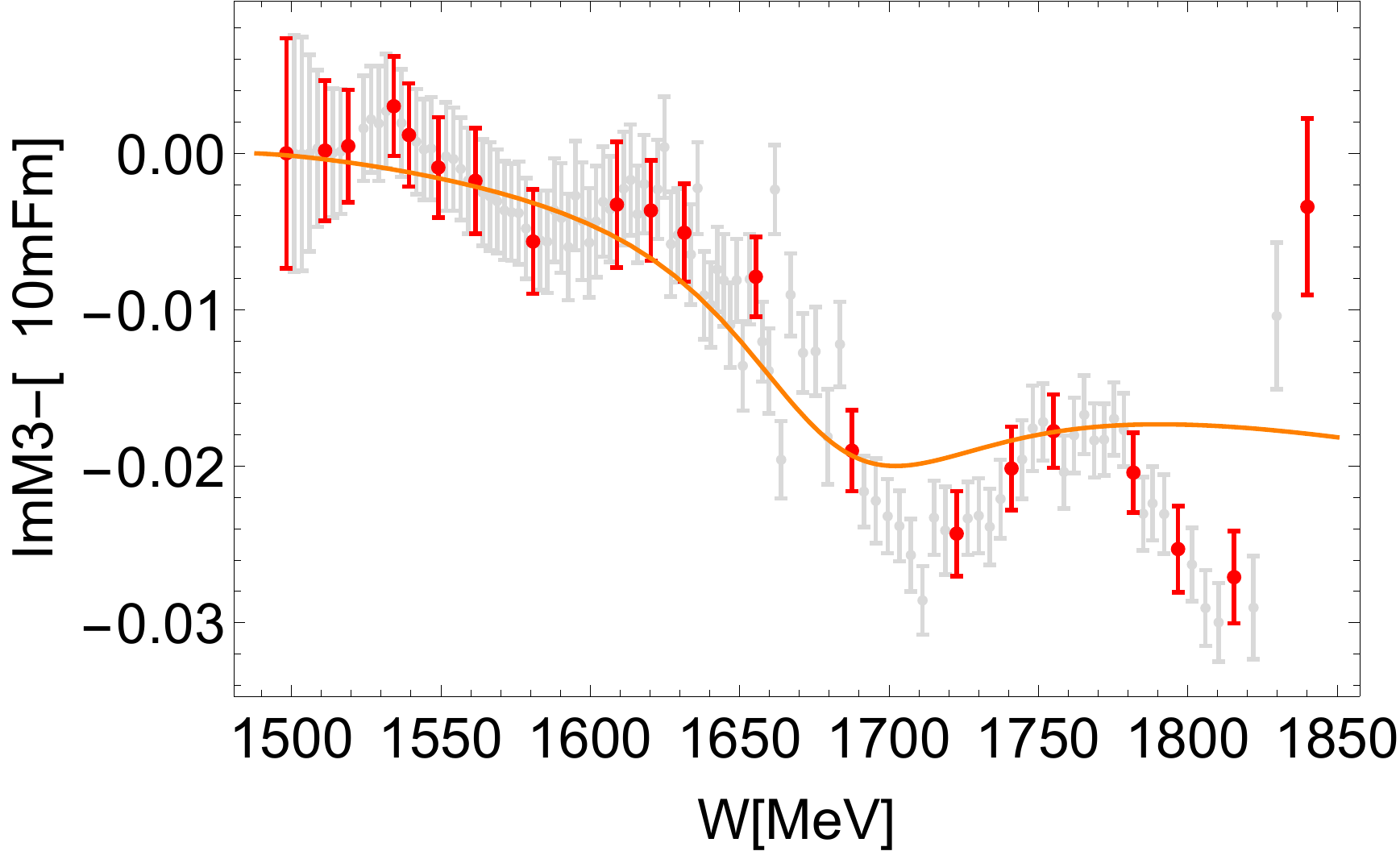}  \\
\caption{\label{Multipoles:Sol2}(Color online) Multipoles for L=0, 1, 2 and 3 partial waves for Sol 2. Grey discrete symbols correspond to Set 1, and red discrete symbols correspond to Set 2. Orange full line is BG2014-2 solution for comparison.      }
\ec
\end{figure}

\newpage
 In Fig.~\ref{Chi2:Sol1}  we repeat the plot of  $\chi^2/ndf$ for the whole process for $Sol \, 1$  from Fig.~\ref{Chi:comparison}, but in addition we give the  $\chi^2/N_{\text{data}}$ for the whole fit, and  $\chi^2/N_{\text{data}}$  for individual observables. The $\chi^2/N_{\text{data}}$  for individual observables is extremely important as it gives one the internal consistency of  used data base. We do not show the similar figure for  $Sol \, 2$  as two figures are practically indistinguishable.
 \begin{figure*}[h!]
\bc
\includegraphics[height=0.22\textwidth,width=0.35\textwidth]{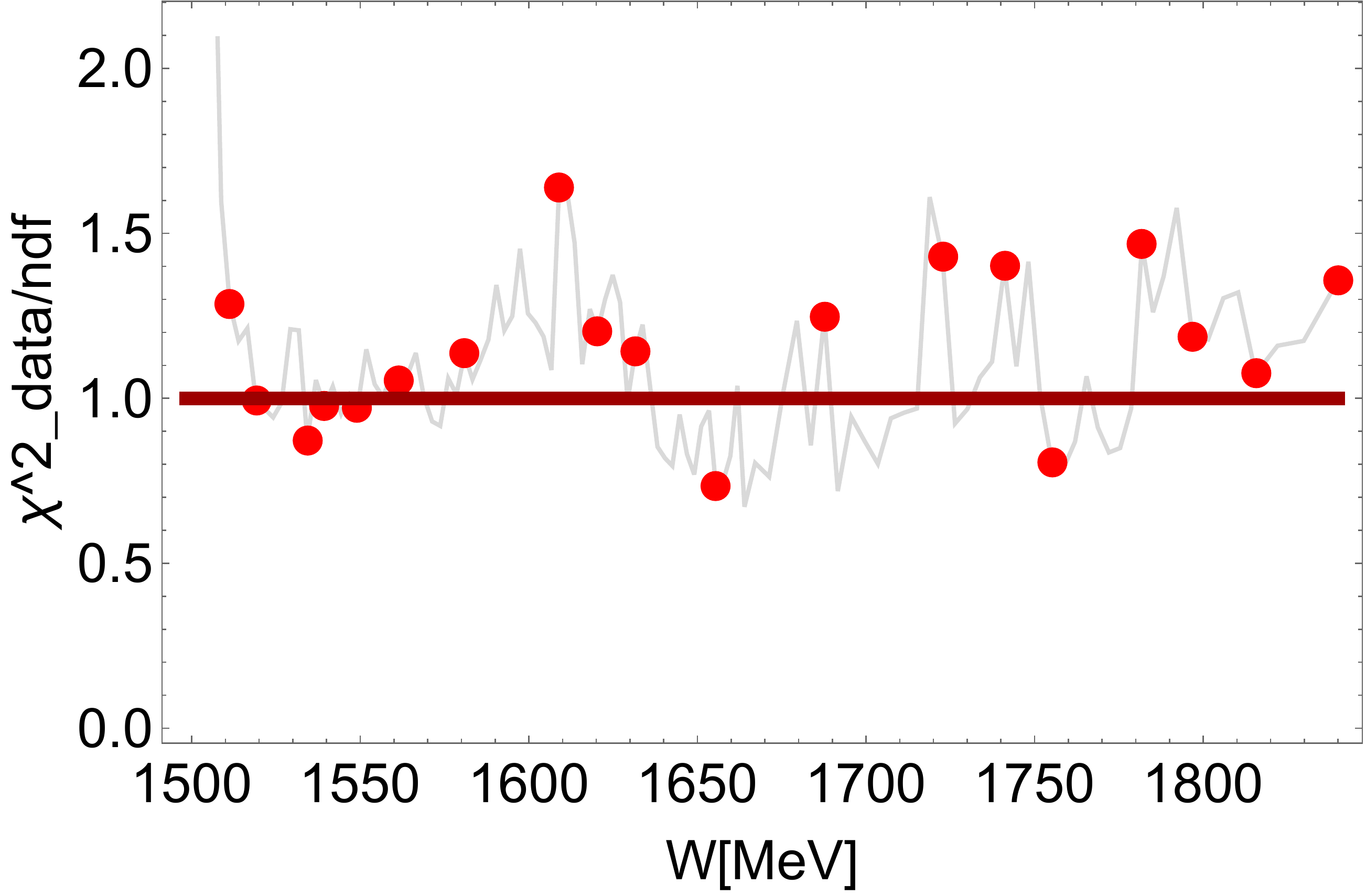} \hspace{0.5cm}
\includegraphics[height=0.22\textwidth,width=0.35\textwidth]{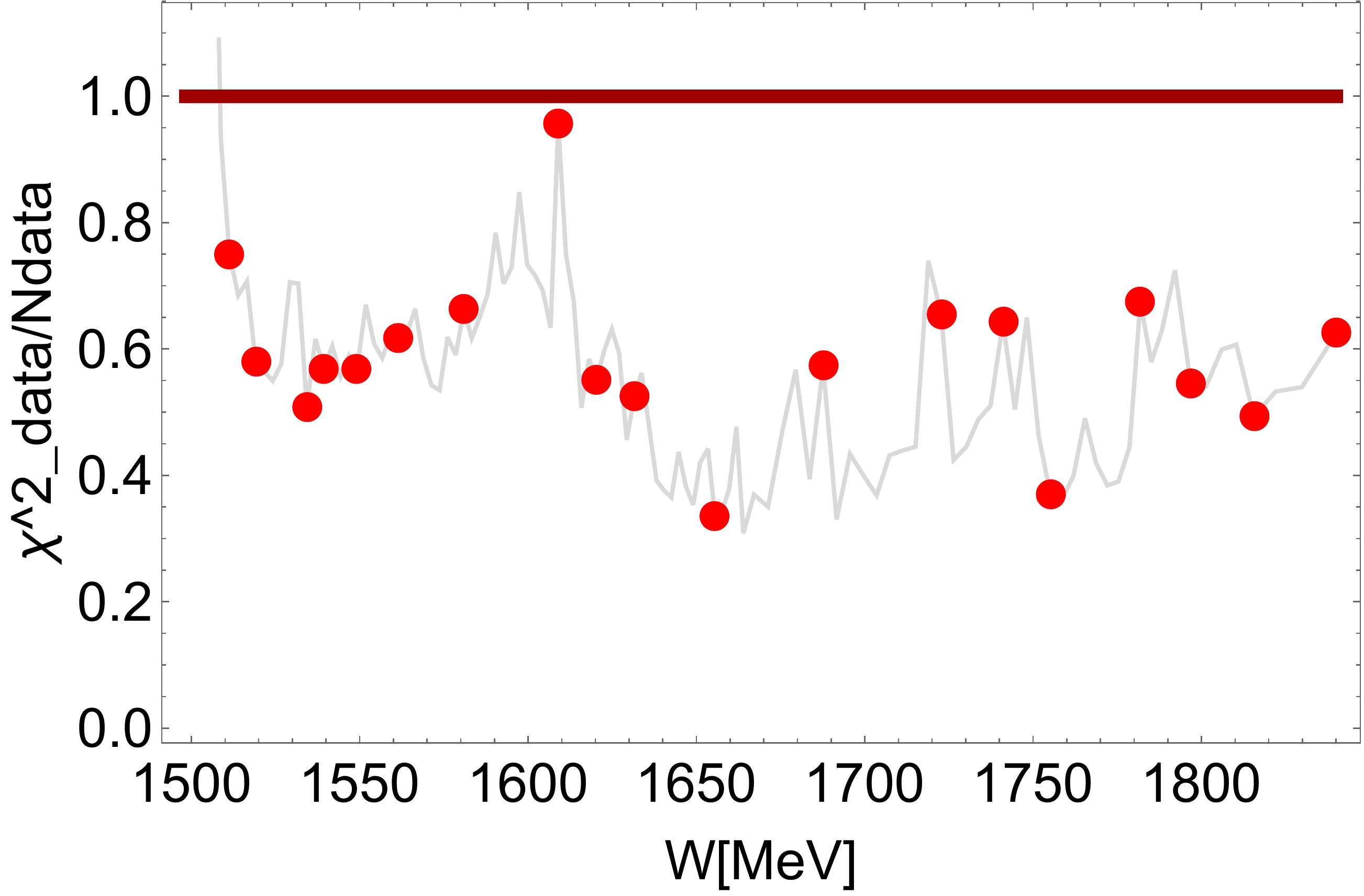} \\
\includegraphics[height=0.22\textwidth,width=0.35\textwidth]{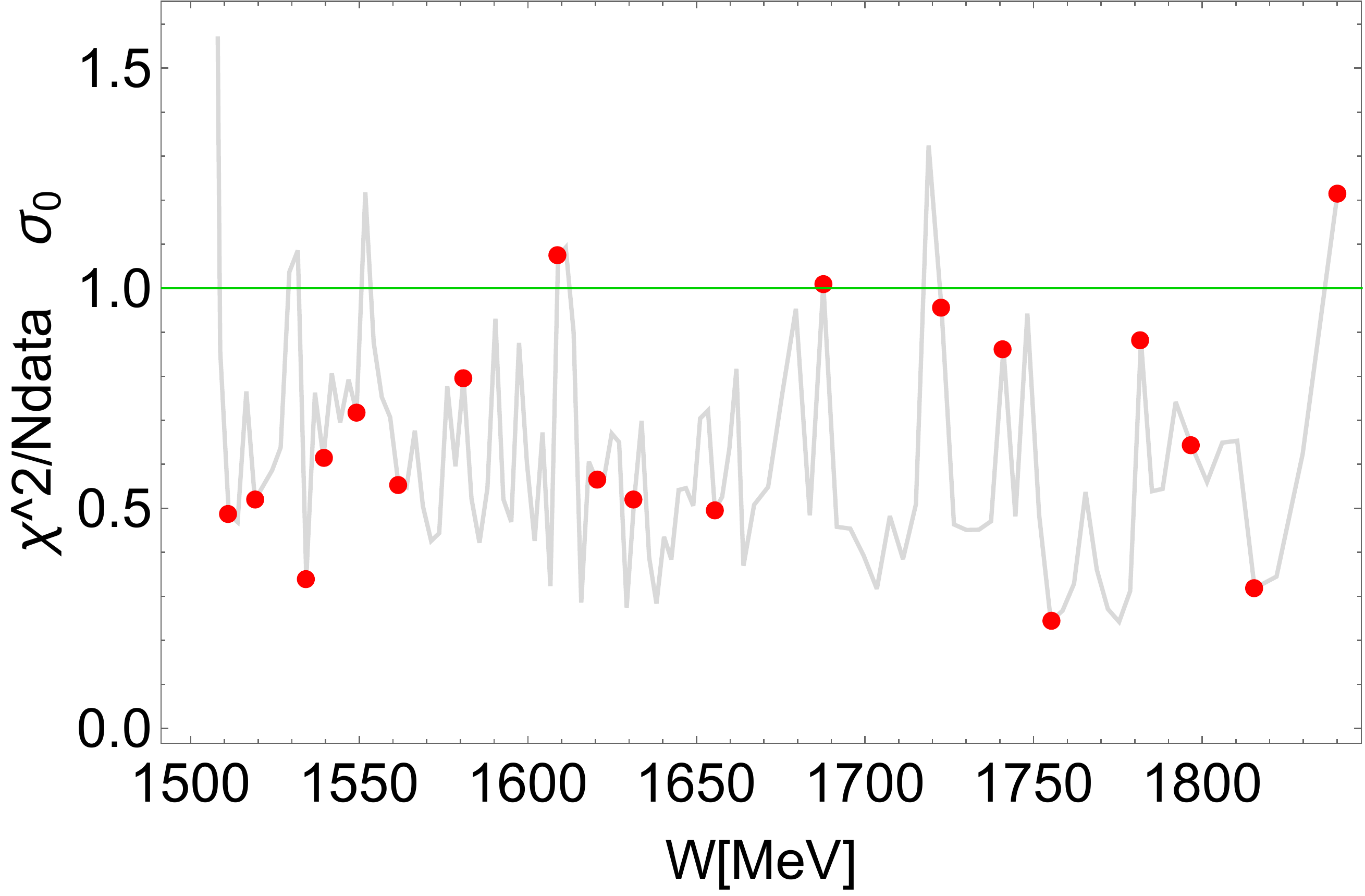} \hspace{0.5cm}
\includegraphics[height=0.22\textwidth,width=0.35\textwidth]{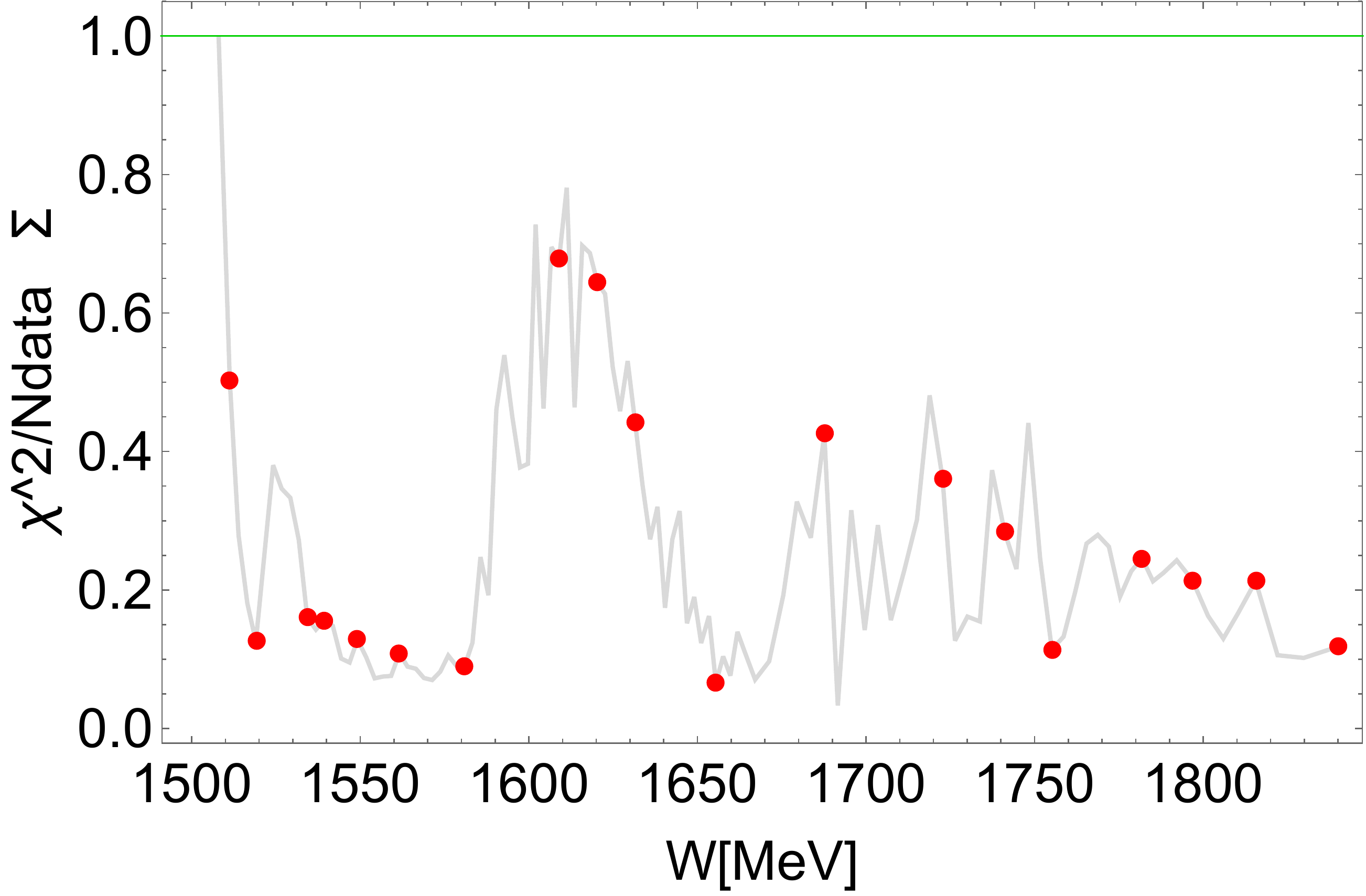} \\
\includegraphics[height=0.22\textwidth,width=0.35\textwidth]{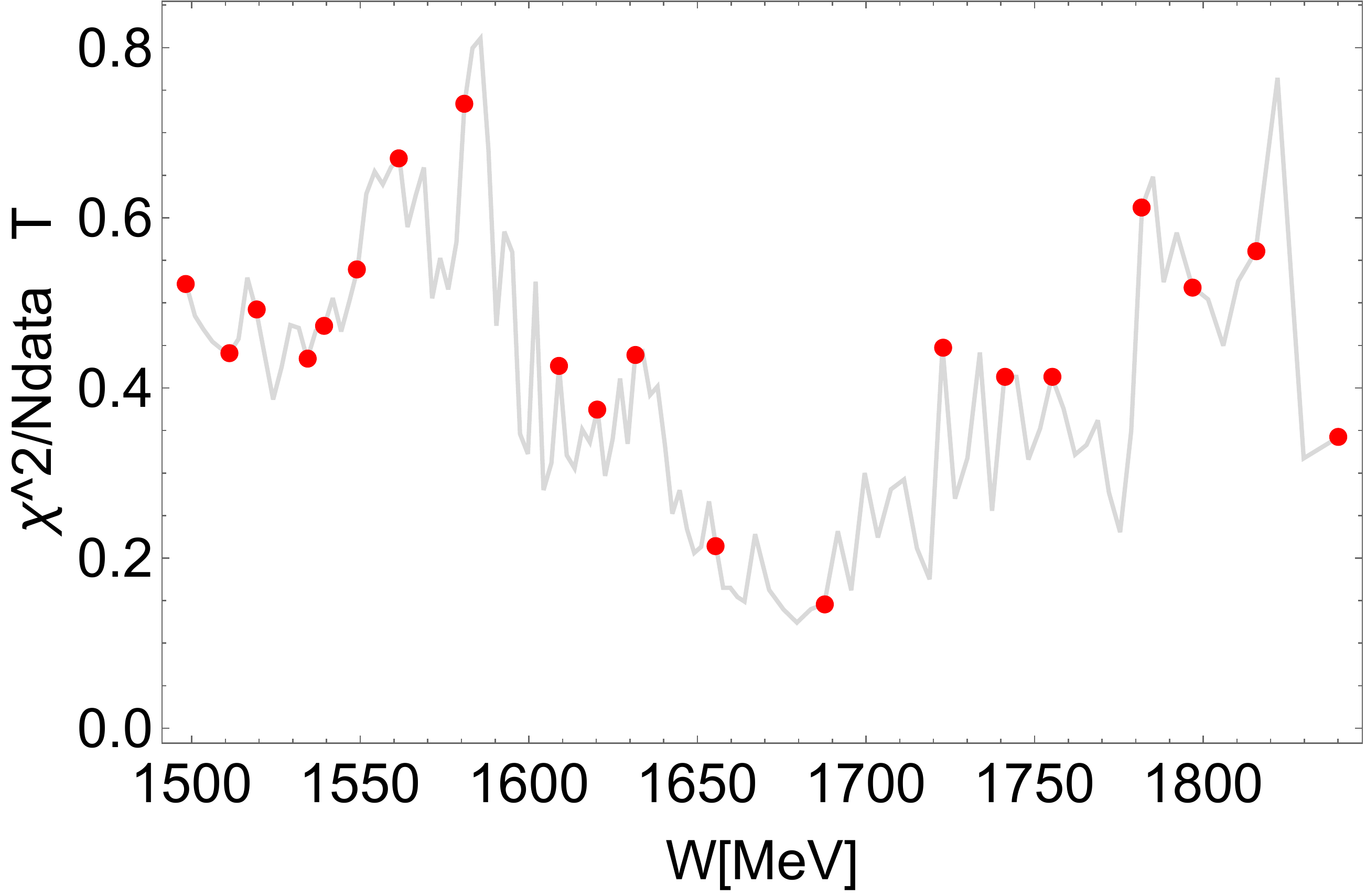} \hspace{0.5cm}
\includegraphics[height=0.22\textwidth,width=0.35\textwidth]{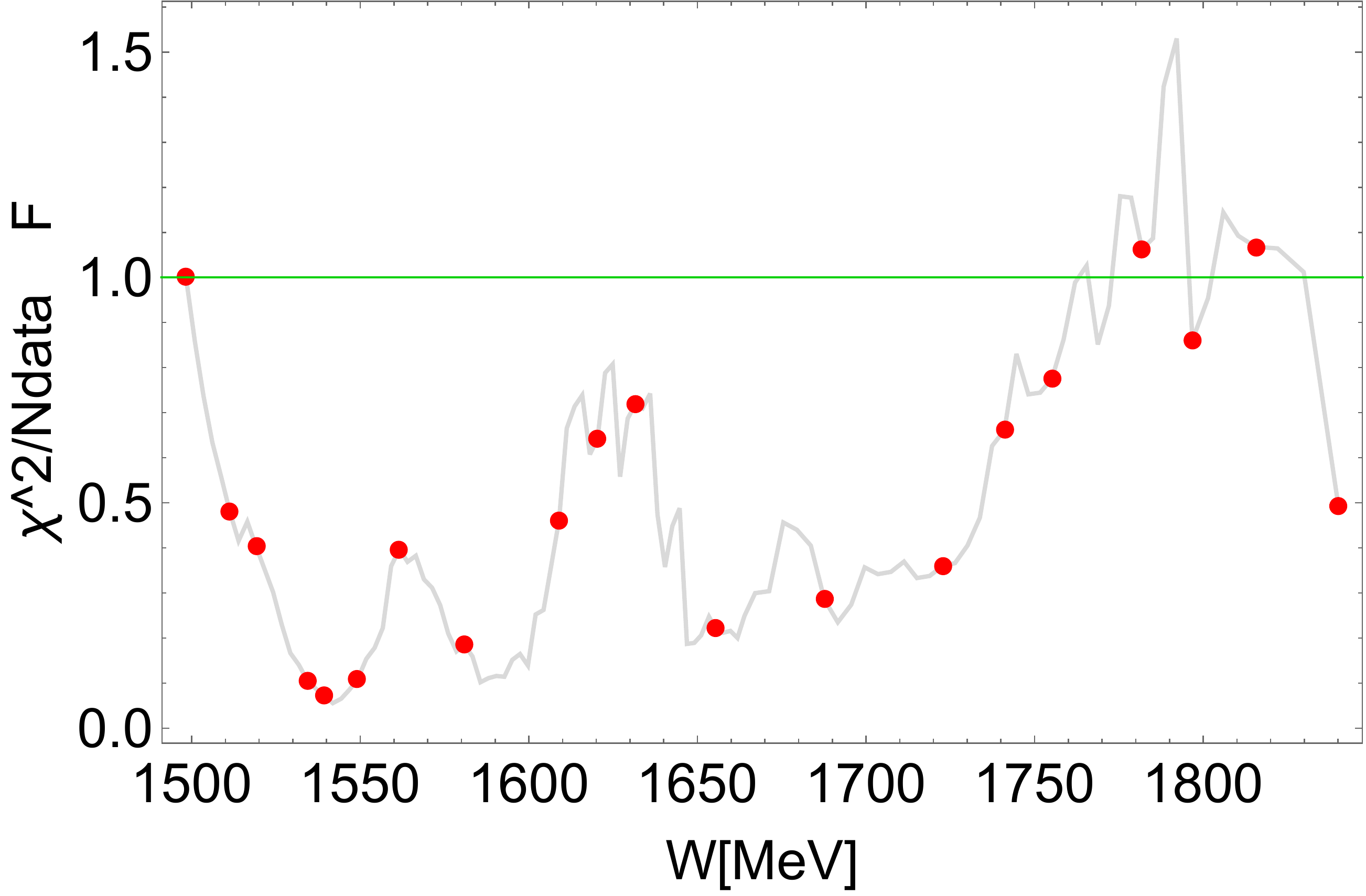} \\
\includegraphics[height=0.22\textwidth,width=0.35\textwidth]{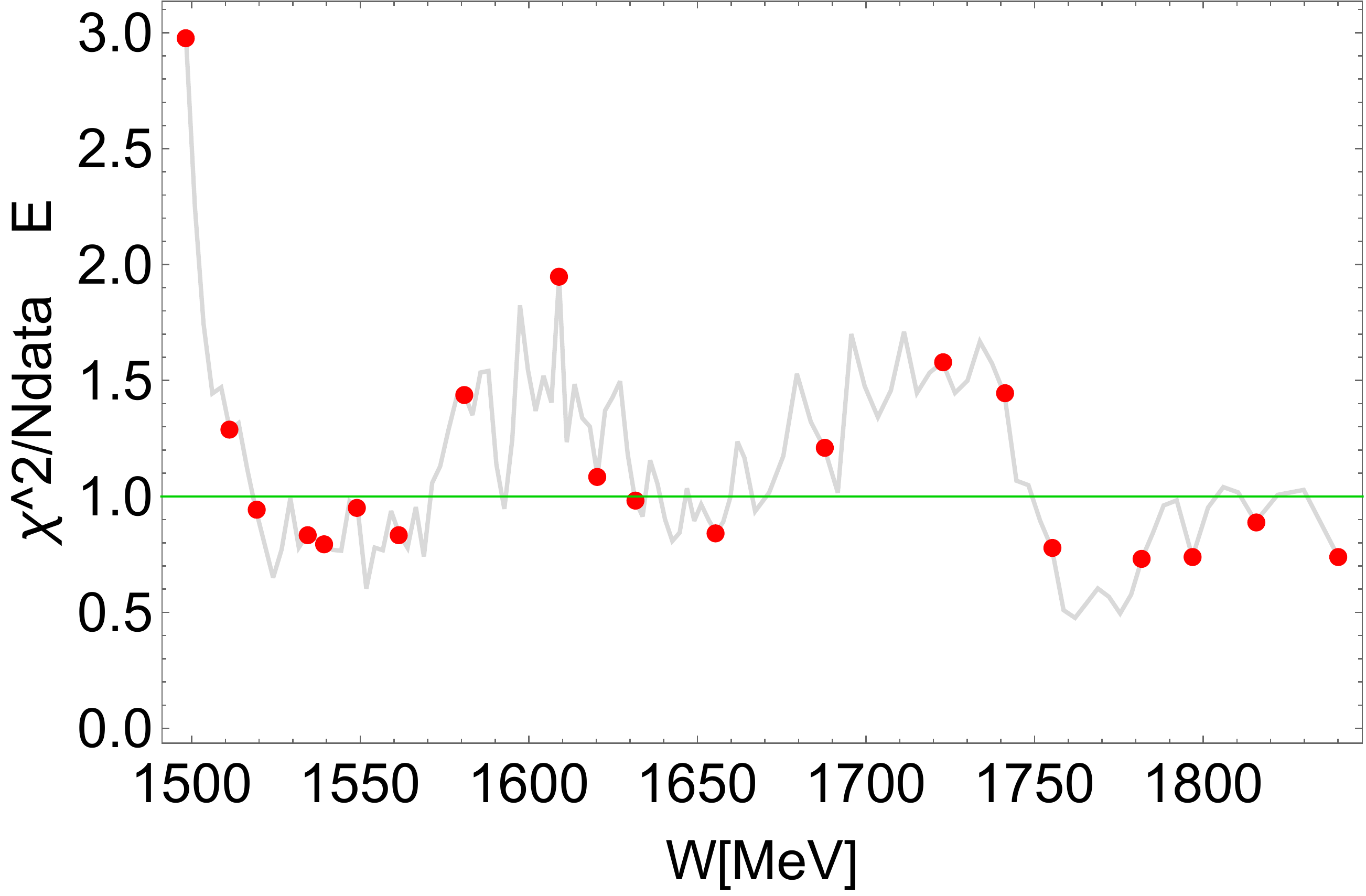} \hspace{0.5cm}
\includegraphics[height=0.22\textwidth,width=0.35\textwidth]{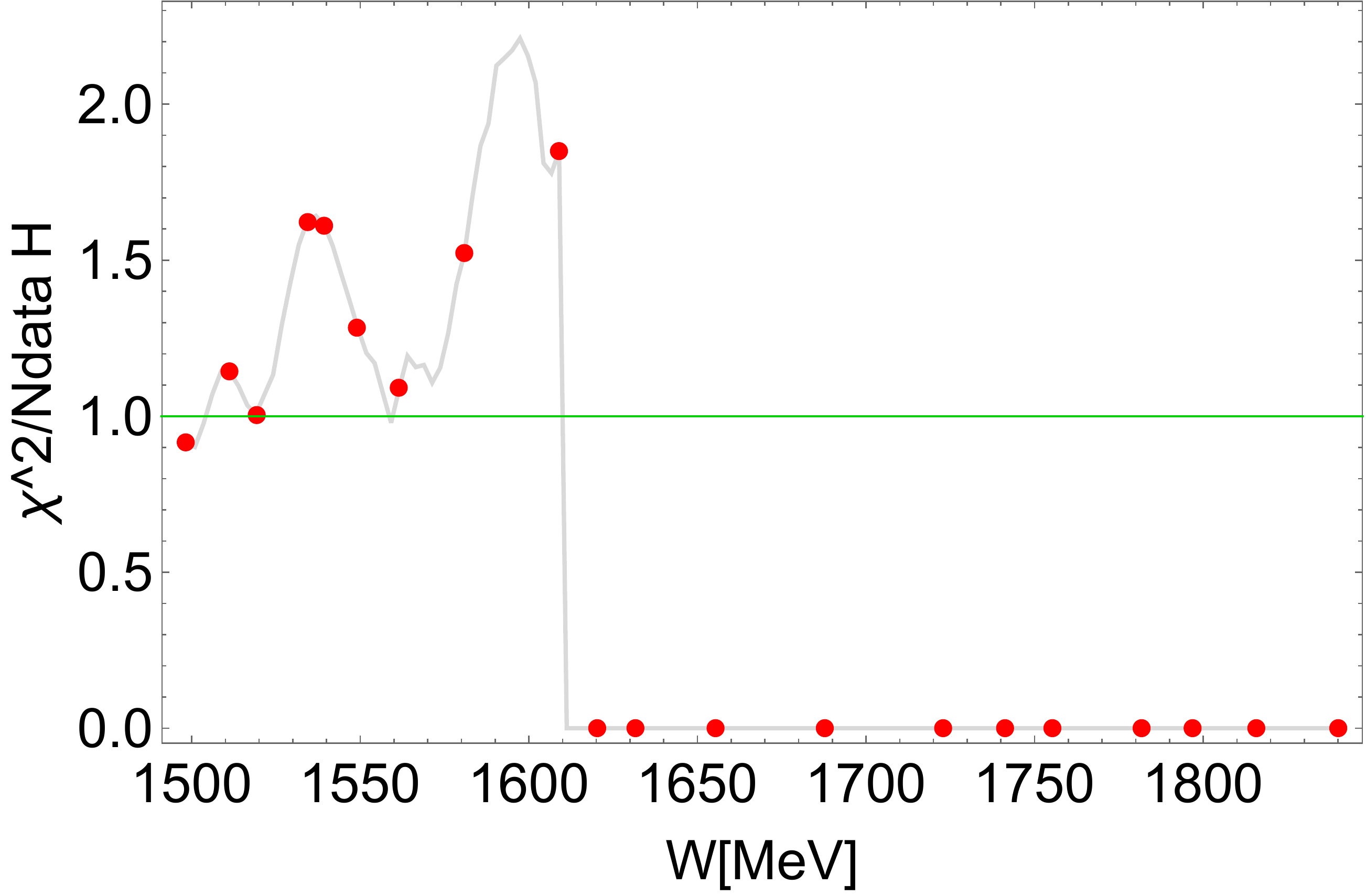} \\
\includegraphics[height=0.22\textwidth,width=0.35\textwidth]{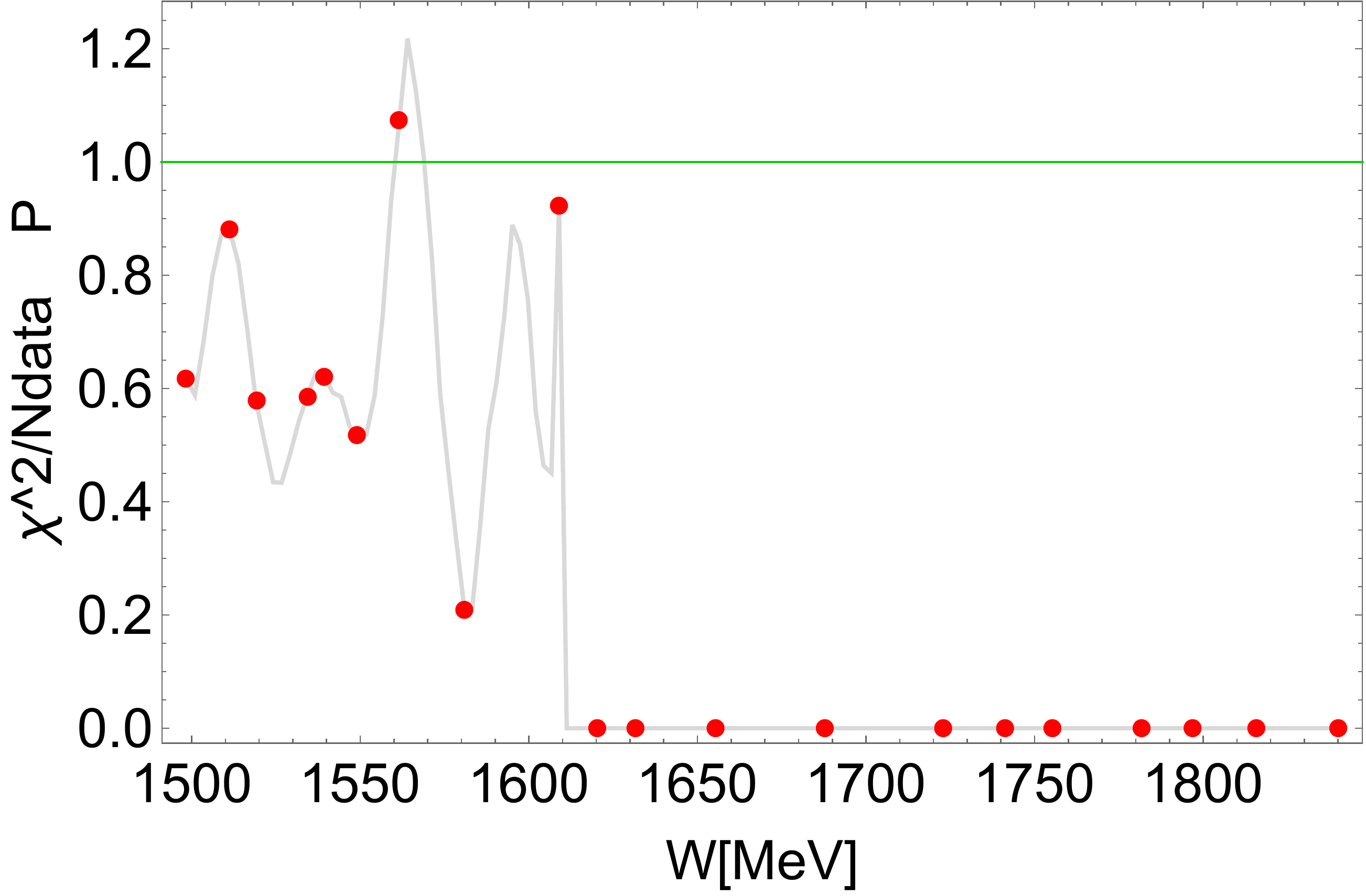} \hspace{0.5cm}
\includegraphics[height=0.22\textwidth,width=0.35\textwidth]{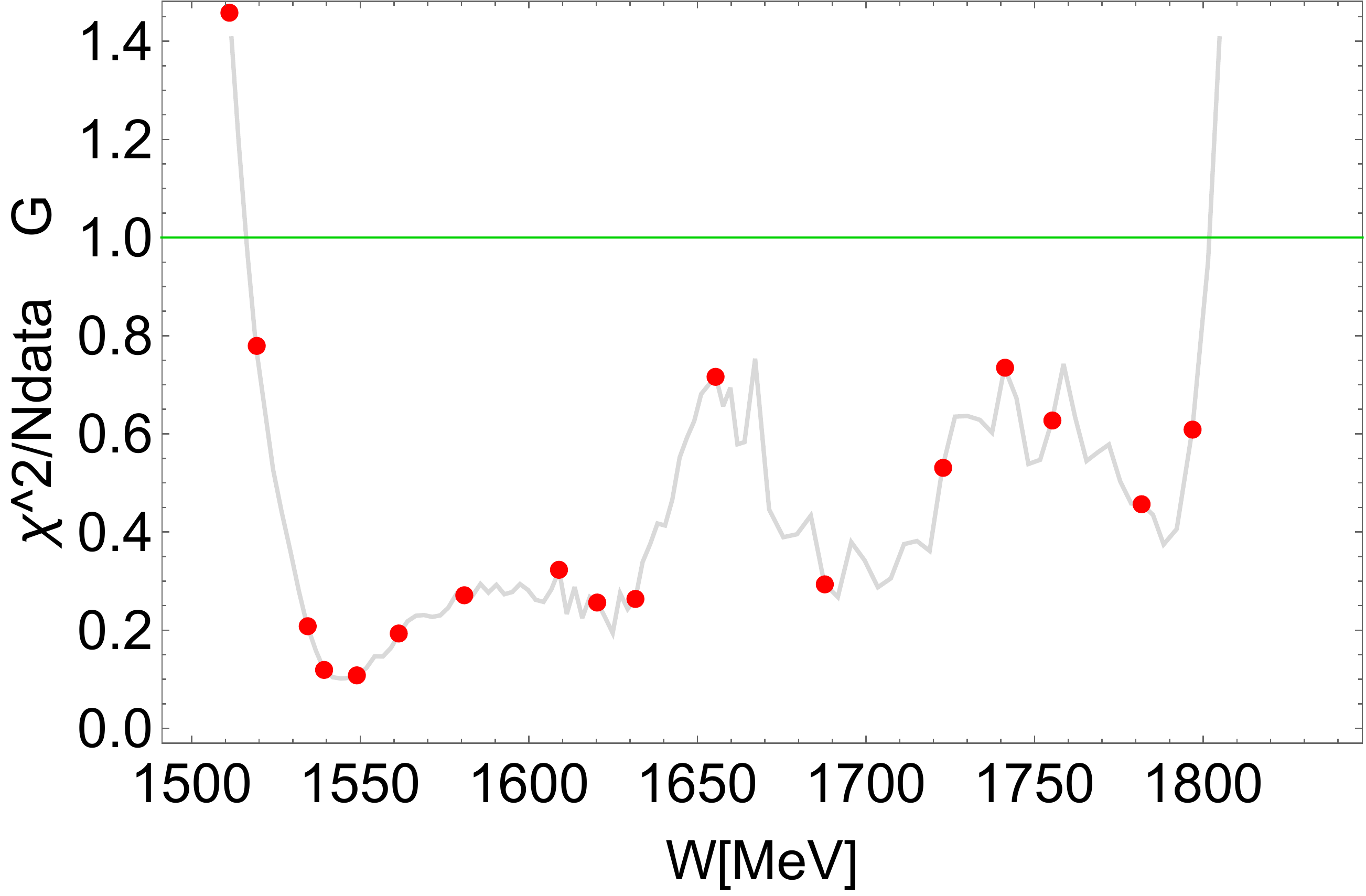} \\
\caption{\label{Chi2:Sol1}(Color online) $\chi^2/N_{\text{data}}$ for Sol 1.    }
\ec
\end{figure*}

\clearpage

However, the fits to the data for both solutions $Sol \, 1$  and $Sol \, 2$ are practically indistinguishable. So, in Figs.~\ref{Sol1:DCS}-\ref{Sol1:H} we give the agreement of $Sol \, 1$ and $Sol \, 2$ with the fitted observables, and compare it to the results of BG2014-2 solution.

\begin{figure*}[h!]
\bc
\includegraphics[width=0.9\textwidth]{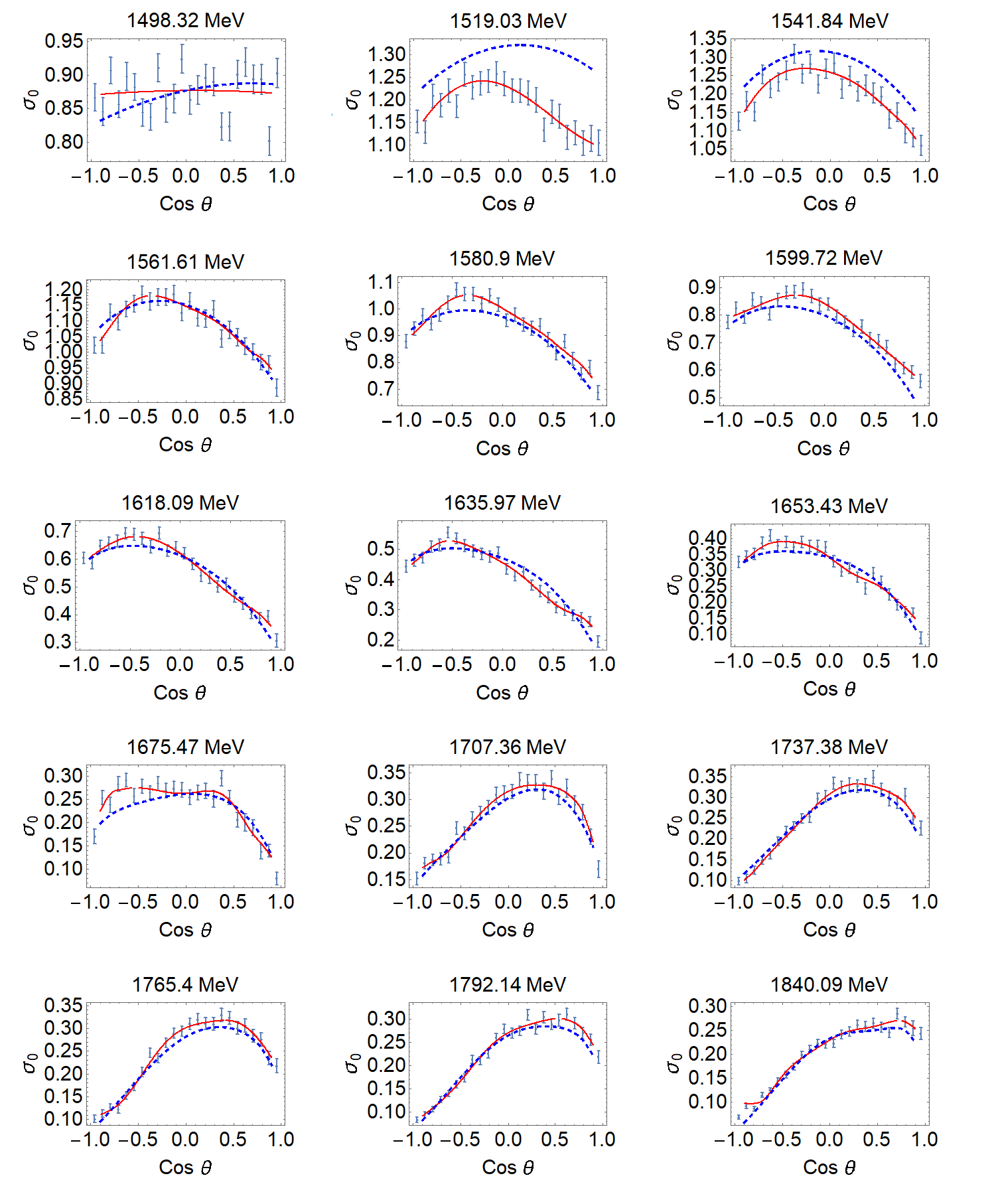}
\caption{\label{Sol1:DCS}(Color online) Comparison of experimental data for $\sigma_0$ (discrete symbols) with results of Sol 1 (red full line) and BG2014-2 fit (blue dashed line) at representative energies.    }
\ec
\end{figure*}

\begin{figure*}[h!]
\bc
\includegraphics[width=0.9\textwidth]{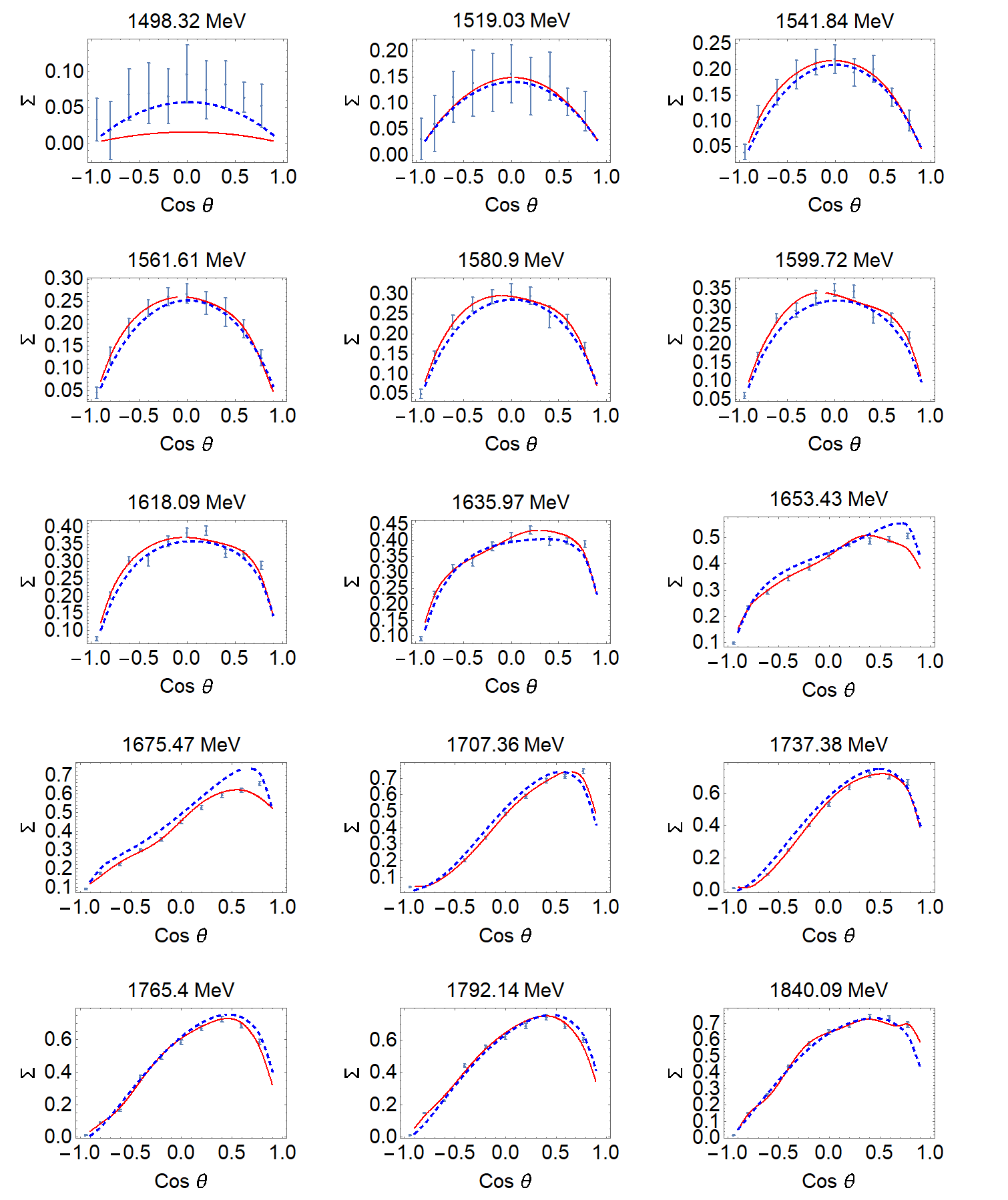}
\caption{\label{Sol1:Sigma}(Color online) Comparison of experimental data for $\Sigma$ (discrete symbols) with results of Sol 1 (red full line) and BG2014-2 fit (blue dashed line) at representative energies.    }
\ec
\end{figure*}

\begin{figure*}[h!]
\bc
\includegraphics[width=0.9\textwidth]{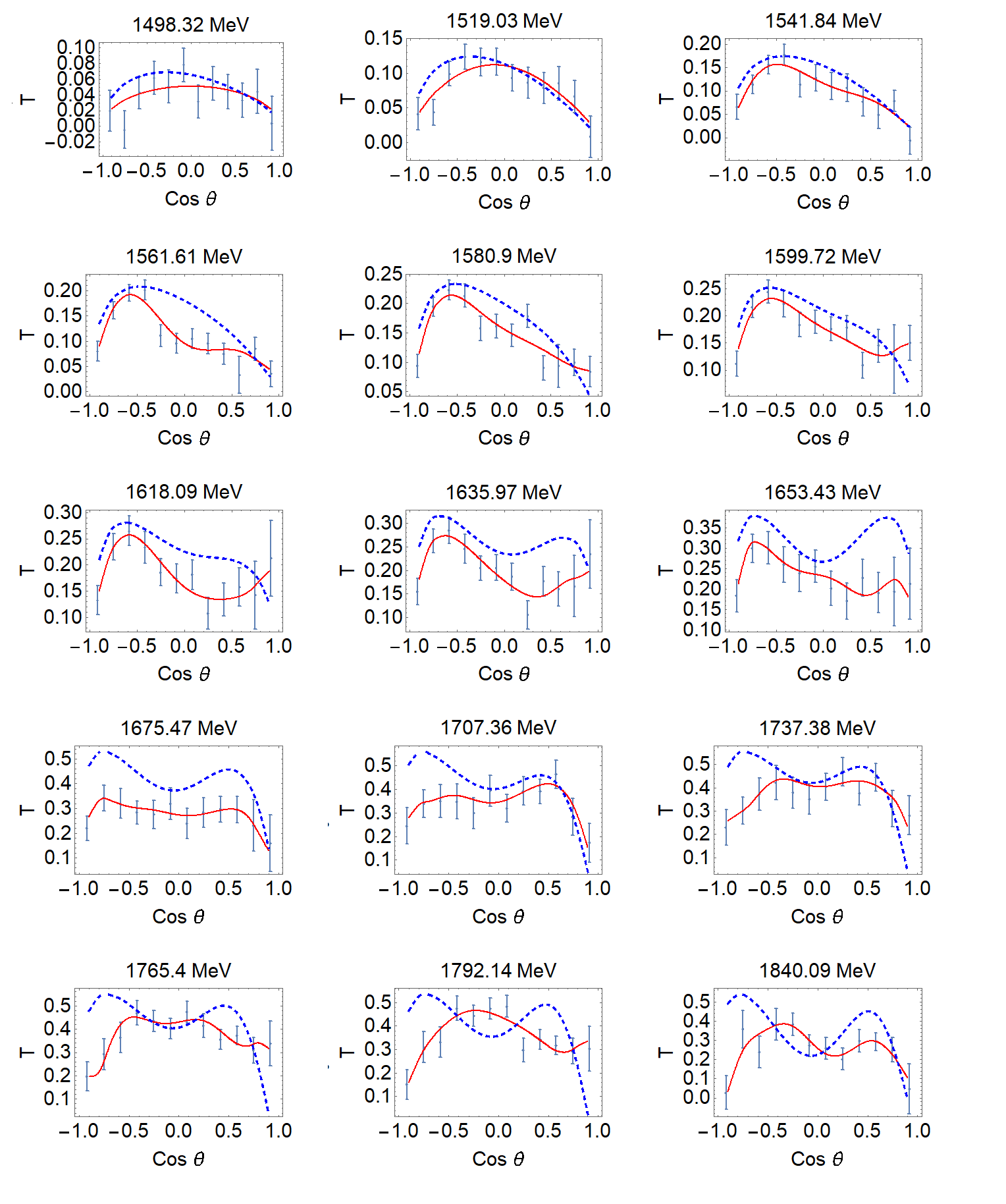}
\caption{\label{Sol1:T}(Color online) Comparison of experimental data for $T$ (discrete symbols) with results of Sol 1 (red full line) and BG2014-2 fit (blue dashed line) at representative energies.    }
\ec
\end{figure*}

\begin{figure*}[h!]
\bc
\includegraphics[width=0.9\textwidth]{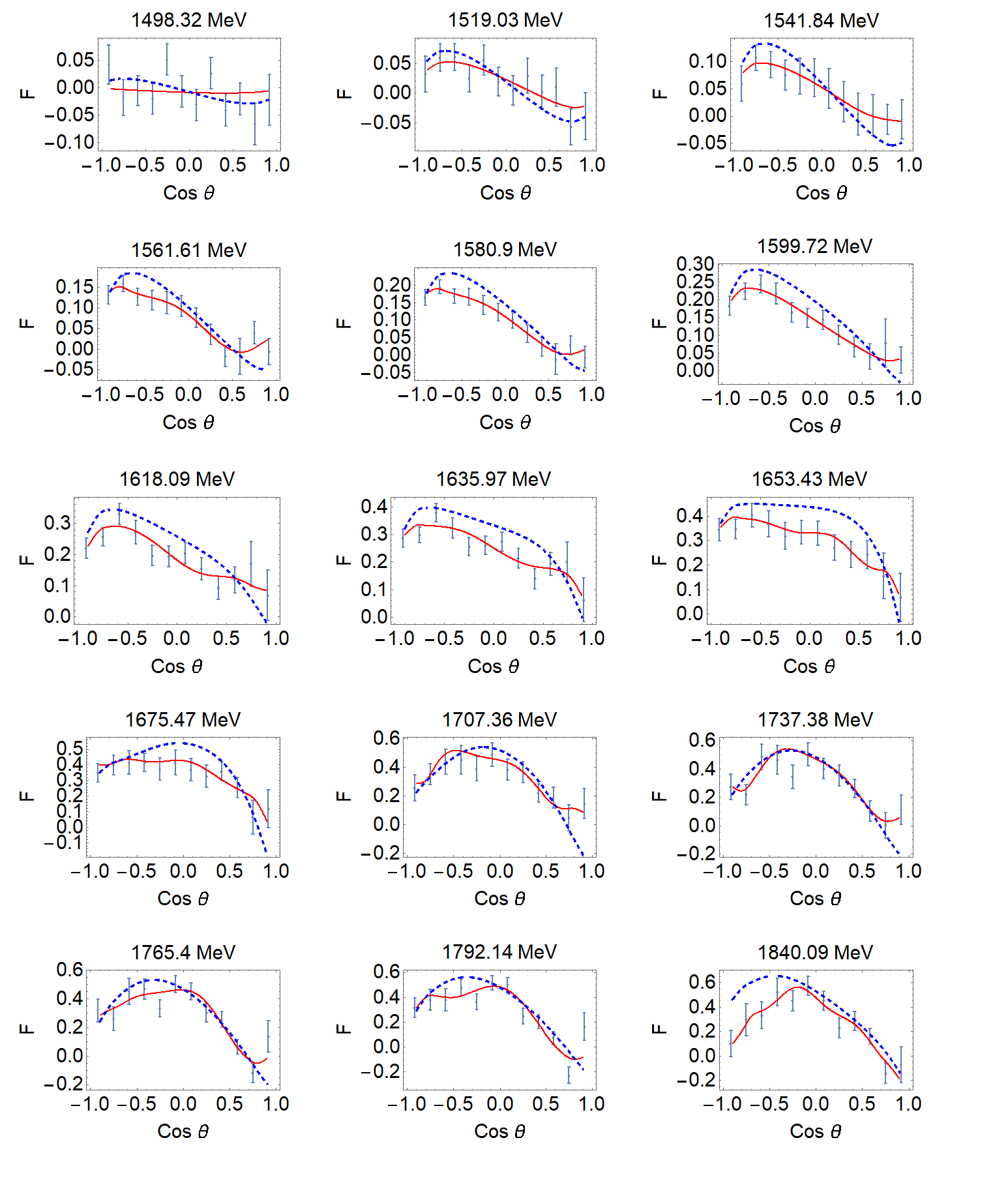}
\caption{\label{Sol1:F}(Color online) Comparison of experimental data for $F$ (discrete symbols) with results of Sol 1 (red full line) and BG2014-2 fit (blue dashed line) at representative energies.    }
\ec
\end{figure*}

\begin{figure*}[h!]
\bc
\includegraphics[width=0.9\textwidth]{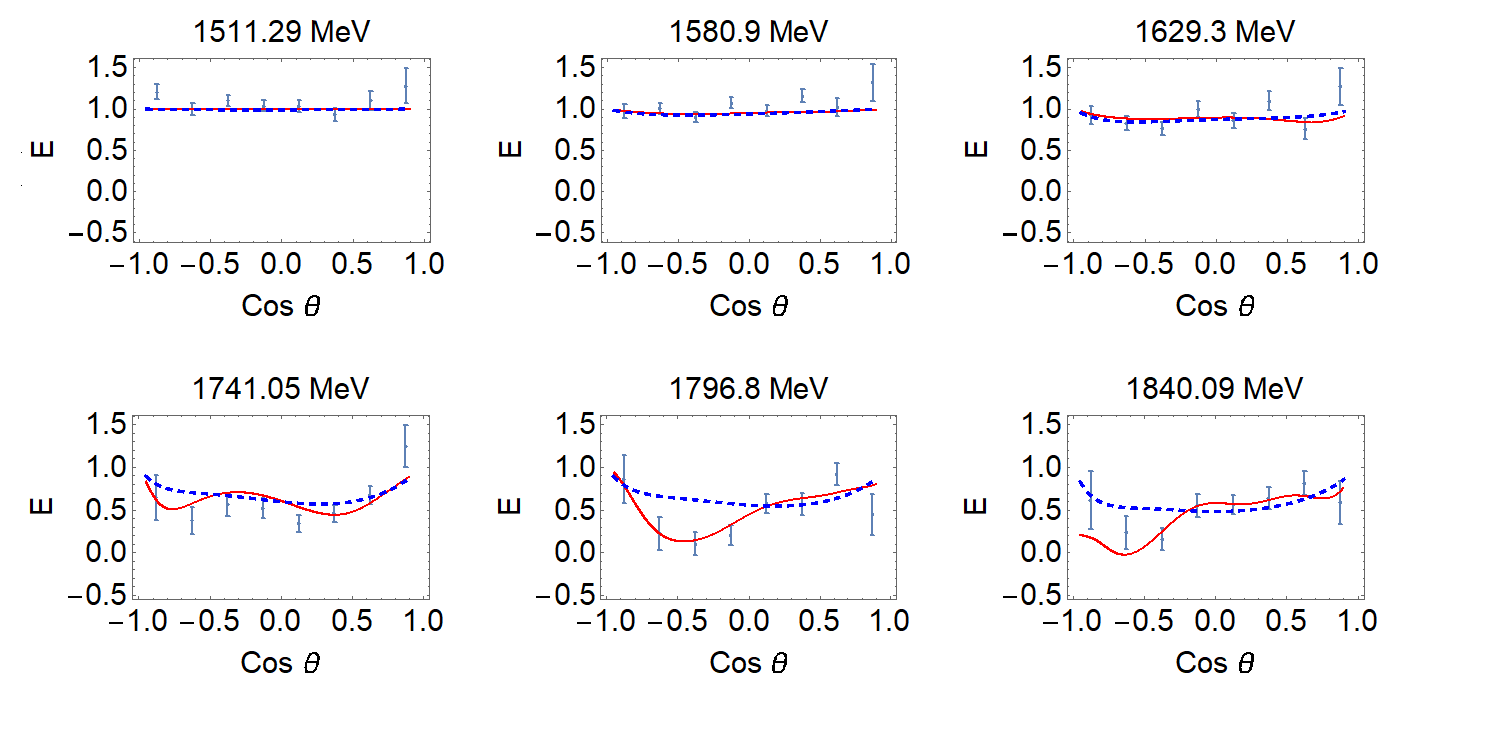}
\caption{\label{Sol1:E}(Color online) Comparison of experimental data for $E$ (discrete symbols) with results of Sol 1 (red full line) and BG2014-2 fit (blue dashed line) at measured energies.    }
\ec
\end{figure*}

\begin{figure*}[h!]
\bc
\includegraphics[width=0.9\textwidth]{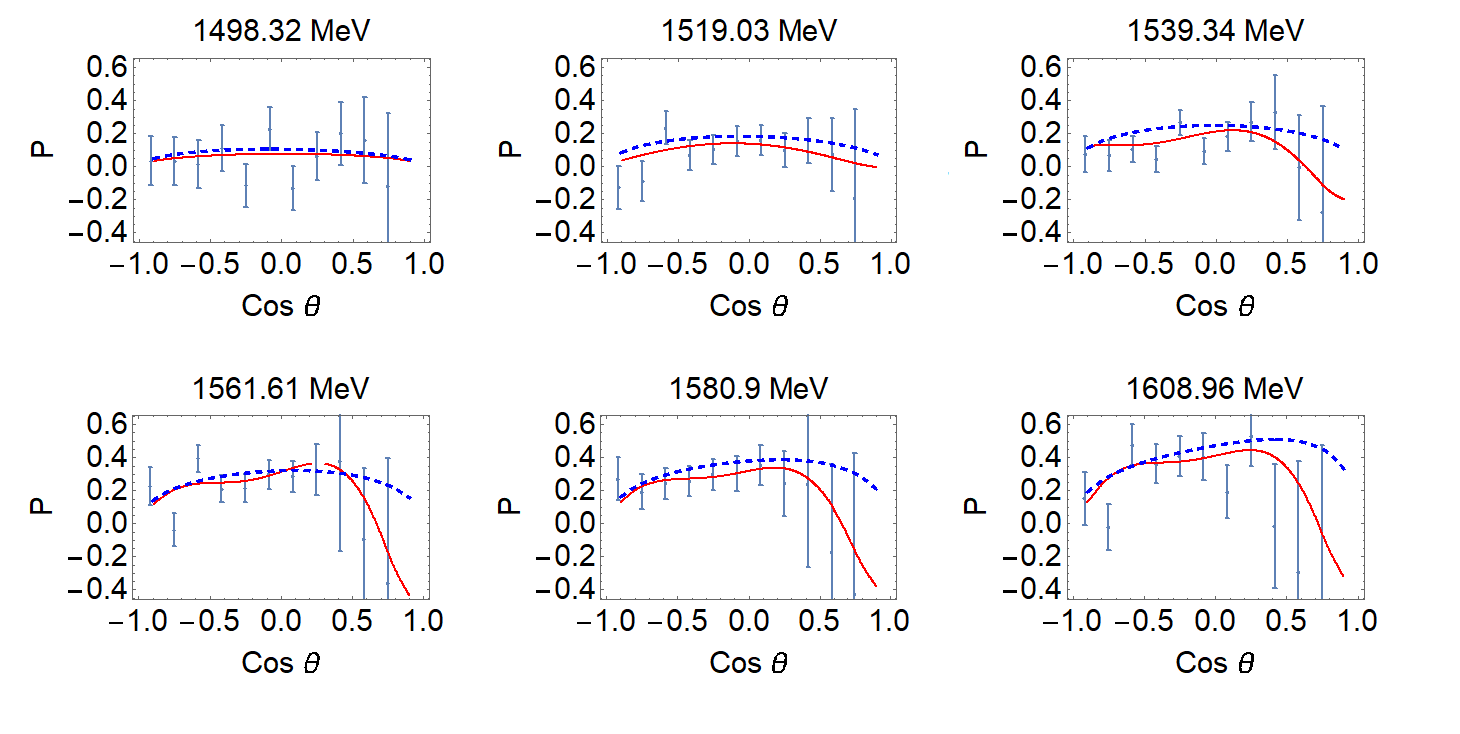}
\caption{\label{Sol1:P}(Color online) Comparison of experimental data for $P$ (discrete symbols) with results of Sol 1 (red full line) and BG2014-2 fit (blue dashed line) at measured energies.    }
\ec
\end{figure*}

\clearpage

\begin{figure*}[h!]
\bc
\includegraphics[width=0.9\textwidth]{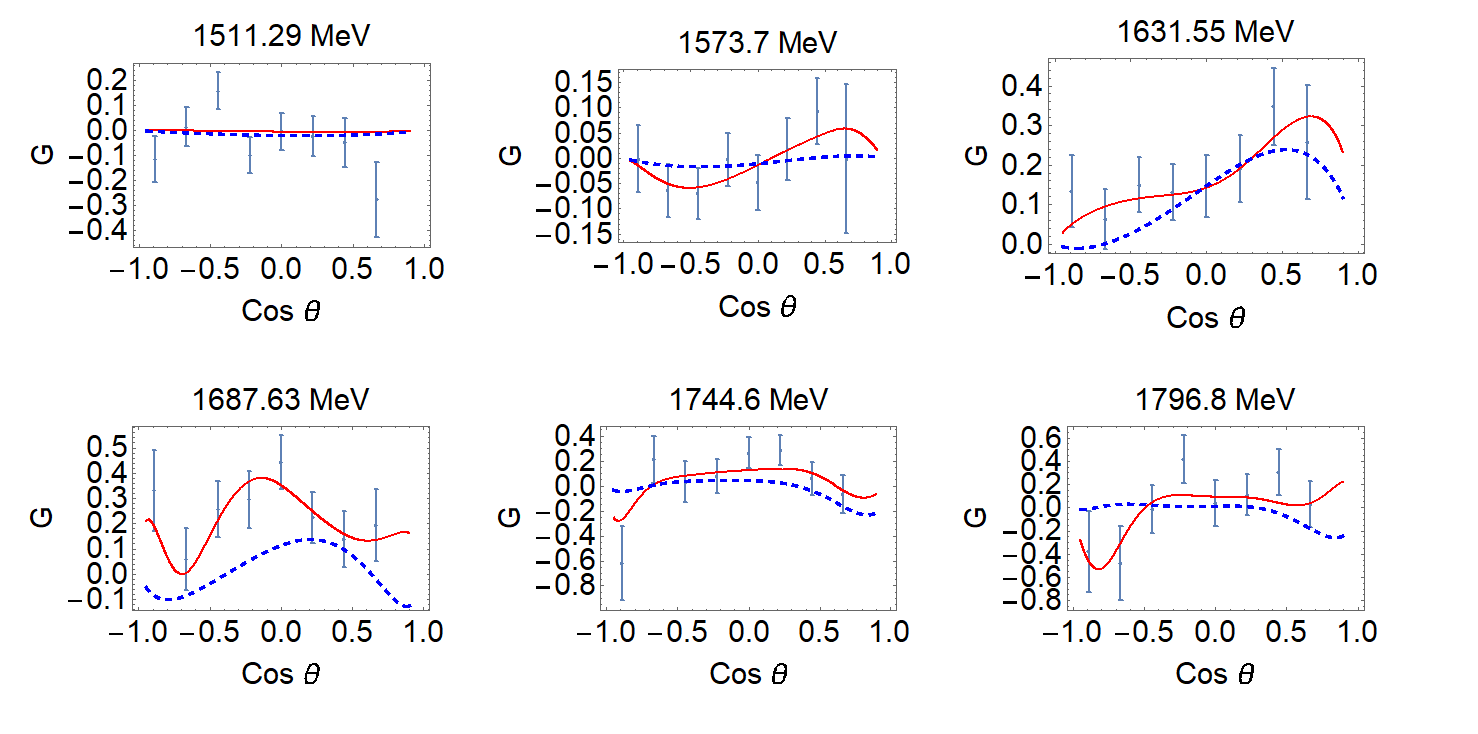}
\caption{\label{Sol1:G}(Color online) Comparison of experimental data for $G$ (discrete symbols) with results of Sol 1 (red full line) and BG2014-2 fit (blue dashed line) at measured energies.    }
\ec
\end{figure*}

\begin{figure*}[h!]
\bc
\includegraphics[width=0.9\textwidth]{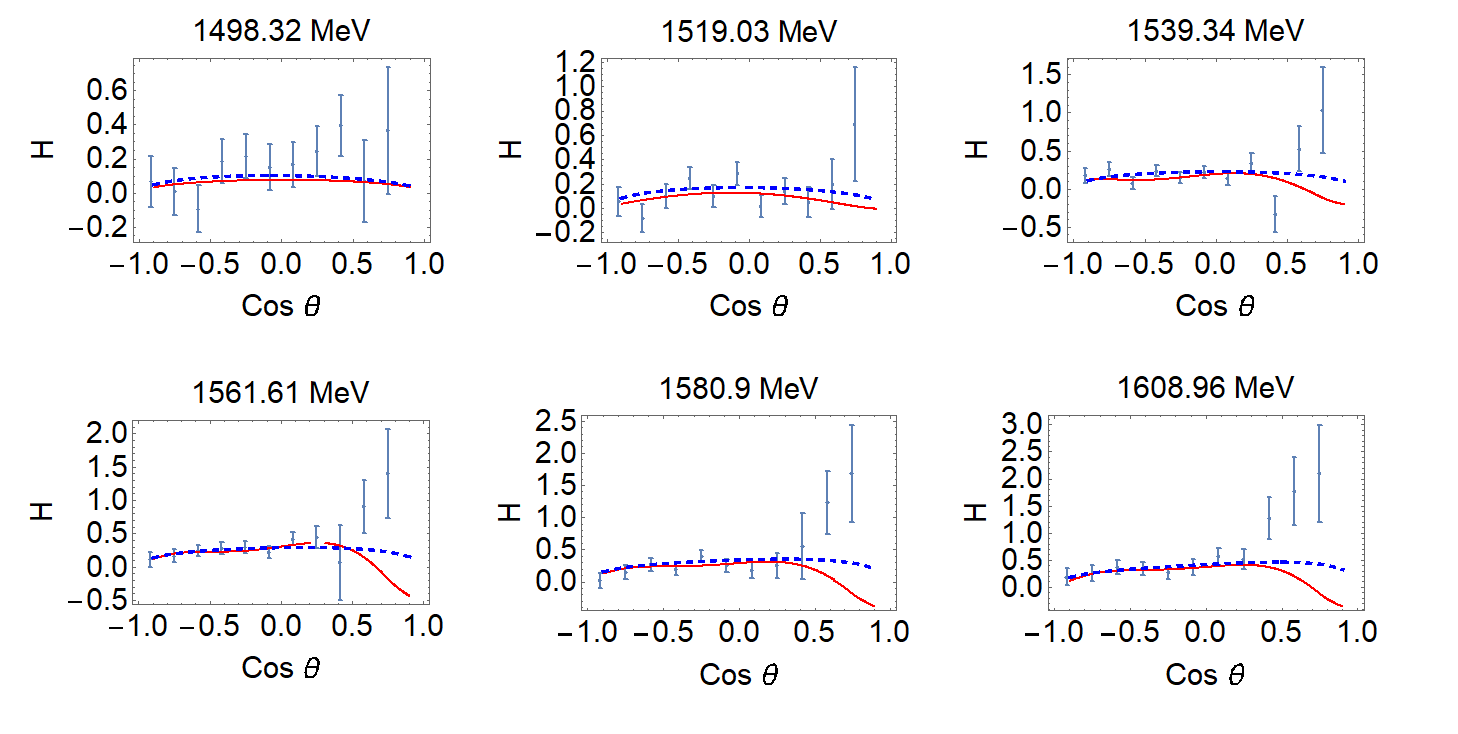}
\caption{\label{Sol1:H}(Color online) Comparison of experimental data for $H$ (discrete symbols) with results of Sol 1 (red full line) and BG2014-2 fit (blue dashed line) at measured energies.    }
\ec
\end{figure*}

 So, we obtained almost identical fits of all observables in the present data base (indistinguishable when plotted, but different below drawing precision when  {a} detailed comparison of numbers is made) with two visibly different sets of multipoles!
\newpage
If the data base  {were more complete},  {the} two sets of  {$\chi_{\text{data}}^2/ \text{ndf}$} would be different {between $Sol \, 1$ and $Sol \, 2$}, and the we could refine $Step \, 1$ to include {the} phase fit as well, very similar as it has been done in fixed-$t$ analyisis\footnote{We could fit the theoretical {BG2014-02} phases with 2D Pietarinen expansion with at  least $N=20$ terms, and then make a global, energy dependent fit of all observables fixing the absolute values of reaction amplitudes to the values of the present fit with only four observables, and  using  {the} Pietarinen expansion coefficients as fitting parameters for improving phases. Then we would go to $Step \, 2$ with improved phases which are connected to BG 2014-2 values only by taking them as initial values.}.  So, improving the precision of existing experiments and {measuring additional observables} to get missing phases is definitely needed to distinguish between  {the} present solutions. On the basis of physics arguments we definitely claim that $Sol \, 1$ is much more favourable, but solely on the basis of measured data $Sol \, 2$ cannot be excluded.
\\ \\ \noindent
However, one thing is important:  {the proposed analysis-scheme} is good enough to  {accomplish the given task} staying only in {a fixed-$W$ representation}.
\\ \\ \noindent
Furthermore, from Figs.~\ref{Multipoles:Sol1}   and  \ref{Multipoles:Sol2} we see:
\begin{itemize}
\item The obtained multipoles are fairly smooth and do not significantly deviate from {the BG2014-02} predictions in the sense that there is no qualitative difference between the two sets of multipoles. They are of the same sign, they have similar shape, they have comparable structure. However, one  sees that both solutions \emph{Sol 1} and \emph{Sol 2}  have notably more structure than the  {energy-dependent BG2014-02} model, and that is to be expected as BG2014-2 is a multi-channel model, and does not ideally fit the data in one particular channel.
\item One does see some apparent discontinuities at certain energies in certain multipoles (i.e. “jump” in  {${\rm Im} E_{2+}$}  at 1687 MeV), but this is the result of  {inconsistencies of the data}, and not of the inability of the proposed {analysis-scheme} to enforce continuity.  Namely, as it has been shown in  {a} former publication~\cite{Svarc2018}, forcing the phase to be {a} continuous function is always resulting with continuity for the complete set of observables measured with sufficient precision ( {in reference~\cite{Svarc2018}, this has been shown} for pseudo-data with infinite precision).  So, if sudden discontinuities appear, they should be solely attributed to the inconsistency in  {the} data itself.
\item  As the low-energy behavior of  {the} multipoles is constrained by the penalty function technique to the $q^{L}$ behaviour, some low energy structure in  {the} multipoles (mostly structures below 1550 MeV) may result from this effect. However, it is clearly visible that  {low-energy} structures are more pronounced for the \emph{Sol 2} which is obtained with {the smoothed nonphysical set of phases}.  This should and will be discussed at length in future research when the pole structure will be analyzed using  {the} Laurent+Pietarinen formalism~\cite{Svarc2013}.
\item We see that both sets of multipoles corresponding to different  {choices} of interpolating techniques (Set 1 –light gray and Set-2 red) are in fair agreement.
\end{itemize}
From Fig.~\ref{Chi2:Sol1} we see:
\begin{itemize}
\item  {The values of} $\chi_{\text{data}}^2/\text{ndf}$  and $\chi_{\text{data}}^2/N_{\text{data}}$  are extremely good  but  notably non-uniform throughout the analyzed energy range. This indicates  {certainly inconsistencies} in data set as it will be discussed later.
\item   {The distributions of $\chi_{\text{data}}^2/N_{\text{data}}$-values} for particular observables notably differ.
     \begin{itemize}
           \item[*] It is uniform and very good (close to 0.5) in the complete energy range for $\sigma_0$.
           \item[*] It is very good  and close to 0.5 in most of the energy range for  {$\Sigma$, $T$, $F$ and $G$}, but each of {the} observables show energy ranges where this quantity suddenly increases:
               \begin{itemize}
                   \item[$\circ$] For  {$F$} it  {rises} from  {an} average value in  {the} ranges 1600-1650 MeV and 1750-1840 MeV; much more for the second range.
                   \item[$\circ$] For  {$T$} it jumps only slightly at lower and higher energies.
                   \item[$\circ$] For  {$G$} it also jumps in the ranges 1600-1650 Mev  and 1750-1840 MeV.
                \end{itemize}
            \item[*] It is uniform in the whole energy range,  but somewhat worse than typical  for  {$E$}.
            \item[*] It is somewhat worse for  {$P$} in the available energy range 1500-1650 MeV.
            \item[*] It notably worse for  {$H$} in the complete measured energy range 1500-1650 MeV.
      \end{itemize}
\item The quantities in  {these figures} indicate that there exist certain  {inconsistencies} among measured data in certain energy ranges. In particular,  {$H$} seems to deviate in the complete measured range and  {$F$} seem to be problematic at higher energies .
\end{itemize}
\noindent
In Figs.~\ref{Sol1:DCS}-\ref{Sol1:H} we  compare the quality of fit of \emph{Sol 1} and  {the theoretical BG014-02} model for all experimental data from Table~\ref{tab:expdata}. We conclude that the quality of fit  {for} \emph{Sol 1} is much better than the one of {BG2014-02}, and this is not surprising as this is a fit, and  {BG2014-02} is  {an energy-dependent} microscopic model. In addition, we have made some tests, and we strongly suspect that the agreement with the data given in Fig.~\ref{Chi2:Sol1} cannot be better even for the free fit. So, this solution is very close to  {the best result one can achieve}.  However, analyzing the details of these figures one can also trace the  {angular- and energy-}ranges which are problematic and have either big  {dissipation} or big uncertainty, and we can  very confidently predict where a particular observable is expected to be.  The need for new measurements is automatically suggested. Immediately, we may recommend that  {$H$ and $F$} should be remeasured towards the end of the measured energy range.  {In addition}, {the} energy range  {of the}  {$P$- and $H$-}observable is much  {smaller}, so we recommend to extend the energy range to at least 1800 MeV.

\section{Summary and Conclusions} \noindent
In this paper we have presented a new data analysis scheme for single-channel pseudoscalar
meson photoproduction. It combines the amplitude analysis CEA/AA of a complete experiment
with the truncated partial wave analysis TPWA of an idealistic case, where all higher partial
waves that cannot be fitted would be completely negligible.
\\ \\ \noindent
The strength of our scheme is its simplicity and minimal reference to any particular
theoretical model. But it is also robust enough that it can always extend the lack of data
by additional theoretical constraints.
\\ \\ \noindent
The possible weaknesses of the  {scheme} are that  {it} requires a lot of
experimental data, and that they should be measured with considerable reliability. The main
opportunities of the {method} are that it enables the direct extraction of resonance parameters
via Laurent-Pietarinen formalism \cite{Svarc2013}, and at the same time gives a direct
possibility to check the consistency of measured data sets. The {scheme} also allows to test
the importance of certain observable to the final result.
\\ \\ \noindent
The  {proposed fit-method yields} a continuous and reliable set of partial waves without
experiencing  {a} strong influence {to} any theoretical model.
\\ \\ \noindent
The new variable  {$P$}, measured by  {the} Bonn group~\cite{Muller:2020plb}, is extremely
important as it helps to pin-down the absolute values of {the} transversity amplitudes in
\emph{Step 1}.
\\ \\ \noindent
The present data set is insufficient to uniquely determine the reaction amplitude phase, so as
an example we  generate  two solutions with almost identical quality of the fit to the data, but
with notably different partial waves. More measurements are needed if one wants to better
specify  {the} pole structure of partial wave solutions.
\\ \\ \noindent
Fitting the relative phase with present data base is futile.
New measurements of well selected observables can improve the analysis a lot. With
them the {analysis scheme} can be extended to include fitting the relative phase too,
so a unique solution could be generated.
\\ \\ \noindent
The method offers the possibility to directly analyze the internal consistency of different data
sets, avoiding the influence of different theoretical models.
\\ \\ \noindent
The separate analysis of $\chi^2/N_{\text{data}}$ for 8 polarization observables in
Fig.~\ref{Chi2:Sol1}  suggest that certain
observables should be remeasured in certain energy ranges, and Figs~\ref{Sol1:DCS}-\ref{Sol1:H}
{imply} the ranges where the consistent data are expected to be.
\\ \\ \noindent
We believe that the central result of our work consists of the fact
that applying CEA/AA in practical data analyses is a very important technique, which should be
employed more and more in the future. The problem of the CEA/AA has been mostly studied as an
isolated mathematical problem in the past, yielding the well-known complete sets of 8
observables. However, the CEA/AA is also quickly applied to real data and it is a numerically
quite well-behaved procedure, due to the fact that it involes only 4 complex numbers for all
energies. The real power of the CEA/AA results emerges once they are combined with the TPWA.
There, they have a great constraining power and make the TPWA, an analysis which is known to
be very badly behaved on its own for the higher
$\ell_{\text{max}}$, a lot more stable. Using the CEA/AA in such a constructive way, we have
been able to derive SE PWA solutions for $\eta$ photoproduction, which have quite controlled
and small discontinuities in their energy dependence, even for the 'small' multipoles
(i.e. all multipoles other than $E_{0+}$).}

\clearpage

\appendix

\section{Photoproduction formalism} \label{sec:PhotoFormalism}

In the following, we collect all aspects of the general photoproduction formalism needed for this work. We consider a $2 \rightarrow 2$-reaction with a spin-$1$ photon and a spin-$\frac{1}{2}$ target nucleon in the initial state and a pseudoscalar meson and a spin-$\frac{1}{2}$ baryon in the final state:
\begin{equation}
\gamma \left( p_{\gamma}; m_{\gamma} \right) + N \left( P_{i}; m_{s_{i}} \right) \longrightarrow \varphi \left( p_{\varphi} \right) + B \left( P_{f};m_{s_{f}} \right) \mathrm{.} \label{eq:ProcessAppMomentaSpins}
\end{equation}
In this expression, the $4$-momenta as well as the variables necessary to label the spin-states have been indicated for each particle. For the reaction of $\eta$ photoproduction studied in this work, the pseudoscalar $\varphi$ is the $\eta$ and the recoil-baryon $B$ is the nucleon $N$. However, other combinations are also possible.

In the following, we collect the customary definitions for the Mandelstam variables $s$, $t$ and $u$. Using $4$-momentum conservation, $p_{\gamma} + P_{i} = p_{\varphi} + P_{f}$, each of these variables can be written in two equivalent forms:
\begin{align}
s &= (p_{\gamma} + P_{i})^{2} = (p_{\varphi} + P_{f})^{2} \mathrm{,} \label{eq:sMandelstam} \\
t &= (p_{\gamma} - p_{\varphi})^{2} = (P_{f} - P_{i})^{2} \mathrm{,} \label{eq:tMandelstam} \\
u &= (p_{\gamma} - P_{f})^{2} = (P_{i} - p_{\varphi})^{2} \mathrm{.} \label{eq:uMandelstam}
\end{align}
Since all particles in the initial- and final state of the reaction~\eqref{eq:ProcessAppMomentaSpins} are assumed to be on the mass-shell, the whole reaction can be described by two independent kinematic invariants. The latter are often chosen to be the pair $(s,t)$.

In case center-of-mass (CMS) coordinates are adopted, the following relations can be established between $(s,t)$ and the center-of-mass energy $W$ and scattering angle $\theta$ of the reaction
\begin{align}
 s &= W^{2}    , \label{eq:sEnergyEquation} \\
 t &= m_{\varphi}^{2} - 2 k \sqrt{m_{\varphi}^{2} + q^{2}} + 2 k q \cos \theta  . \label{eq:tAngleEquation}
\end{align}
Here, $k$ and $q$ are the absolute values of the CMS $3$-momenta for the photon and the meson, respectively. Both of these variables can be expressed in terms of $W$ and the masses of the initial- and final state particles. Therefore, it is seen that the reaction can be described equivalently in terms of $(W, \theta)$. Furthermore, the phase-space factor for the considered $2 \rightarrow 2$-reaction is defined as $\rho = q / k$.

The spins of the particles in the initial- and final-state of photoproduction~\eqref{eq:ProcessAppMomentaSpins} imply a general decomposition for the reaction amplitude. This decomposition has been found by Chew, Goldberger, Low and Nambu (CGLN)~\cite{CGLN} and it reads:
\begin{equation}
\mathcal{F} = \chi_{m_{s_{f}}}^{\dagger} \left( i  \vec{\sigma} \cdot \hat{\epsilon}\; F_{1} + \vec{\sigma} \cdot \hat{q}
  \; \vec{\sigma} \cdot \hat{k} \times \hat{\epsilon}\; F_{2} + i  \vec{\sigma} \cdot \hat{k}\; \hat{q}
  \cdot \hat{\epsilon}\; F_{3} + i \vec{\sigma} \cdot \hat{q} \; \hat{q} \cdot \hat{\epsilon}\; F_{4}  \right) \hspace*{2pt} \chi_{m_{s_{i}}} \mathrm{.} \label{eq:FullAmplitudeCGLN}
\end{equation}
Here, $\hat{k}$ and $\hat{q}$ are normalized CMS 3-momenta, $\hat{\epsilon}$ is the normalized photon polarization-vector and $\chi_{m_{s_{i}}}$, $\chi_{m_{s_{f}}}$ are Pauli-spinors. The complex amplitudes $F_{1}, \ldots, F_{4}$ depend on $(W, \theta)$ and are called CGLN-amplitudes. Once this set of $4$ amplitudes is determined, the full dynamics of the process is known.

The axis of spin-quantization chosen for the initial-state nucleon and the final-state baryon in the decomposition~\eqref{eq:FullAmplitudeCGLN} coincides with the $\hat{z}$-axis in the CMS. However, other choices are also feasible, which then lead to different but equivalent systems composed of $4$ spin-amplitudes. For instance, it is possible to introduce so-called {\it transversity amplitudes} $b_{1}, \ldots, b_{4}$ by rotating the spin-quantization axis to the direction normal to the so-called reaction-plane. The latter is defined as the plane spanned by the CMS $3$-momenta $\vec{k}$ and $\vec{q}$. Using the conventions employed implicitly in the work of Chiang and Tabakin~\cite{Chiang:1996em}, one arrives at the following set of linear and invertible relations between transversity- and CGLN-amplitudes:
\begin{align}
 b_{1} \left( W, \theta\right) &= - b_{3} \left( W, \theta\right)
  - \frac{1}{\sqrt{2}} \sin \theta \left[ F_{3} \left( W, \theta\right) e^{- i \frac{\theta}{2}} + F_{4} \left( W, \theta\right) e^{ i \frac{\theta}{2}} \right] \mathrm{,} \label{eq:b1BasicForm} \\
 b_{2} \left( W, \theta\right) &= - b_{4} \left( W, \theta\right)
  +  \frac{1}{\sqrt{2}} \sin \theta \left[ F_{3} \left( W, \theta\right) e^{i \frac{\theta}{2}} + F_{4} \left( W, \theta\right) e^{- i \frac{\theta}{2}} \right] \mathrm{,} \label{eq:b2BasicForm} \\
 b_{3} \left( W, \theta\right) &= \frac{i}{\sqrt{2}} \left[ F_{1} \left( W, \theta\right) e^{- i \frac{\theta}{2}} -  F_{2} \left( W, \theta\right) e^{ i \frac{\theta}{2}} \right] \mathrm{,} \label{eq:b3BasicForm} \\
 b_{4} \left( W, \theta\right) &= \frac{i}{\sqrt{2}} \left[ F_{1} \left( W, \theta\right) e^{i \frac{\theta}{2}} -  F_{2} \left( W, \theta\right) e^{- i \frac{\theta}{2}} \right] \mathrm{.} \label{eq:b4BasicForm}
\end{align}
The transversity basis greatly simplifies the definitions of polarization observables (see further below) and is therefore generally used as a starting point for the discussion of complete-experiment problems (see appendices~\ref{sec:CEA} and~\ref{sec:TPWA}). Due to these mathematical advantages, this basis is also used in the discussion in the main text (section~\ref{sec:ProposalAndResults}).

In order to extract information on the properties of resonances, one has to analyze partial waves. In this work, we adopt the well-known expansion of the CGLN-amplitudes into electric and magnetic multipoles, which reads~\cite{CGLN,Sandorfi2011}
\begin{align}
F_{1} \left( W, \theta \right) &= \sum \limits_{\ell = 0}^{\infty} \Big\{ \left[ \ell M_{\ell+} \left( W \right) + E_{\ell+} \left( W \right) \right] P_{\ell+1}^{'} \left( \cos \theta \right) \nonumber \\
 & \quad \quad \quad + \left[ \left( \ell+1 \right) M_{\ell-} \left( W \right) + E_{\ell-} \left( W \right) \right] P_{\ell-1}^{'} \left( \cos \theta \right) \Big\} \mathrm{,} \label{eq:MultExpF1} \\
F_{2} \left( W, \theta \right) &= \sum \limits_{\ell = 1}^{\infty} \left[ \left( \ell+1 \right) M_{\ell+} \left( W \right) + \ell M_{\ell-} \left( W \right) \right] P_{\ell}^{'} \left( \cos \theta \right) \mathrm{,} \label{eq:MultExpF2} \\
F_{3} \left( W, \theta \right) &= \sum \limits_{\ell = 1}^{\infty} \Big\{ \left[ E_{\ell+} \left( W \right) - M_{\ell+} \left( W \right) \right] P_{\ell+1}^{''} \left( \cos \theta \right) \nonumber \\
 & \quad \quad \quad + \left[ E_{\ell-} \left( W \right) + M_{\ell-} \left( W \right) \right] P_{\ell-1}^{''} \left( \cos \theta \right) \big\} \mathrm{,} \label{eq:MultExpF3} \\
F_{4} \left( W, \theta \right) &= \sum \limits_{\ell = 2}^{\infty} \left[ M_{\ell+} \left( W \right) - E_{\ell+} \left( W \right) - M_{\ell-} \left( W \right) - E_{\ell-} \left( W \right) \right] P_{\ell}^{''} \left( \cos \theta \right) \mathrm{.} \label{eq:MultExpF4}
\end{align}
The multipoles can be assigned to definite conserved spin-parity quantum numbers $J^{P}$. In particular, resonances with spin $J = \left| \ell \pm \frac{1}{2} \right|$ couple to the multipoles $E_{\ell \pm}$ and $M_{\ell \pm}$.

The multipole expansion of the CGLN-amplitudes $F_{i}$ is formally inverted by the following well-known set of projection-integrals~\cite{Ball:1960baa,YannickPhD}:
\begin{align}
M_{\ell+} &= \frac{1}{2 \left( \ell+1 \right)} \int_{-1}^{1} dx \left[ F_{1} P_{\ell} \left( x \right) - F_{2} P_{\ell+1} \left( x \right) - F_{3} \frac{P_{\ell-1} \left( x \right) - P_{\ell+1} \left( x \right)}{ 2\ell + 1} \right] \mathrm{,} \label{eq:MlplusProjection} \\
E_{\ell+} &= \frac{1}{2 \left( \ell+1 \right)} \int_{-1}^{1} dx \bigg[ F_{1} P_{\ell} \left( x \right) - F_{2} P_{\ell+1} \left( x \right) + \ell F_{3} \frac{P_{\ell-1} \left( x \right) - P_{\ell+1} \left( x \right)}{ 2\ell + 1} \nonumber \\
 & \quad \quad \quad \quad \quad \quad \quad \quad + \left( \ell+1 \right) F_{4} \frac{P_{\ell} \left( x \right) - P_{\ell+2} \left( x \right)}{ 2\ell + 3} \bigg] \mathrm{,} \label{eq:ElplusProjection} \\
M_{\ell-} &= \frac{1}{2 \ell} \int_{-1}^{1} dx \left[ - F_{1} P_{\ell} \left( x \right) + F_{2} P_{\ell-1} \left( x \right) + F_{3} \frac{P_{\ell-1} \left( x \right) - P_{\ell+1} \left( x \right)}{ 2\ell + 1} \right] \mathrm{,} \label{eq:MlminusProjection} \\
E_{\ell-} &= \frac{1}{2 \ell} \int_{-1}^{1} dx \bigg[ F_{1} P_{\ell} \left( x \right) - F_{2} P_{\ell-1} \left( x \right) - \left( \ell+1 \right) F_{3} \frac{P_{\ell-1} \left( x \right) - P_{\ell+1} \left( x \right)}{ 2\ell + 1} \nonumber \\
 & \quad \quad \quad \quad \quad \enspace - \ell F_{4} \frac{P_{\ell-2} \left( x \right) - P_{\ell} \left( x \right)}{ 2\ell - 1} \bigg] \mathrm{.} \label{eq:ElminusProjection}
\end{align}
In these projection-equations, one has $x = \cos \theta$. Polarization observables in pseudoscalar meson photoproduction are generically defined as dimensionless asymmetries among differential cross sections for different beam-, target- and recoil polarization states
\begin{equation}
\Ocal = \frac{\beta \left[ \left( \frac{d \sigma}{d \Omega} \right)^{\left( B_{1}, T_{1}, R_{1} \right)} - \left( \frac{d \sigma}{d \Omega} \right)^{\left( B_{2}, T_{2}, R_{2} \right)} \right]}{\sigma_{0}} \mathrm{.} \label{eq:ObservableDefinitionGeneric}
\end{equation}
The factor $\beta$ has been introduced in the work by Sandorfi
et al.~\cite{Sandorfi2011} for consistency and it takes the value $\beta = \frac{1}{2}$ for observables which involve only beam- and target
polarization and $\beta = 1$ for quantities with recoil polarization. The unpolarized cross section $\sigma_{0}$ always assumes the form of the sum of the two polarization configurations:
\begin{equation}
\enspace \sigma_{0} = \beta \left[ \left( \frac{d \sigma}{d \Omega} \right)^{\left( B_{1}, T_{1}, R_{1} \right)} + \left( \frac{d \sigma}{d \Omega} \right)^{\left( B_{2}, T_{2}, R_{2} \right)} \right] \mathrm{.} \label{eq:UnpolarizedCrossSectionDenominator}
\end{equation}
The dimensioned asymmetry $\sigma_{0} \Ocal$ is often called a {\it profile function} \cite{Chiang:1996em,YannickDiploma,YannickPhD} and it is distinguished by a hat-mark on the $\Ocal$:
\begin{equation}
\hat{\Ocal} = \beta \left[ \left( \frac{d \sigma}{d \Omega} \right)^{\left( B_{1}, T_{1}, R_{1} \right)} - \left( \frac{d \sigma}{d \Omega} \right)^{\left( B_{2}, T_{2}, R_{2} \right)} \right] \mathrm{.} \label{eq:ProfileFunctionDefinition}
\end{equation}
For the photoproduction of a single pseudoscalar meson, there exist in total $16$ polarization observables~\cite{Sandorfi2011}, which include also the unpolarized cross section $\sigma_{0}$ and which can be further divided into the four groups of single-spin observables ($\mathcal{S}$), beam-target- ($\mathcal{BT}$), beam-recoil- ($\mathcal{BR}$) and target-recoil ($\mathcal{TR}$) observables~\cite{Barker:1975bp,Chiang:1996em}. Each group is composed of $4$ observables. When expressed in terms of the transversity amplitudes $b_{i}$, the $16$ observables take the following shape
\begin{equation}
 \hat{\Ocal}^{\alpha} \left( W, \theta \right) = \frac{1}{2} \sum_{i,j = 1}^{4} b_{i}^{\ast} \left( W, \theta \right) \Gamma^{\alpha}_{ij} b_{j} \left( W, \theta \right) , \hspace*{2.5pt} \alpha = 1,\ldots,16 . \label{eq:ObservablesTransversityAlgebraicForm}
\end{equation}
The matrices $\Gamma^{\alpha}$ represent a complete and orthogonal set of $4 \times 4$ Dirac-matrices, which are all by themselves hermitean and unitary. Thus, the observables $\hat{\Ocal}^{\alpha}$ are bilinear hermitean forms in the transversity amplitudes. The matrices have been listed by Chiang and Tabakin~\cite{Chiang:1996em} and they can also be found in the appendices of the works~\cite{YannickDiploma,YannickPhD}. Their algebraic properties imply useful quadratic constraints among the observables $\hat{\Ocal}^{\alpha}$ known as the (generalized) {\it Fierz identities}~\cite{Chiang:1996em}. A listing of the $16$ quantities~\eqref{eq:ObservablesTransversityAlgebraicForm} is expressed in terms of moduli $\left|  b_{i} \right|$ and relative phases $\phi_{ij} := \phi_{i} - \phi_{j}$ of the transversity amplitudes in Table~\ref{tab:PhotoproductionObservables}.

\begin{table}[t]
 \begin{center}
 \begin{tabular}{lr}
 \hline
 \hline
  Observable  &  Group  \\
  \hline
  $\sigma_{0} = \frac{1}{2} \left( \left| b_{1} \right|^{2} + \left| b_{2} \right|^{2} + \left| b_{3} \right|^{2} + \left| b_{4} \right|^{2} \right)$  &     \\
  $\hat{\Sigma} = \frac{1}{2} \left( - \left| b_{1} \right|^{2} - \left| b_{2} \right|^{2} + \left| b_{3} \right|^{2} + \left| b_{4} \right|^{2} \right)$  &   $\mathcal{S}$ \\
  $\hat{T} = \frac{1}{2} \left( \left| b_{1} \right|^{2} - \left| b_{2} \right|^{2} - \left| b_{3} \right|^{2} + \left| b_{4} \right|^{2} \right)$  &     \\
  $\hat{P} = \frac{1}{2} \left( - \left| b_{1} \right|^{2} + \left| b_{2} \right|^{2} - \left| b_{3} \right|^{2} + \left| b_{4} \right|^{2} \right)$  &     \\
  \hline
   $\hat{E}  = \mathrm{Re} \left[ - b_{3}^{\ast} b_{1} - b_{4}^{\ast} b_{2} \right]  = - \left| b_{1} \right| \left| b_{3} \right| \cos \phi_{13} - \left| b_{2} \right| \left| b_{4} \right| \cos \phi_{24}$  &  \\
   $\hat{F} = \mathrm{Im} \left[ b_{3}^{\ast} b_{1} - b_{4}^{\ast} b_{2} \right] = \left| b_{1} \right| \left| b_{3} \right| \sin \phi_{13} - \left| b_{2} \right| \left| b_{4} \right| \sin \phi_{24}  $ &  $\mathcal{BT} $ \\
   $\hat{G} = \mathrm{Im} \left[ - b_{3}^{\ast} b_{1} - b_{4}^{\ast} b_{2} \right] = - \left| b_{1} \right| \left| b_{3} \right| \sin \phi_{13} - \left| b_{2} \right| \left| b_{4} \right| \sin \phi_{24} $  &  \\
   $ \hat{H} = \mathrm{Re} \left[ b_{3}^{\ast} b_{1} - b_{4}^{\ast} b_{2} \right] = \left| b_{1} \right| \left| b_{3} \right| \cos \phi_{13} - \left| b_{2} \right| \left| b_{4} \right| \cos \phi_{24} $  &   \\
   \hline
   $\hat{C}_{x'}  = \mathrm{Im} \left[ - b_{4}^{\ast} b_{1} + b_{3}^{\ast} b_{2} \right]  = - \left| b_{1} \right| \left| b_{4} \right| \sin \phi_{14} + \left| b_{2} \right| \left| b_{3} \right| \sin \phi_{23}  $ &  \\
   $\hat{C}_{z'} = \mathrm{Re} \left[ - b_{4}^{\ast} b_{1} - b_{3}^{\ast} b_{2} \right] = - \left| b_{1} \right| \left| b_{4} \right| \cos \phi_{14} - \left| b_{2} \right| \left| b_{3} \right| \cos \phi_{23} $  &  $\mathcal{BR}$   \\
   $\hat{O}_{x'} = \mathrm{Re} \left[ - b_{4}^{\ast} b_{1} + b_{3}^{\ast} b_{2} \right] = - \left| b_{1} \right| \left| b_{4} \right| \cos \phi_{14} + \left| b_{2} \right| \left| b_{3} \right| \cos \phi_{23} $  &  \\
   $\hat{O}_{z'} = \mathrm{Im} \left[ b_{4}^{\ast} b_{1} + b_{3}^{\ast} b_{2} \right] = \left| b_{1} \right| \left| b_{4} \right| \sin \phi_{14} + \left| b_{2} \right| \left| b_{3} \right| \sin \phi_{23} $  &   \\
   \hline
   $\hat{L}_{x'} = \mathrm{Im} \left[ - b_{2}^{\ast} b_{1} - b_{4}^{\ast} b_{3} \right] = - \left| b_{1} \right| \left| b_{2} \right| \sin \phi_{12} - \left| b_{3} \right| \left| b_{4} \right| \sin \phi_{34}$  &   \\
   $\hat{L}_{z'}  = \mathrm{Re} \left[ - b_{2}^{\ast} b_{1} - b_{4}^{\ast} b_{3} \right]  = - \left| b_{1} \right| \left| b_{2} \right| \cos \phi_{12} - \left| b_{3} \right| \left| b_{4} \right| \cos \phi_{34}$  &  $\mathcal{TR}$  \\
   $\hat{T}_{x'} = \mathrm{Re} \left[ b_{2}^{\ast} b_{1} - b_{4}^{\ast} b_{3} \right] = \left| b_{1} \right| \left| b_{2} \right| \cos \phi_{12} - \left| b_{3} \right| \left| b_{4} \right| \cos \phi_{34}$  &    \\
   $\hat{T}_{z'} = \mathrm{Im} \left[ - b_{2}^{\ast} b_{1} + b_{4}^{\ast} b_{3} \right] = - \left| b_{1} \right| \left| b_{2} \right| \sin \phi_{12} + \left| b_{3} \right| \left| b_{4} \right| \sin \phi_{34}$ &   \\
   \hline
   \hline
 \end{tabular}
 \end{center}
 \caption{The definitions of the $16$ polarization ob\-serva\-bles of pseudoscalar meson photoproduction in terms of transversity amplitudes $b_{i}$ (cf. ref.~\cite{Chiang:1996em}) are collected here. Expressions are given both in terms of moduli and relative phases of the amplitudes and in terms of real- and imaginary parts of bilinear products of amplitudes. Furthermore, the phase-space factor $\rho$ has been suppressed in the given expressions. The four different groups of observables are indicated as well. The sign-conventions for the ob\-serva\-bles are consistent with reference~\cite{YannickPhD}.}
 \label{tab:PhotoproductionObservables}
\end{table}

\clearpage

\section{CEA/AA and TPWA } \label{sec:CEAandTPWAandNewMethod}
\begin{figure}[h]
\begin{overpic}[width=0.99\textwidth,trim=0 95 0 65,clip]%
      {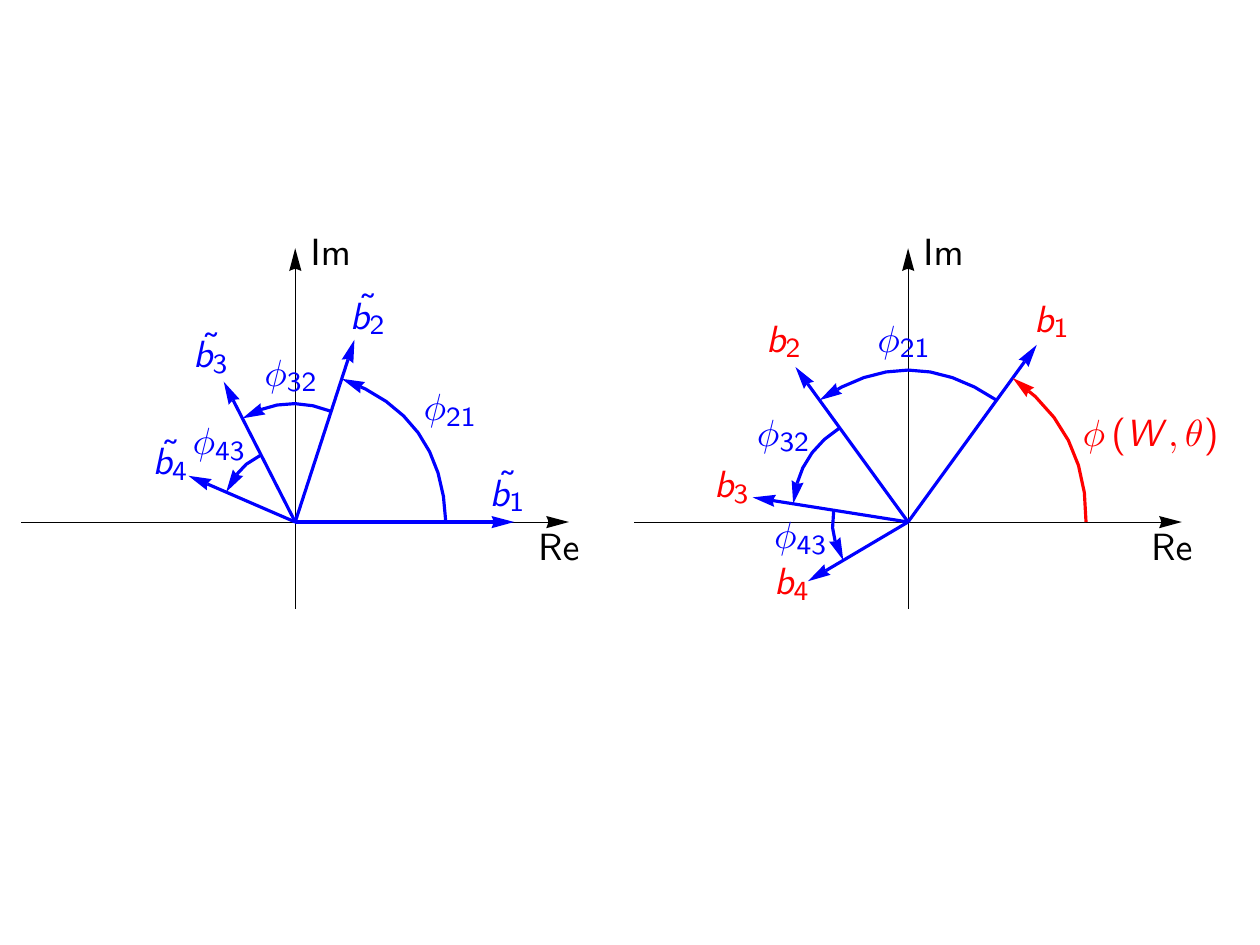}
\end{overpic}
\caption{ (Color online) The photoproduction amplitudes in the transversity-basis are shown as an arrangement of $4$ complex numbers. The schematic is taken over from reference \cite{YannickPhD}. The plots serve to illustrate the possible solutions of the CEA/AA. Left: The reduced amplitudes $ \tilde{b}_{i} $, defined by the phase-contraint~$\mathrm{Im} \left[ \tilde{b}_{1} \right] = 0, \hspace*{2pt} \mathrm{Re} \left[ \tilde{b}_{1} \right] \geq 0$, are plotted. A possible choice of three relative-phase angles is indicated. Right: The \textit{true} solution for the actual transversity amplitudes $ b_{i} $ is shown, and it is obtained from the $ \tilde{b}_{i} $ via a rotation by the overall phase $ \phi \left(W, \theta \right) $ shown in red.}
\label{fig:CompareReducedAmplitudesToActualAmplitudes}
\end{figure}

\noindent
This appendix compiles the definitions and mathematical details of both, the complete-experiment analysis/amplitude analysis (CEA/AA) and the truncated partial-wave analysis (TPWA). Then, both methods are compared and the analysis-method proposed in this work emerges as a synergy of the two.
\\ \\ \noindent
The CEA/AA represents the method to extract the $4$ spin-amplitudes, for instance the transversity amplitudes $b_{1}, \ldots, b_{4}$, from a subset of the $16$ observables collected in Table~\ref{tab:PhotoproductionObservables}. Due to the structure of the expressions~\ref{eq:ObservablesTransversityAlgebraicForm} as sums span over {\it bilinear amplitude products} $b_{i}^{\ast} b_{j}$, the amplitudes can only be extracted uniquely up to one unknown overall phase~\cite{Chiang:1996em,Svarc2018} which is a real function that can depend on the full reaction-kinematics, i.e. on $(W, \theta)$.
The final goal of the analysis is to obtain $4$ amplitudes in the complex plane, with $4$ uniquely defined moduli and $3$ relative-phase angles. This 'rigid' amplitude arrangement is however free to rotate as a full entity in the complex plane with energy and angle dependent phase. See Figure~\ref{fig:CompareReducedAmplitudesToActualAmplitudes} for an illustration.

The choice of variables in terms of which to parameterize the amplitude arrangement is in principle not unique for the CEA/AA. Since one has $4$ complex amplitudes and $1$ unknown overall phase, the number of independent real variables in the choice always has to amount to $8 - 1 = 7$.
Usually, one chooses four moduli of the $b_{i}$ plus three suitably chosen relative-phases, for instance
\begin{equation}
 \left| b_{1} \right| , \hspace*{2pt} \left| b_{2} \right| , \hspace*{2pt} \left| b_{3} \right| , \hspace*{2pt} \left| b_{4} \right| , \hspace*{2pt} \phi_{21} , \hspace*{2pt} \phi_{32} , \hspace*{2pt} \phi_{43}  . \label{eq:CEAModPhaseParameters}
\end{equation}
However, in numeric data analyses, the parametrization in terms of moduli and relative phases can lead to difficulties caused by the logarithmic singularity which enters the procedure once complex exponentials have to be inverted. Alternatively, one can also think about parametrizing the CEA/AA in terms of the {\it phase-rotation functions} $e^{i \phi_{jk}}$, i.e. to use the set of variables
\begin{equation}
 \left| b_{1} \right| , \hspace*{2pt} \left| b_{2} \right| , \hspace*{2pt} \left| b_{3} \right| , \hspace*{2pt} \left| b_{4} \right| , \hspace*{2pt} e^{i \phi_{21}} = \frac{\left| b_{1} \right|}{\left| b_{2} \right|} \frac{b_{2}}{b_{1}} , \hspace*{2pt} e^{i \phi_{32}} = \frac{\left| b_{2} \right|}{\left| b_{3} \right|} \frac{b_{3}}{b_{2}} , \hspace*{2pt} e^{i \phi_{43}} = \frac{\left| b_{3} \right|}{\left| b_{4} \right|} \frac{b_{4}}{b_{3}}  . \label{eq:CEAPhaseRotFunctParameters}
\end{equation}
This removes the difficulty of having to invert exponentials. However, this is bought at the disadvantage of having increased the number of real degrees of freedom artificially, since the functions $e^{i \phi_{jk}}$ have both a real- and an imaginary part. Still, parametrizations in terms of phase-rotation functions are used in the main text (section~\ref{sec:ProposalAndResults}) in order to 'smoothen' phase-information coming from a PWA-model.

\begin{figure}[h!]
\begin{overpic}[width=0.99\textwidth,trim=0 90 0 55,clip]%
      {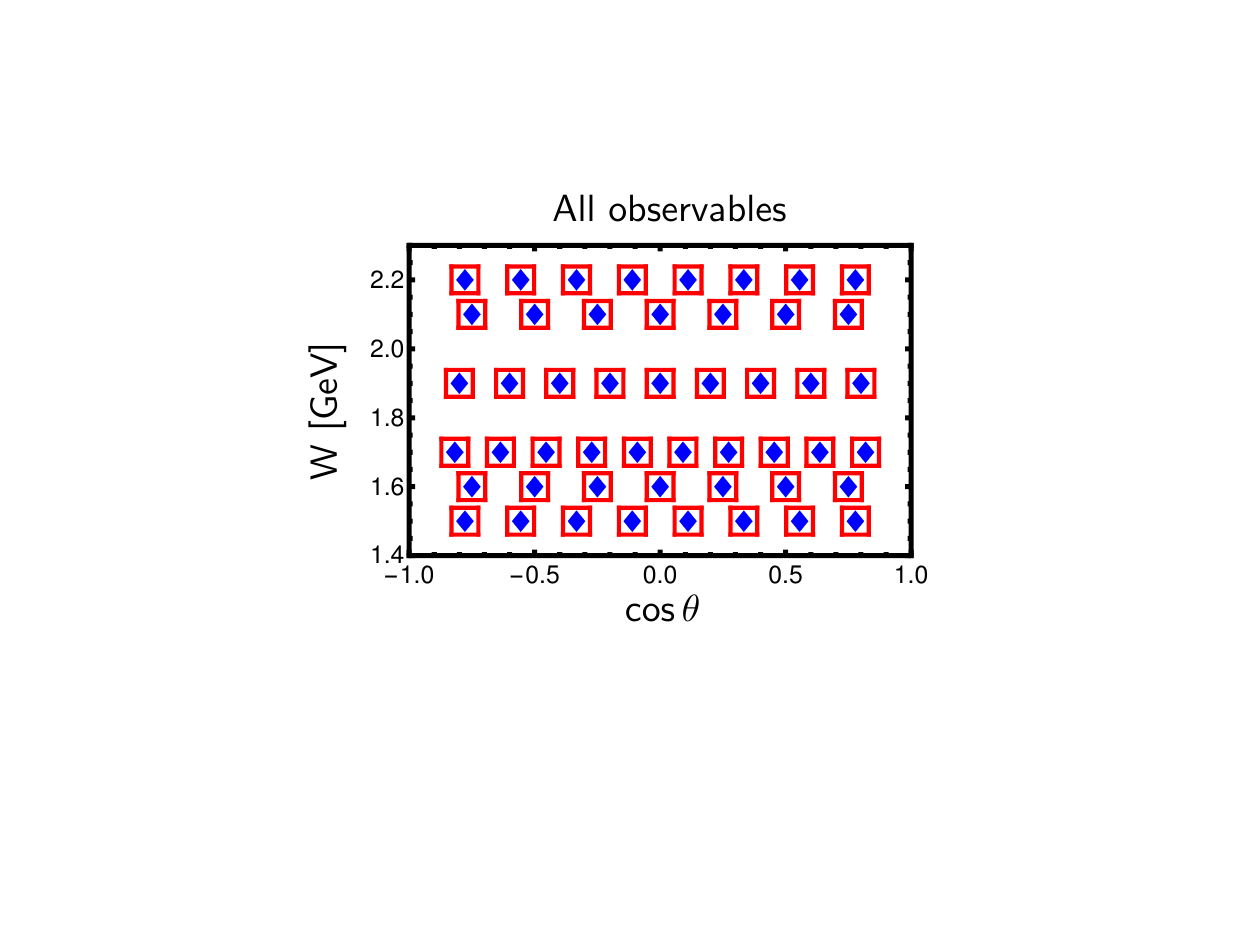}
\end{overpic}
\caption{A schematic illustration for a specific binning of points in phase-space (blue blue polygons) for the CEA/AA. The kinematic binning has to agree for all observables. The CEA/AA then acts on each point individually (illustrated by the red boxes) and therefore all the observables involved in the analysis have to be brought to the same binning. (Color Online)}
\label{fig:CEAPhaseSpacePlot}
\end{figure}

\newpage
Whatever choice one makes to parametrize the amplitudes, the CEA/AA is always a numerical (or algebraic) procedure which takes place at {\it one isolated point} in $\left( W, \theta \right)$ individually. This means that in case one wishes to perform the CEA/AA for a collection of observables over a wider kinematic region, the kinematic binning of all these observables has to be brought to a match over this common region. The situation is illustrated in Figure~\ref{fig:CEAPhaseSpacePlot}.
\\ \\ \noindent
 Consequently, the result of the CEA/AA, i.e. the $7$ variables parametrizing the transversity amplitudes with a fixed overall phase, is also returned as a set of 'discrete data' in complex-space. In other words, the standard CEA/AA without any constraints returns a discrete \emph{but} not necessarily continuous set of points. The direct consequence is that partial waves with physical meaning cannot be extracted~\cite{Workman:2016irf} without further imposing an overall phase provided by a theoretical model.
\\ \\ \noindent
The TPWA denotes the procedure of extracting a finite set of photoproduction multipoles from experimental data by  introducing continuous angular decomposition of amplitudes over Legendre polynomials. In practice, the multipole-expansion defined by equations~\eqref{eq:MultExpF1} to~\eqref{eq:MultExpF4} is truncated at some finite angular-momentum $\ell_{\mathrm{max}}$. Inserting this truncation into the definitions of the $16$ polarization observables shown in Table~\ref{tab:PhotoproductionObservables} yields the mathematical parametrization lying at the heart of the analysis. \\ \\ \noindent
\\ \\ \noindent
The TPWA parametrization can be expressed in a concise form. Choosing to express the emerging angular dependence of the polarization observables $\hat{\Ocal}^{\alpha}$ in terms of associated Legendre polynomials, one arrives at the following form (cf~\cite{Tiator:2011tu,Grushin,Wunderlich:2016imj,YannickPhD}):
\begin{align}
\hat{\Ocal}^{\alpha} \left( W, \theta \right) &= \frac{q}{k} \hspace*{3pt} \sum \limits_{n = \beta_{\alpha}}^{2 \ell_{\mathrm{max}} + \beta_{\alpha} + \gamma_{\alpha}} \left(a_{L}\right)_{n}^{\hat{\Ocal}^{\alpha}} \left( W \right) P^{\beta_{\alpha}}_{n} \left( \cos \theta \right) , \hspace*{2.5pt} \alpha = 1,\ldots,16 \mathrm{,}  \label{eq:LowEAssocLegStandardParametrization1} \\
\left(a_{L}\right)_{n}^{\hat{\Ocal}^{\alpha}} \left( W \right) &= \left< \mathcal{M}_{\ell_{\mathrm{max}}} \left( W \right) \right| \left( \mathcal{C}_{L}\right)_{n}^{\hat{\Ocal}^{\alpha}} \left| \mathcal{M}_{\ell_{\mathrm{max}}} \left( W \right) \right> \mathrm{.} \label{eq:LowEAssocLegStandardParametrization2}
\end{align}

The {\it Legendre coefficients} $\left(a_{L}\right)_{n}^{\hat{\Ocal}^{\alpha}}$ become bilinear hermitean forms defined by a certain set of matrices $\left( \mathcal{C}_{L}\right)_{n}^{\hat{\Ocal}^{\alpha}}$ (such matrices are given explicitly, for the group $\mathcal{S}$- and $\mathcal{BT}$-observables, in the appendix of reference~\cite{YannickPhD}). The multipoles are organized into the $4 \ell_{\mathrm{max}}$-dimensional complex vector $\left| \mathcal{M}_{\ell_{\mathrm{max}}}  \right>$ according to the convention
\begin{equation}
 \left| \mathcal{M}_{\ell_{\mathrm{max}}} \right> = \left[ E_{0+}, E_{1+}, M_{1+}, M_{1-}, E_{2+}, E_{2-}, \ldots , M_{\ell_{\mathrm{max}}-} \right]^{T} \mathrm{.} \label{eq:MultipoleVectorIntro}
\end{equation}
The quantities $\beta_{\alpha}$ and $\gamma_{\alpha}$ in equations~\eqref{eq:LowEAssocLegStandardParametrization1} and~\eqref{eq:LowEAssocLegStandardParametrization2} are constants which define the precise form of the TPWA for each observable. These constants can be found for instance in references~\cite{Wunderlich:2016imj,YannickPhD}.

\begin{figure}[h]
\begin{overpic}[width=0.99\textwidth,trim=0 90 0 55,clip]%
      {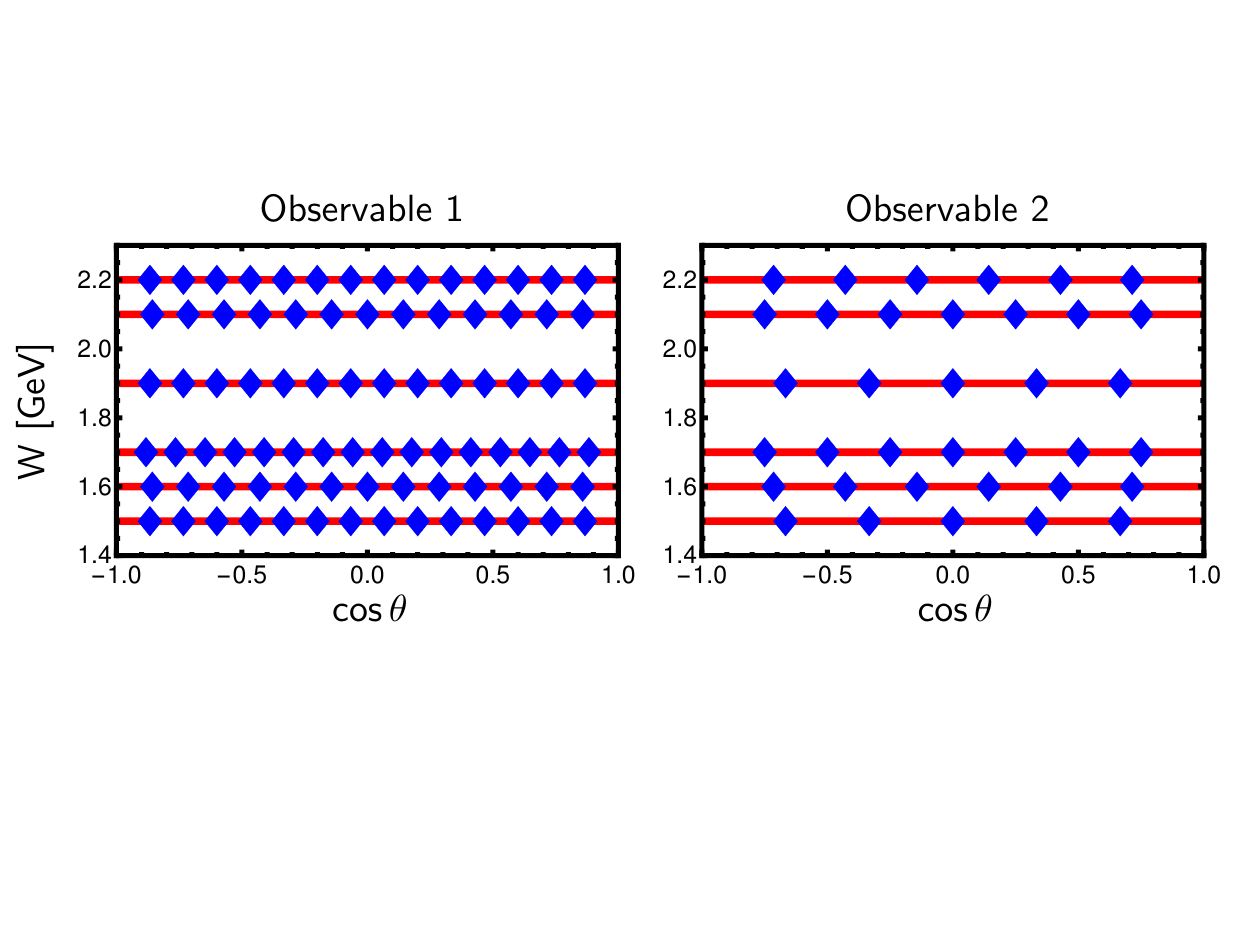}
\end{overpic}
\caption{The plots show schematic illustrations for the kinematic situation in the TPWA. Figures are shown for two different hypothetical observables. Further observables are not shown, but surely present in the TPWA. The kinematic binning for all the datapoints (blue polygons) does not have to fully agree between all observables. However, the energy-binning has to be the same for all datasets. The TPWA (represented by red solid lines) introduces a continuous dependence on the angular variable $\cos \theta$. Compare this to Figure~\ref{fig:CEAPhaseSpacePlot}. (Color Online)}
\label{fig:TPWAPhaseSpacePlots}
\end{figure}

For the TPWA, all observables have to be prepared with a common energy-binning. However, since this method for extracting amplitudes actually parametrizes the angular dependence continuously (cf. equation~\eqref{eq:LowEAssocLegStandardParametrization1}), the angular binnings of the observables can be different. The TPWA then returns a continuous function in angle for each of the discrete energy-bins. However, continuity in energy is another matter, and has been discussed elsewhere~\cite{Svarc2018}. The kinematic situation is illustrated in Figure~\ref{fig:TPWAPhaseSpacePlots}.
\\ \\ \noindent
Note that the CEA/AA and the TPWA \emph{are not} equivalent procedures and will not lead to identical results. This becomes especially apparent once one compares the {\it complete sets of observables}~\cite{Chiang:1996em,Wunderlich:2014xya}, i.e. minimal subsets of all polarization observables which allow for an unambiguous extraction of the complex amplitudes (or multipoles), valid for both analysis-procedures. The differences among and the most important characteristics of the CEA/AA and the TPWA are listed in the following:
\begin{itemize}
 \item \underline{CEA/AA:}
 \begin{itemize}
 \item[i.)] Kinematic regime: the CEA/AA takes place at individual points in the $2$-dimensional space $(W, \theta)$ spanned by the energy $W$ and scattering angle $\theta$.
 \item[ii.)] In the CEA/AA, one has $4$ complex amplitudes while one overall phase $\phi (W, \theta)$ is not known. This results in $8 - 1 = 7$ real independent variables.
 \item[iii.)] A mathematcal complete set for the CEA/AA is given by $8$ carefully chosen observables~\cite{Chiang:1996em,Nakayama:2018yzw}. In addition to the $4$ observables from the group $\mathcal{S}$ (cf. Table~\ref{tab:PhotoproductionObservables}), one has to select $4$ double-polarization observables which must not belong to the same group. This becomes apparent once one considers for instance the $4$ $\mathcal{BT}$-observables listed in Table~\ref{tab:PhotoproductionObservables}: all $4$ observables only contain information on two relative phases, $\phi_{13}$ and $\phi_{24}$. Thus, even in case all $\mathcal{BT}$-observables were measured, at least one connecting relative phase, for instance $\phi_{12}$, remains unknown, which results in a continuous ambiguity. Therefore, at least $2$ observables must be chosen from a third group, e.g. the $\mathcal{BR}$-observables. Further rules for the selection of complete sets can be found in references~\cite{Chiang:1996em,Nakayama:2018yzw}. \\
 Recent studies~\cite{Vrancx:2013pza,Vrancx:2014yja,Nys:2015kqa,Ireland:2010bi} demonstrate the fact that the completeness of the minimal complete sets of $8$ is lost once measurement-errors of realistic sizes are introduced. Then, in order to recover a unique solution for the amplitudes, the considered complete set of $8$ has to be enlarged.
 \item[iv.)] In case a mathematically {\it in}complete set of observables has been selected for the CEA/AA, in most cases this results in only an additional $2$-fold discrete ambiguity (in case the $4$ double-polarization measurements are not taken from the same group).
 \item[v.)] In a world without measurement uncertainties, the CEA/AA yields an exact representation of the photoproduction $\mathcal{T}$-matrix (up to one overall phase). This is accomplished by extracting $4$ complex numbers, independently of the considered energy-region. The phase of one of the $4$ complex numbers has to be constrained, e.g. by demanding this number to be real and positive.
 \end{itemize}
 \item \underline{TPWA:}
 \begin{itemize}
 \item[i.)] Kinematic regime: the TPWA is performed at an individual point in $W$, but over a whole distribution in the angular variable $\theta$ (or $\cos \theta$).
 \item[ii.)] In the TPWA, one has $4 \ell_{\text{max}}$ complex multipoles while one energy-dependent overall phase $\phi (W)$ is not known. This results in $8 \ell_{\text{max}}  - 1$ real independent variables in case the phase $\phi (W)$ is fixed in some way.
 \item[iii.)] A mathematical complete set for the TPWA is given by minimally $4$ observables~\cite{TPWAall,YannickPhD}. However, these complete sets of $4$ can only be found in numerical simulations. On the other hand, an algebraic solution-theory exists for complete sets composed of $5$ carefully chosen observables~\cite{Omelaenko,Grushin,Wunderlich:2014xya,YannickPhD}. The minimal mathematical complete sets mentioned here loose their validity once measurement-errors of realistic sizes are introduced (cf. section~5.5 of reference~\cite{YannickPhD}) and then have to be enlarged in order to facilitate a unique solution.
 \item[iv.)] In case a mathematically {\it in}complete set of observables has been selected for the TPWA, one obtains an exact $2$-fold discrete ambiguity called {\it double ambiguity}~\cite{Omelaenko,Wunderlich:2014xya,YannickPhD}, but also a number of (possible) approximate {\it accidental ambiguities} exists, which scales as $4^{2 \ell_{\text{max}}} - 2$~\cite{Omelaenko,YannickPhD}. These discrete ambiguities can cause severe stability-problems for TPWAs performed with higher truncation orders, i.e. $\ell_{\text{max}} > 2$ (cf. the appendices of reference~\cite{YannickPhD}).
 \item[v.)] In a world without measurement uncertainties, the TPWA already contains an inherent systematic error, due to the fact that it only yields an {\it approximation} of the photoproduction $\mathcal{T}$-matrix for any finite $\ell_{\mathrm{max}}$. For higher energies, one generally has to choose a higher truncation order $\ell_{\mathrm{max}}$, which can result in an increased numerical instability.
 \end{itemize}
 \item[$\Rightarrow$] In this work, we combine both analysis-procedures and use the amplitudes resulting from a CEA/AA in order to resolve the instability-problems of the TPWA, which exist mainly for higher truncation orders.
\end{itemize}


%
 As it is explained in details in the main text (section~\ref{sec:ProposalAndResults}), a CEA/AA with smooth, analytic phases originating from a theoretical model is used as a penalty function in the two-step process in order to ensure the continuity and to increase the stability of the TPWA. In this way, the advantages of both methods have been combined, and a synergy is created which produces a reliable, and very precise description of the data, while at the same time additional theoretical requirements like a good threshold behaviour are obeyed.

\end{document}